\documentclass[fleqn,usenatbib]{mnras}
\usepackage{newtxtext,newtxmath}
\usepackage[T1]{fontenc}
\DeclareRobustCommand{\VAN}[3]{#2}
\let\VANthebibliography\thebibliography
\def\thebibliography{\DeclareRobustCommand{\VAN}[3]{##3}\VANthebibliography}

\usepackage{graphicx}
\usepackage{amsmath}
\usepackage{xspace}

\newcommand{\flamingo}{FLAMINGO\xspace}
\newcommand{\swift}{{\sc{swift}}\xspace}
\newcommand{\hM}[1]{10^{#1}\, \mathrm{M_\odot}}
\newcommand{\M}{M_{\mathrm{500c}}}
\newcommand{\Mm}{M_{\mathrm{200m}}}
\newcommand{\Mdmo}{M_{\mathrm{500c,DMO}}}
\newcommand{\radc}{R_{\mathrm{500c}}}
\newcommand{\radm}{R_{\mathrm{200m}}}
\newcommand{\radcc}{R_{\mathrm{2500c}}}
\newcommand{\ksc}[2]{k\,{#1}\,{#2}\,h\mathrm{\,Mpc^{-1}}}
\newcommand{\Pmm}{P_\mathrm{mm}}

\newcommand{\Pmhi}{P_{\mathrm{mh},i,\Delta}}
\newcommand{\Pmnh}{P_{\mathrm{mnh},\Delta}}
\newcommand{\Pmhtwo}{P_{\mathrm{mh,200m}}}
\newcommand{\Pmhitwo}{P_{\mathrm{mh},i,\mathrm{200m}}}
\newcommand{\Pmhifive}{P_{\mathrm{mh},i,\mathrm{500c}}}

\newcommand{\Pmmp}{P'_\mathrm{mm}}
\newcommand{\Pmhp}{P'_{\mathrm{mh},\Delta}}

\newcommand{\Pmnhp}{P'_{\mathrm{mnh},\Delta}}
\newcommand{\Pmmpp}{P''_\mathrm{mm}}
\newcommand{\Pmhpp}{P''_{\mathrm{mh},\Delta}}

\newcommand{\Pmnhpp}{P''_{\mathrm{mnh},\Delta}}
\newcommand{\Pmhitwop}{P'_{\mathrm{mh},i,\mathrm{200m}}}
\newcommand{\Pmhifivep}{P'_{\mathrm{mh},i,\mathrm{500c}}}
\newcommand{\Pmhitwentyfivep}{P'_{\mathrm{mh},i,\mathrm{2500c}}}
\newcommand{\Pmnhtwop}{P'_{\mathrm{mnh,200m}}}
\newcommand{\Pmhitwopp}{P''_{\mathrm{mh},i,\mathrm{200m}}}
\newcommand{\PmAi}{P_{\mathrm{mA},i}}
\newcommand{\PmAip}{P'_{\mathrm{mA},i}}
\newcommand{\bM}{b_{i,\Delta}}
\newcommand{\bMitwo}{b_{i,\mathrm{200m}}}
\newcommand{\bMifive}{b_{i,\mathrm{500c}}}
\newcommand{\bMitwfive}{b_{i,\mathrm{2500c}}}
\newcommand{\bMcomb}{b_{\mathrm{5x500c}}}
\newcommand{\fM}{f_{\mathrm{M},i,\Delta}}
\newcommand{\fMitwo}{f_{\mathrm{M},i,\mathrm{200m}}}
\newcommand{\fMifive}{f_{\mathrm{M},i,\mathrm{500c}}}
\newcommand{\fMitwfive}{f_{\mathrm{M},i,\mathrm{2500c}}}
\newcommand{\fMitwop}{f'_{\mathrm{M},i,\mathrm{200m}}}
\newcommand{\fMifivep}{f'_{\mathrm{M},i,\mathrm{500c}}}
\newcommand{\fMitwfivep}{f'_{\mathrm{M},i,\mathrm{2500c}}}
\newcommand{\fMp}{f'_{\mathrm{M},i,\Delta}}
\newcommand{\fMcomb}{f_{\mathrm{M},\mathrm{5x500c}}}
\newcommand{\fret}{f_{\mathrm{ret},i,\Delta}}
\newcommand{\frettwo}{f_{\mathrm{ret},i,\mathrm{200m}}}
\newcommand{\fretfive}{f_{\mathrm{ret},i,\mathrm{500c}}}
\newcommand{\fretcomb}{{f_\mathrm{ret,5x500c}}}
\newcommand{\fbi}{\bar{f}_{\mathrm{b},i,\Delta}}
\newcommand{\fbci}{f_{\mathrm{bc},i,\Delta}}

\title[Resummation model applied to FLAMINGO]{The resummation model in FLAMINGO: precisely predicting matter power suppression from observed halo baryon fractions}

\author[M. P. van Daalen et al.]{
Marcel P. van Daalen,$^{1}$\thanks{E-mail: daalen@strw.leidenuniv.nl}
Ioannis Koutalios,$^{1}$
Jeger C. Broxterman,$^{1,2}$ Bart J.H. Wolfs,$^{1}$ John C. Helly,$^{3}$\newauthor
\,Matthieu Schaller$^{2,1}$ and Joop Schaye$^{1}$
\\
$^{1}$Leiden Observatory, Leiden University, PO Box 9513, NL-2300 RA Leiden, The Netherlands\\
$^{2}$Lorentz Institute for Theoretical Physics, Leiden University, PO Box 9506, NL-2300 RA Leiden, The Netherlands\\
$^{3}$Institute for Computational Cosmology, Department of Physics, University of Durham, South Road, Durham, DH1 3LE, UK}

\date{Accepted XXX. Received YYY; in original form ZZZ}

\pubyear{\the\year{}}

\begin{document}
\label{firstpage}
\pagerange{\pageref{firstpage}--\pageref{lastpage}}
\maketitle

\begin{abstract}
In order to derive unbiased cosmological parameters from Stage-IV surveys, we need models that can predict the matter power spectrum for at least $\ksc{\lesssim}{10}$ with percent-level accuracy. The main challenge in this endeavour is that baryonic feedback significantly redistributes matter on large scales, but to an unknown degree. Here, we present an improved version of the ``resummation'' model, which maps observed halo baryon fractions of massive haloes ($\M\gtrsim\hM{12.5}$) to a flexible suppression signal -- i.e.\ the ratio of baryonic to dark-matter-only (DMO) matter power spectra -- using zero free parameters. We calibrate this model to the \flamingo hydrodynamical simulations, obtaining a typical accuracy of $\lesssim 1\%$ for $\ksc{\leq}{10}$ given mean halo baryon fractions within the spherical overdensity radii $\radc$ and $\radm$. When only those within $\radc$ are available, we still obtain $\lesssim 2\%$ accuracy. We show that given small-scale stellar mass fractions, the model can be extended to yield $\lesssim 3\%$ accurate suppression signals for all scales measured ($\ksc{\leq}{25}$). We also extend the model to redshifts $z>0$. Central to the model is a seemingly mass-independent and feedback-independent relation that allows observed halo masses to be mapped to equivalent DMO halo masses using only observed mean halo baryon fractions, to $\lesssim 1\%$ accuracy. This relation can also be used to retrieve the DMO halo mass function from observed halo masses and baryon fractions with percent-level accuracy, without any assumptions on the strength of feedback. A Python package implementing the resummation model is made publicly available.
\end{abstract}




\section{Introduction}
\label{sec:introduction}
One of the main goals of cosmology -- to precisely constrain the parameters that determine the Universe we live in -- is now more within reach than ever, thanks to nascent and upcoming Stage-IV surveys like ESA's Euclid\footnote{\url{https://www.euclid-ec.org/}}, Vera C. Rubin Observatory's LSST\footnote{\url{https://rubinobservatory.org/}}, the Roman wide-field survey\footnote{\url{https://roman.gsfc.nasa.gov/}} and the Simons Observatory \footnote{\url{https://simonsobservatory.org/}}. These surveys will measure matter clustering to unprecedented precision, which can in principle be connected to the cosmology of our Universe. In order to predict measures like the matter power spectrum with $\sim 1\%$ precision for $\ksc{\lesssim}{10}$, as required to extract all available information from these observations, dark-matter-only (DMO) two-point clustering emulators based on large suites of N-body simulations have been developed. These include \texttt{FrankenEmu} \citep{Heitmann2014}, \texttt{EuclidEmulator2} \citep{Euclid2021}, \texttt{BACCO} \citep{Angulo2021}, \texttt{CosmicEmu} \citep{Moran2023} and \texttt{CSST Emulator} \citep{Chen2025}, and are continuously improved to yield higher accuracy, extend to larger wave numbers $k$, and cover more diverse cosmologies.

However, as \citet{vanDaalen2011} and \citet{Semboloni2011} first put forth, the main challenge in correctly interpreting measurements of the clustering of matter and coupling these to a cosmology in an unbiased manner, is that the baryonic feedback processes associated with galaxy formation redistribute matter on large scales. One often-studied measure of this is the ratio of the matter power spectrum of a realistic scenario (i.e.\ one including baryons) to a power spectrum predicted for a DMO Universe with the same cosmology. Since on the most relevant scales for weak lensing surveys ($\ksc{\lesssim}{10}$), baryons smooth the matter distribution and therefore suppress clustering, we will refer to this ratio $P_\mathrm{hydro}(k)/P_\mathrm{DMO}(k)$ as the ``power suppression signal'' -- even though on smaller scales, corresponding to inner haloes and galaxies, baryons greatly enhance the power relative to DMO. While the accuracy, volumes and resolution with which galaxies are modelled in simulations have immensely improved since this systematic bias was discovered, galaxy formation is not a solved problem, and a lot of freedom remains in how processes like feedback from active galactic nuclei (AGN) redistribute gas, how much gas is ejected, and out to what scales.

Fortunately, the wealth of data measuring the distributions of gas or matter as a whole, collected in recent years, has opened up many avenues for constraining the effects of baryons with observations. These now include hot gas fractions and cluster profiles measured in X-ray \citep[e.g.][]{Akino2022,Grandis2024,Ferreira2024,Bahar2024}, the kinetic Sunyaev-Zel'dovich effect \citep[kSZ, e.g.][]{Hadzhiyska2025a}, the thermal Sunyaev-Zel'dovich effect with weak-lensing-calibrated halo masses \citep[tSZ, e.g.][]{To2024,Dalal2025}, and, increasingly, combinations of multiple such probes, often combined with cosmic shear, galaxy-galaxy or CMB lensing \citep[e.g.][]{Troster2022,Schneider2022,Chen2023,Arico2023,Garcia-Garcia2024,Bigwood2024,McCarthy2025,Sunseri2025,Kovac2025,Hadzhiyska2025b}. 
Since very recently, it has even become possible to probe the effects of feedback through the dispersion measure as probed by fast radio bursts \citep[FRBs, e.g.][]{Leung2025,ReischkeHagstotz2025}, as first proposed by \citet{Reischke2023}. All these constraints, certainly when used in combination, have already ruled out a number of models. Despite this, the precise magnitude of baryonic effects in our Universe is still highly uncertain, especially on small scales, and can differ a lot per study, with results depending on the type of observable, the instrument, and the method used to interpret the data. To get the most out of large-scale clustering surveys, we therefore need both more observations and more precise models.

Central to these latter efforts are hydrodynamical simulations, particularly suites that explore a range in feedback strengths and implementations in relevant volumes. These include cosmo-OWLS \citep{LeBrun2014}, BAHAMAS \citep{McCarthy2017}, CAMELS \citep{Villaescusa-Navarro2021}, ANTILLES \citep{Salcido2023}, MillenniumTNG \citep{Hernandez-Aguayo2023}, \flamingo \citep{Schaye2023,Kugel2023}, FABLE and XFABLE \citep{Martin-Alvarez2025,Bigwood2025}. In the context of matter clustering, such simulations can not only be used to quantify the effects of feedback, but also to test modelling and mitigation strategies for the effects of baryons, and, for example, develop measures for the redistribution of gas on large scales \citep[e.g.][]{Gebhardt2024,Medlock2025b,Sharma2025}. As argued by \citet{vanDaalen2020}, one such very informative quantity is the mean halo baryon fraction, particularly on the group scale ($\sim\hM{14}$). This quantity can by itself accurately predict the suppression signal across a wide range of feedback strengths and implementations, at least on scales $\ksc{\lesssim}{1}$. It is the baryon fraction -- i.e.\ the mass in gas and stars within some radius around halo centres relative to the total mass therein -- rather than just the gas fraction, that reveals how much baryonic matter has been driven \emph{out} of the halo, and redistributed to larger scales. After all, a scenario in which massive haloes overcool and lock much of their gas content in stars, and another in which AGN feedback is highly effective, may have similar gas fractions, but will have very different baryon fractions, and very different large-scale matter distributions. Only at fixed stellar fraction, then, are gas fractions a measure of feedback. Observations of halo gas profiles and gas content are therefore best combined with optical and infrared surveys measuring the stellar content of haloes whenever possible, such as done in e.g.\ \citet{Akino2022} and \citet{Dev2024}.

Many modelling efforts and mitigation strategies concerning the effects of baryons on clustering, based on the results of hydrodynamical simulations, currently exist \citep[e.g.][]{Semboloni2011,Semboloni2013,Zentner2013,Huang2021,Salcido2023,Schaller2025,Kammerer2025}, including the popular \textsc{hmcode} \citep{Mead2015,Mead2021}, the highly successful baryonification/baryon-correction method approaches \citep[e.g.][]{SchneiderTeyssier2015,Arico2020,Arico2021,Schneider2019,Schneider2025}, and direct modifications of the power spectral shape \citep[e.g.][]{AmonEfstathiou2022,Preston2023,SchallerSchaye2025}. However, while these are of great use, even when based on the results of hydrodynamical simulations, \emph{all} of these methods need to either assume limited functional forms for density profiles or the suppression signal itself, have free parameters that need to be marginalised over, or both. This often limits their flexibility, i.e.\ their ability to model suppression signals they were not designed for or calibrated to, and/or increases the uncertainties in the derived set of cosmological parameters. The former is an issue as the theory of galaxy formation is still being expanded with new processes that may play an active role in redistributing gas in ways that were not previously modelled. Examples include bipolar jets \citep[e.g.][]{Dave2019,Schaye2023} and cosmic ray feedback, which \citet{QuataertHopkins2025} recently argued may have a particularly large effect on the gas distribution around and beyond the virial radii of groups.

With all of this in mind, it is worthwhile to consider whether the effects that baryons have on haloes and the matter distribution at large scales can be modelled more agnostically, while simultaneously keeping the number of free parameters to a minimum. Recently, \citet{vanLoonvanDaalen2024} investigated the role of haloes of different masses in shaping the matter power spectrum, with and without feedback, building on the work of \citet{Semboloni2011}, \citet{Semboloni2013} and \citet{vanDaalenSchaye2015}. Their findings demonstrated why the baryon fractions of group-size haloes of mass $\M\approx\hM{14}$ is strongly correlated with the suppression of power on large scales: these haloes have about the optimal combination of mass and halo bias to provide the dominant contribution to the total power on scales $\ksc{\lesssim}{2}$ \citep[a result that was further explored by][]{Salcido2023}. Additionally, the baryon fraction in haloes of adjacent mass scales, which provide the second-most dominant contribution, can reasonably be expected to correlate with the baryon fractions of the $\hM{14}$ haloes themselves. This offered an explanation for the strong correlation between power suppression and baryon fractions of massive groups found by \citet{vanDaalen2020}, though it also pointed to some of its limits. Firstly, on smaller scales ($\ksc{\gtrsim}{1}$), contributions of less massive haloes and those from matter \emph{inside} massive haloes all play an increasingly strong role in shaping the power spectrum, and so the correlation is weakened. Secondly, if feedback operates in such a way that the baryon fractions at, say, haloes with mass $\hM{13}$ and $\hM{15}$ are not as tightly correlated with those of $\hM{14}$ haloes, for example because AGN rapidly transition to a different feedback mode as a function of mass, then the correlation with the power suppression breaks down on even larger scales.

Because of this, \citet{vanLoonvanDaalen2024} proposed a new method, dubbed the ``resummation model''. The basis of the model is simple: since we can decompose the matter power spectrum into a sum of haloes cross-correlated with all matter, and a cross-term for the remaining matter not in haloes, knowing how the mass fraction in each component changes should tell us how the total signal changes -- provided the shape of the redistribution beyond haloes is unimportant. In this paper, with the help of the \flamingo simulations, we develop the proof-of-concept version of the resummation method into a fully-fledged model, capable of predicting the power suppression signal to percent-level accuracy on all relevant Fourier scales $k$ given observed halo baryon fractions -- without free parameters, and without assuming a profile for the matter distribution or power spectral shape.

This paper is structured as follows. In Section~\ref{sec:methods}, we introduce the \flamingo simulations and the resummation model, including several improvements and extensions relative to \citet{vanLoonvanDaalen2024}. Then, in Section~\ref{sec:results}, we demonstrate the seemingly-universal relation between the mass ratio of a hydrodynamical halo and its DMO equivalent as a function of its baryon fraction, regardless of its mass, cosmology or feedback implementation. This relation, presented in \S\ref{subsec:fretvsfbc}, is one of our main results. Then, continuing Section~\ref{sec:results}, we test the model's ability to predict the power suppression signal of all \flamingo simulations, given various levels of information about the mean baryon fractions of haloes, showing it achieves percent-level accuracy. Next, in Section~\ref{sec:discussion}, we discuss our findings and the strengths and limitations of the model, predict the suppression signal of our Universe given observed baryon fractions within $\radc$, and show that it can also be used to accurately recover the DMO halo mass function without additional assumptions or parameters, given those same baryon fractions. Finally, we summarise our work in Section~\ref{sec:summary}.

A Python package, \texttt{resummation}, along with the necessary DMO cross spectra and other quantities for the \flamingo simulations, is made publicly available.

\section{Methods}
\label{sec:methods}
\subsection{Simulations}
\label{subsec:sims}
The \flamingo simulations, described in detail in \citet{Schaye2023} and \citet{Kugel2023}, are a suite of large-volume, cosmological hydrodynamical simulations, calibrated to simultaneously reproduce both the $z=0$ galaxy stellar mass function (SMF) and gas fractions ($f_\mathrm{gas}$) of massive groups and clusters in the redshift range $z=0.1-0.3$.\footnote{We note that the gas fractions used to calibrate \flamingo were based on pre-eROSITA X-ray and weak-lensing data, and that more recent X-ray as well as kSZ data suggest lower gas fractions are required. We return to this point in \S\ref{subsec:discussion_obs}.} The simulations were performed with the \swift code \citep{Schaller2024}, using the SPHENIX formulation \citep{Borrow2022} of Smoothed Particle Hydrodynamics (SPH) and the $\delta f$ method for evolving neutrinos of \citet{Elbers2021}. The initial conditions (ICs) were generated using the \textsc{MonofonIC} code \citep{Hahn2021,Elbers2022} using a 3-fluid formalism with a separate transfer function or each of the dark matter, gas, and neutrinos. The ICs used partially fixed modes \citep{AnguloPontzen2016}, setting the amplitudes of modes with $(kL)^2<1025$ to the mean variance, where $L$ is the side length of the simulated volume and $k$ is the wave number of a mode. The multi-fluid approach \citep{Rampf2021} used by \textsc{MonofonIC} allows for a perfect match of the gravity-only and full-hydrodynamics power spectrum on the largest scales.

The subgrid prescriptions of \flamingo include element-by-element radiative cooling and heating \citep{PloeckingerSchaye2020}, star formation \citep{SchayeDallaVecchia2008}, stellar mass loss \citep{Wiersma2009b,Schaye2015}, stellar feedback \citep{Chaikin2022,Chaikin2023}, and the growth of supermassive black holes \citep{SpringelDiMatteoHernquist2005,BoothSchaye2009,Bahe2022}. Feedback from active galactic nuclei (AGN) is implemented either thermally following \citet{BoothSchaye2009}, which is the case for most simulations, or following the kinetic jet-based prescription of \citet{Husko2022}, which is the case for the two simulations with ``Jet'' in their name.

The simulation suite has several properties beneficial to our investigation. Firstly, they vary the parameters governing stellar and AGN feedback in such a way as to reproduce the SMF and/or $f_\mathrm{gas}$ data shifted up or down by $\pm N\sigma$ -- where $\sigma$ is the relevant error on the observational data -- following the procedure detailed in \citet{Kugel2023}. Secondly, they vary the cosmological parameters of the simulations relative to the fiducial ``3x2pt + all external constraints'' Dark Energy Survey year 3 results of \citet{DES2022}. Thirdly, every hydrodynamical \flamingo simulation has a dark-matter-only (DMO) counterpart, with the same cosmology and initial conditions. By considering simulations with different feedback implementations and cosmologies, and comparing their matter power spectra to that of their DMO counterparts, we can investigate a range of baryonic suppression signals, and explore the performance of the resummation model in varying scenarios.

The \flamingo simulations we use, along with their physical and numerical parameters, are listed in Table~2 of \citet{Schaye2023}. Their DMO counterparts are listed in their Table~3, and the cosmologies probed are listed in their Table~4. The fiducial simulation is dubbed ``L1\_m9''. All simulations use periodic volumes $1\,\mathrm{Gpc}$ on a side and $1800^3$ particles per species (plus $1000^3$ neutrino particles), except the flagship simulation which is $2.8\,\mathrm{Gpc}$ on a side while having equal mass resolution ($5040^3$ particles per species, plus $2800^3$ neutrino particles). All other simulations used are variations of L1\_m9, with their names indicating whether they use AGN jets, the shifted datasets they were calibrated to, and/or their cosmologies. Besides the baryonic variations presented in \citet{Schaye2023}, we also include two later additions to the \flamingo suite introduced in \citet{McCarthy2025}. The first is an unphysical run called ``No cooling'' in which no net radiative cooling is allowed (though radiative cooling is allowed to offset any heating), and thus no star formation or feedback processes take place. The second is a run with very strong AGN feedback in the LS8 cosmology, called ``LS8\_fgas$-8\sigma$''.

Finally, we note that we use the terms ``hydrodynamical simulations'' and ``baryonic simulations'' interchangeably to indicate simulations that contain gas and baryon physics in addition to dark matter.

\subsection{Resummation model}
\label{subsec:resum_descr}
The prototypical resummation model for predicting the suppression of the matter power spectrum by baryonic feedback was presented in \citet{vanLoonvanDaalen2024}. In the current work, we extend and improve the model, and apply it to \flamingo. For completeness, we describe the full model here, and note where we deviate from the model description of \citet{vanLoonvanDaalen2024}, which is otherwise reproduced below. On top of this, compared to this previous work we have added some clarifications and summaries.

Let us call the cross power of haloes in mass bin $i$ and all matter, $\Pmhi(k)$. Here we will define haloes as regions with an overdensity $\Delta$ relative to the critical density $\rho_\mathrm{crit}$ or mean density $\rho_\mathrm{mean}$, e.g.\ $\Delta=\mathrm{500,crit}$ or $\Delta=\mathrm{200,mean}$, centred on a peak in the local density field.\footnote{We will abbreviate such overdensities as ``500c'' and ``200m''.} Each halo mass bin will contain some fraction of the total mass in a volume, let us call these fractions $\fM$.

We will refer to the matter-matter auto-power spectrum as $\Pmm$(k). The halo cross terms are assumed to be unnormalised with respect to mass -- the large-scale (linear) halo bias of haloes in mass bin $i$, then, is given by 
\begin{equation}
\label{eq:bias}
\bM=\lim_{k\rightarrow 0}\frac{\Pmhi(k)}{\Pmm(k)}.
\end{equation}
In practice, we will calculate the bias by averaging the power ratio on large scales ($\ksc{\lesssim}{0.13}$), where it is roughly constant (see Appendix~\ref{sec:app_bias}).

We can write the matter auto-power spectrum as a sum over all halo-matter cross terms, plus a cross term between all matter and the matter \emph{not} in haloes. This necessarily includes matter in unresolved haloes (see also Appendix~\ref{sec:app_resolution}). Since haloes are highly biased, this last cross term is expected to be dominated by matter just outside haloes. We will refer to the cross term of all matter with matter not in (resolved) haloes as $P_{\mathrm{mnh},\Delta}(k)$ (note that this term, too, depends on our choice of overdensity region). By definition, then, these cross-power terms must satisfy:
\begin{equation}
\label{eq:powersum}
\Pmm(k)=\Pmnh(k)+\sum_i \fM\Pmhi(k),
\end{equation}
where the sum is over all halo mass bins, and thus gives the total contribution of matter in (resolved) haloes.\footnote{Note that for consistency, we could also have defined the matter--non-halo cross term as $(1-\sum_i \fM)\Pmnh(k)$, but for the sake of conciseness we have absorbed the mass fraction of non-halo matter into $\Pmnh(k)$.}

\begin{figure*}
\includegraphics[width=\columnwidth, trim=17mm 1mm 25mm 12mm]{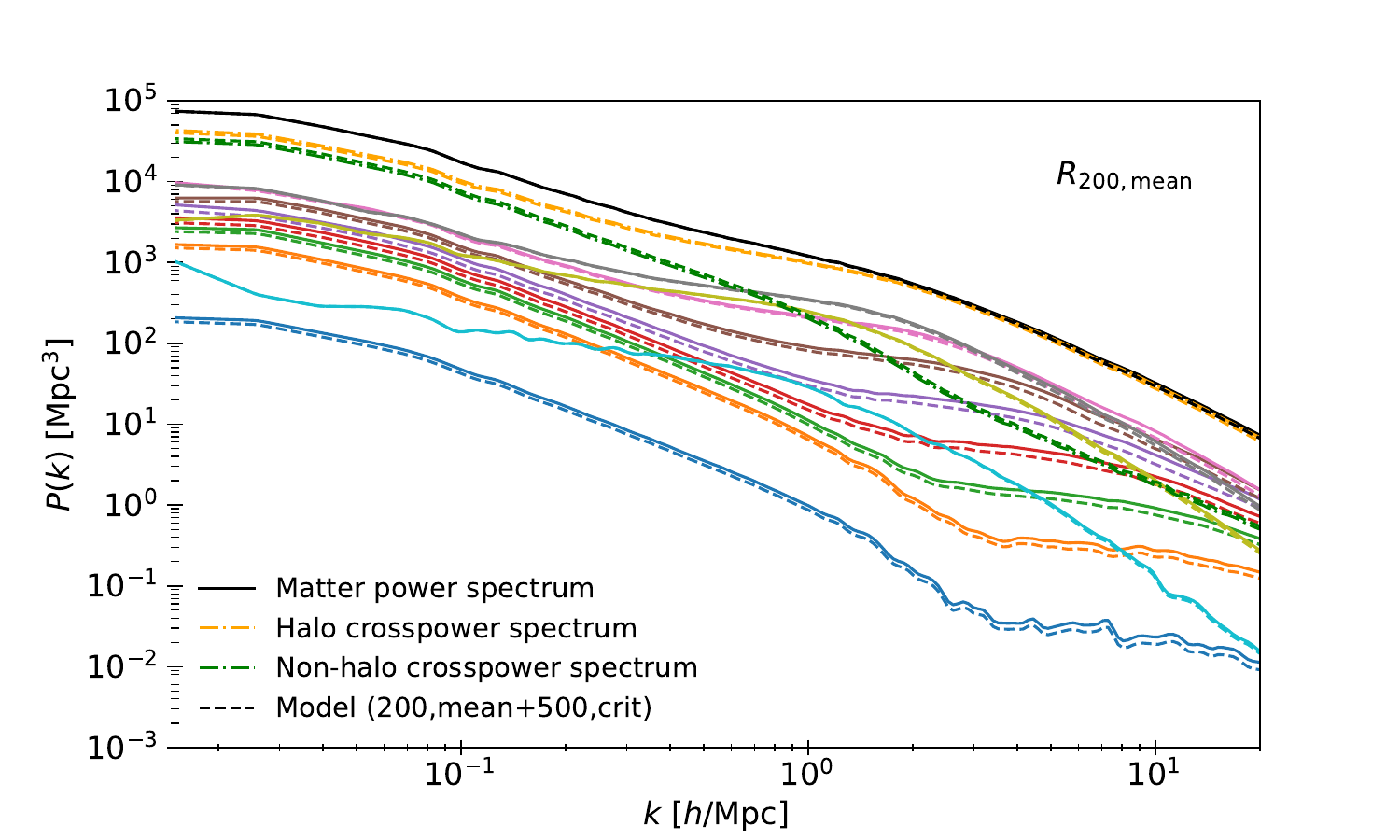}
\includegraphics[width=\columnwidth, trim=8mm 1mm 34mm 12mm]{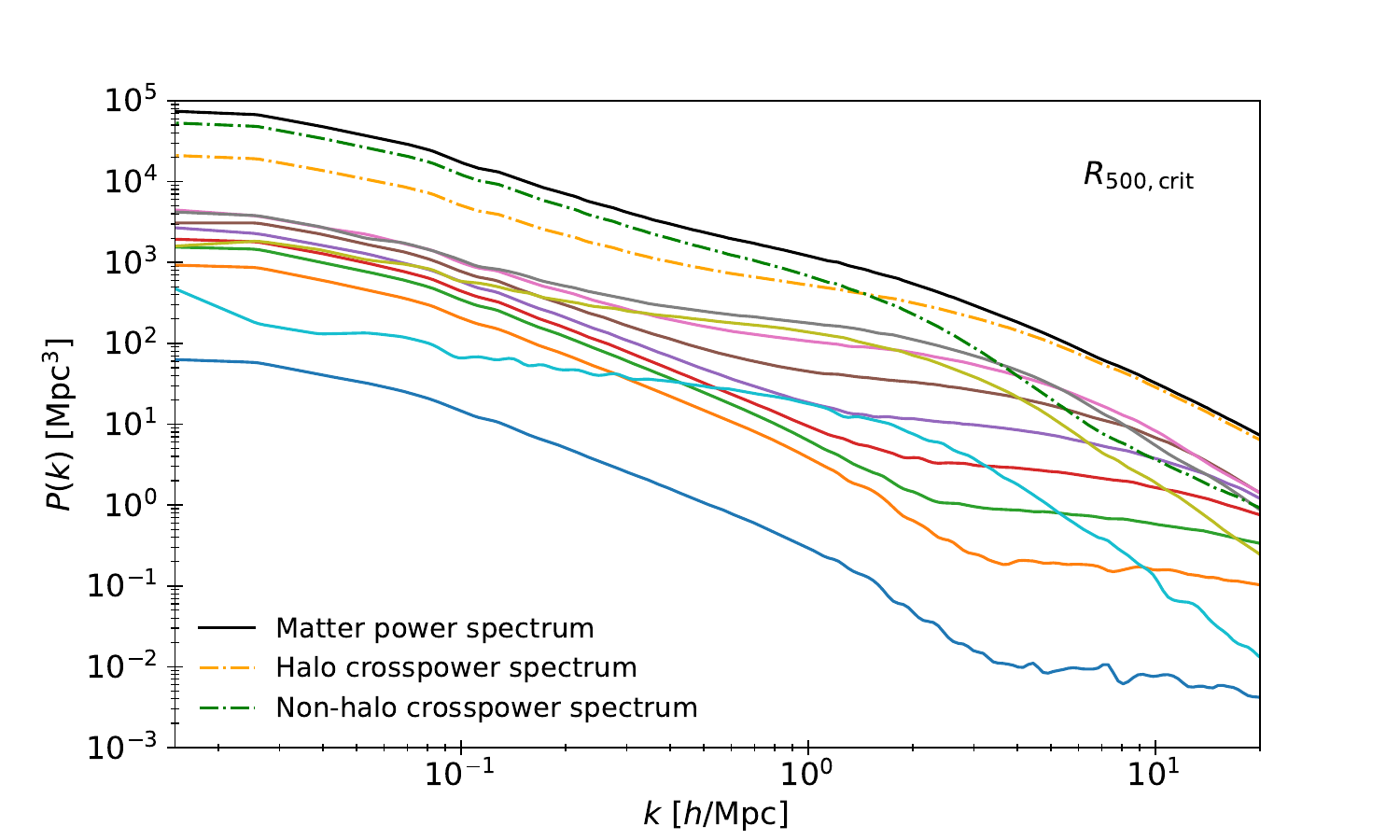}\\
\includegraphics[width=\textwidth, trim=0mm 16mm 0mm 0mm]{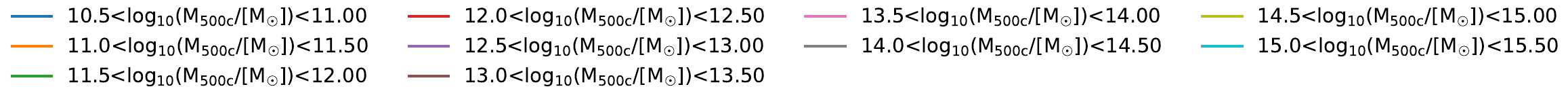}
\caption{The halo-matter cross power components for two different overdensity regions for L1\_m9\_DMO. Here and elsewhere halo mass selection is consistently done based on $\M$. \textit{Left:} The cross power for mass inside overdensity regions with $\Delta=\mathrm{200m}$ with all matter. Also shown are the total halo-matter cross power (dot-dashed orange), the cross power of matter not in (resolved) haloes and all matter (dot-dashed green), and the model predictions for L1\_m9 based on these cross-power components and the measured baryon fractions (dashed lines). \textit{Right:} As on the left-hand side, but for $\Delta=\mathrm{500c}$ and without model predictions. Note that as $\Delta$ increases, the radius decreases, and the total halo contribution becomes smaller while the non-halo contribution becomes larger.}
\label{fig:crosspower}
\end{figure*}

All quantities considered here can be readily calculated from dark-matter-only simulations. Fig.~\ref{fig:crosspower} shows these quantities for L1\_m9\_DMO (see \S\ref{subsec:crosspower}). For $\Delta=\mathrm{200m}$ at $z=0$, we find that only $40.5\%$ of the mass in this simulation resides in resolved haloes, yet due to its higher bias, its contribution to the matter power spectrum dominates over the non-halo contribution on all scales. We find that using $0.5\,\mathrm{dex}$ $\M$ mass bins works well, and therefore use these in what follows, unless noted otherwise. Using smaller, $0.2\,\mathrm{dex}$ mass bins yielded no significant improvement, but we have not experimented with wider bins.

\subsubsection{Accounting for mass loss}
\label{subsec:modelmassloss}
Let us now consider what happens to these different terms as matter is redistributed by the processes associated with galaxy formation. While these processes change halo profiles -- for example by gas cooling to small scales and forming stars, contracting the inner dark matter halo -- the main effect on large scales is caused by removing mass from clustered regions. Therefore, on sufficiently large scales we can approximate the effects of galaxy formation by scaling the mass fractions of haloes by the mass they retained -- that is, the ratio of the mass of a feedback-affected halo and its DMO equivalent.

Let us call the mean retained mass fraction for haloes in mass bin $i$, $\fret$. We can thus write $\fret\equiv M_{\mathrm{hydro},i,\Delta}/M_{\mathrm{DMO},i,\Delta}$. Let us further assume, just for the moment, that the total matter distribution we cross-correlate with is fixed to the DMO distribution, meaning we temporarily rescale the halo component decoupled from the matter that constitutes it. The total contribution to the matter-matter power spectrum of feedback-affected haloes then becomes:
\begin{align}
\nonumber
\Pmhp(k)&\equiv\sum_i \fret\fM\Pmhi(k)\\
\label{eq:halopowerprime}
&\equiv\sum_i \fMp\Pmhi(k),
\end{align}
where the prime indicates a correction for retained mass.

The mass removed from haloes has to go somewhere, in such a way that the total mass in the volume is conserved. One option is to add this mass to a linear power component -- however, the ejected mass is expected to stay around haloes, and therefore to still cluster more strongly than linear. We thus add the mass to the non-halo component, which likely was already dominated by mass around haloes (though we return to this assumption in \S\ref{subsec:resum_z}). Herein the resummation model differs from the standard halo model approach. One wrinkle is that we are removing mass from biased regions -- we therefore need to take bias into account to ensure that the power at low $k$ after modelling galaxy formation still converges to the original value. The ratio of the corrected non-halo cross-contribution to the original non-halo cross-contribution by definition satisfies:
\begin{align}
\label{eq:nonhalopower_k}
\frac{\Pmnhp(k)}{\Pmnh(k)}&=\frac{\Pmmp(k)-\sum_i \fret\fM\Pmhi(k)}{\Pmm(k)-\sum_i \fM\Pmhi(k)}\\
\nonumber
&=\frac{\Pmmp(k)/\Pmm(k)-\sum_i \fMp\Pmhi(k)/\Pmm(k)}{1-\sum_i \fM\Pmhi(k)/\Pmm(k)}.
\end{align}
If we now consider the low-$k$ limit of this expression, and demand that $\lim_{k\rightarrow 0}\Pmmp(k)/\Pmm(k)=1$, we find:
\begin{equation}
\label{eq:nonhalopower_corr}
\frac{\Pmnhp}{\Pmnh}=\frac{1-\sum_i \fret\fM\bM}{1-\sum_i \fM\bM}.
\end{equation}
We thus see that to preserve the power on the largest scale when redistributing mass, we need to not just conserve mass, but also the sum over the product of mass and bias. Our model for the total ``once-corrected'' matter power spectrum is then:
\begin{align}
\nonumber
\Pmmp(k)&=\Pmnhp(k)+\Pmhp(k)\\
\label{eq:modelP}
&=\frac{1-\sum_i \fret\fM\bM}{1-\sum_i \fM\bM}\Pmnh(k)\,\,+\\
\nonumber
&\phantom{=}\,\,\sum_i \fret\fM\Pmhi(k).
\end{align}

Finally, we have to drop our temporary assumption that the total matter distribution that we cross-correlate with is held fixed. Fortunately, at this point we already know how the contributions from both the halo and non-halo matter distributions transform, and therefore how the total matter contribution transforms. Using double primes to indicate a correction for halo mass loss in both matter components that make up the cross power, we find:
\begin{equation}
\label{eq:modelq}
\frac{\Pmmpp}{\Pmm}=\left(\frac{\Pmmp}{\Pmm}\right)^2\equiv q_\Delta^2.
\end{equation}
The fully-corrected halo-matter cross power term then becomes $\Pmhpp=q_\Delta\Pmhp$, and similarly $\Pmnhpp=q_\Delta\Pmnhp$.\\

Summarising what we have presented so far, this model takes dark-matter-only halo mass fractions, linear biases and (cross-)power spectra, and combines them with the mean fraction of mass retained by haloes that have undergone galaxy formation, relative to their dark-matter-only equivalent, to predict the change in the total matter power spectrum. This mean fraction of mass retained, $\fret$, can be calculated as the total mass ratio of (matched) haloes in a hydrodynamical simulation and its DMO equivalent.\footnote{The halo overdensity radii of hydrodynamical and DMO haloes will in general be different -- however, we will demonstrate in \S\ref{subsec:fretvsfbc} that this can be fully accounted for.} Observationally, $\fret$ can in principle be derived from the mean observed baryon fractions of haloes of a certain mass, $\bar{f}_{\mathrm{b},i,\Delta}$, relative to the cosmic baryon fraction $\Omega_\mathrm{b}/\Omega_\mathrm{m}$. Specifically, under the assumption that all matter removed from the halo was baryonic matter, one would find the retained total matter fraction to be:
\begin{equation}
\label{eq:fbc}
\fbci\equiv\frac{1-\Omega_\mathrm{b}/\Omega_\mathrm{m}}{1-\fbi},
\end{equation}
where $\fbi=M_{\mathrm{bar},i,\Delta}/M_{\mathrm{tot},i,\Delta}$ is the mean baryon fraction measured in haloes in mass bin $i$. This can be straightforwardly derived assuming that the ``original'' baryon fraction of each halo is $\Omega_\mathrm{b}/\Omega_\mathrm{m}$, and that the retained cold dark matter (CDM) fraction, equal to $\fret/\fbci$, is unity (see Appendix~\ref{sec:app_deriv} for a derivation).\footnote{Note that whenever we calculate the cosmic baryon fraction, we use a definition of $\Omega_\mathrm{m}$ that includes baryonic and dark matter, but not neutrinos; i.e.\ $\Omega_\mathrm{m}=\Omega_\mathrm{cdm}+\Omega_\mathrm{b}$.}

However, in reality the dark matter will respond gravitationally to the loss of baryonic mass, and additional mass will be lost as the halo relaxes (see also \S\ref{subsec:discussion_fret}). \citet{vanLoonvanDaalen2024} noted that a linear relation between the retained mass fraction and the halo baryon fraction, with a slope above unity, reproduced the results of the cosmo-OWLS \citep{LeBrun2014} and BAHAMAS \citep{McCarthy2017} simulations well. However, the larger volumes of \flamingo allow us to explore this relation with far better statistics, and to larger mass scales. We find that the following functional form is able to accurately fit all \flamingo haloes with masses $M_{\mathrm{h},i,\mathrm{500c}}\gtrsim 10^{12}\,\mathrm{M}_\odot$, for different values of $\Delta$:
\begin{equation}
\label{eq:fretained}
\fret= c-b(1-\fbci)^a.
\end{equation}
Crucially, as we will show in \S\ref{subsec:fretvsfbc}, where we fit this equation to \flamingo, this relation can link the baryon fraction measured at a \emph{retained} total mass (which is what we observe, using lensing) to a retained fraction at a DMO total mass -- hence no shifting of mass bins is necessary in applying this relationship to equations~\eqref{eq:powersum}-\eqref{eq:nonhalopower_corr}.

For lower-mass haloes, $M_{\mathrm{h},i,\mathrm{500c}}\leq \hM{12}$, the scatter in $\fret$ for the simulated haloes is large\footnote{We note that at m9 resolution, a mass of $\hM{12}$ corresponds to fewer than 150 DMO particles.}, and additionally the baryon fraction is more challenging to measure observationally. However, the changes in the contributions of these haloes due to baryonic effects are negligible on the scales explored here (see also \S\ref{subsec:barfrac_response}), and the power suppression is equally well reproduced if we apply the fit to $\fret$ to these masses as well or if we fix the retained fractions for these halo masses to unity. Here, we choose to do the former, but we have checked that our results at $z=0$ are practically indistinguishable when setting $\fret=1$ for low-mass haloes.

\subsubsection{Combining overdensity regions}
\label{subsec:modelannulus}
In principle, we now have all the ingredients to apply this model in practice for a given choice of overdensity region, using the mean baryon fraction measured inside that region for different halo mass bins to rescale and re-sum DMO cross-power spectra. However, if these measurements are available for multiple overdensity regions -- which is now feasible through combinations of X-ray observations, kSZ and/or tSZ measurements, optical/infrared observations and, in the near future, FRBs -- then we can combine this information to model the power spectra more accurately. As we will show, doing so essentially does away with the implicit approximation that the halo profiles are fixed, and greatly improves the results.

Say that we measure the total (retained) mass and baryon fraction of a sample of haloes for both $\Delta=\mathrm{500c}$ and $\Delta=\mathrm{200m}$. We can then calculate the retained fraction for each region with equation~\eqref{eq:fretained}. The net fraction of the ``original'' DMO mass that has been removed from $\radc$, but not from $\radm$, is then $\frettwo-\fretfive$. As shown in \citet{vanLoonvanDaalen2024}, this fraction is quite significant for massive haloes. We will refer to the region $\radc\,<r<\,\radm$ as the halo annulus, and will indicate it with subscript $A$.\footnote{Note that $\radc$ and $\radm$ can be replaced by any other two regions in which the halo baryon fraction was measured, such as $R_\mathrm{200c}$ and $R_\mathrm{50c}$.}

One thing that was not taken into account in \citet{vanLoonvanDaalen2024}, is that conservation of the product of mass and bias does not just apply to mass that is added to the non-halo matter, but generally, to mass moved between any two regions. While we have checked that for the regions we have chosen here the distinction between mass-bias conservation or simply mass conservation leads to negligible differences, as the $\radc$ and $\radm$ regions have a virtually identical bias, we still adapt equations~(10) and (11) from \citet{vanLoonvanDaalen2024} to include the bias correctly. The cross-power contribution of matter in these regions is then given by:
\begin{equation}
\PmAi(k)=\frac{\fMitwo\Pmhitwo(k)-\fMifive\Pmhifive(k)}{\bMitwo\fMitwo-\bMifive\fMifive},
\end{equation}
and the feedback-corrected annulus cross-contribution is then:
\begin{align}
\nonumber
\PmAip(k)&=[\frettwo\bMitwo\fMitwo\,-\\
\label{eq:annulus1}
&\phantom{=}\,\,\fretfive\bMifive\fMifive]\PmAi(k)\\
\nonumber
&\equiv[\fMitwop\bMitwo-\fMifivep\bMifive]\PmAi(k).
\end{align}
When the annulus contribution can be calculated, we replace the regular $\Delta=\mathrm{200m}$ cross power by:
\begin{equation}
\label{eq:annulus3}
\Pmhitwop=\Pmhifivep+\PmAip.
\end{equation}
Note that no cross terms exist between the annulus and the inner region, or between any two regions in the resummation model: by framing the power spectrum as a sum of halo-matter cross terms, all correlations are with the matter distribution as a whole, which is taken as fixed until the final step. As before, the fully-corrected power contribution is then $\Pmhitwopp=q_\mathrm{200m}\Pmhitwop$, with $q_\mathrm{200m}=(\Pmnhtwop+\Pmhitwop)/\Pmm$, which corrects the matter distribution for the changes made to all its components.

\subsubsection{Pushing to smaller scales}
\label{subsec:r2500c_stars}
The resummation model is not limited to including only contributions from within two overdensity radii: any region for which halo baryon fractions can be measured or assumed can be folded into the model prediction. Neither is it limited to only predicting power \emph{suppression}: as the region considered becomes smaller, eventually we reach galactic scales, where we expect the retained mass to exceed unity even for low-mass haloes. In what follows, we will demonstrate how such smaller scales can be included.

Say that we additionally obtain halo gas and stellar fractions (and therefore baryon fractions) within overdensity radii $\radcc$. We can then consider four regions for every halo: $r<\radcc$, $\radcc<r<\radc$, $\radc<r<\radm$ and $\radm<r$. Extending equations \eqref{eq:annulus1}--\eqref{eq:annulus3}, but omitting the explicit dependence on $k$ for conciseness, we can write:
\begin{align}
\nonumber
\Pmhitwop&=\Pmhitwentyfivep+P'_{\mathrm{mA},i,\mathrm{500c-2500c}}+P'_{\mathrm{mA},i,\mathrm{200m-500c}}\\
\nonumber
&=\fMitwfivep P_{\mathrm{mh},i,\mathrm{2500c}}+\\
\nonumber
&\phantom{=}\,\,\frac{\fMifivep\bMifive-\fMitwfivep\bMitwfive}{\fMifive\bMifive-\fMitwfive\bMitwfive}\times\\
\label{eq:annulus4}
&\phantom{=}\,\,[\fMifive P_{\mathrm{mh},i,\mathrm{500c}}-\fMitwfive P_{\mathrm{mh},i,\mathrm{2500c}}]+\\
\nonumber
&\phantom{=}\,\,\frac{\fMitwop\bMitwo-\fMifivep\bMifive}{\fMitwo\bMitwo-\fMifive\bMifive}\times\\
\nonumber
&\phantom{=}\,\,[\fMitwo P_{\mathrm{mh},i,\mathrm{200m}}-\fMifive P_{\mathrm{mh},i,\mathrm{500c}}].
\end{align}
Note that, again, there are no cross terms between these different regions, as we take the cross power between these regions and the full matter distribution, which is held fixed until the final step.

On small scales, the matter power spectrum increases more steeply in baryonic simulations than in dark-matter-only simulations, due to gas cooling into stars, forming galaxies, and the contraction of the inner dark matter halo in response. Going beyond what was done in \citet{vanLoonvanDaalen2024}, we can crudely include the effect of this on larger scales ($\ksc{\sim}{10}$) in the resummation model as follows. First, we determine the cross power spectrum of halo \emph{centres} with all matter, $P_{\mathrm{hc},i}$, in the DMO simulation, where each halo centre is represented by a single particle with mass equal to the $\M$ mass of the halo. We then make the approximation that all stellar mass within a region, e.g.\ $\radcc$, is located at its centre. Under the (implicit) additional assumption that the stellar mass of a halo is roughly proportional to its $\M$ \emph{within a mass bin}, we can then rescale the halo centre power spectrum to the stellar mass contribution to the power spectrum:
\begin{equation}
\label{eq:stellar_contr}
P_{*,i,\Delta}'(k)=f_{*,i,\Delta}\frac{f_{\mathrm{ret},i,\Delta}f_{\mathrm{M},i,\Delta}}{\fMifive}P_{\mathrm{hc},i}(k),
\end{equation}
where $f_{*,i,\Delta}$ is the fraction of the mass in stars relative to the total halo mass.
Choosing $\Delta=\mathrm{2500c}$, the first term in equation~\eqref{eq:annulus4} can be split in a stellar and non-stellar component, and thus becomes:
\begin{equation}
\label{eq:stellarterm}
\Pmhitwentyfivep\rightarrow (1-f_{*,i,\mathrm{2500c}})P_{\mathrm{mh},i,\mathrm{2500c}}'+P_{*,i,\mathrm{2500c}}'.
\end{equation}
Conceptually, the complete model is then a combination of these terms, outward from halo centres: a central stellar component, the non-stellar matter of the central overdensity region, any number of halo annuli, and finally the contribution of matter not in (resolved) haloes.

\subsubsection{Model summary}
\label{subsec:model_summary}
To summarise, these are the steps one goes through in order to apply the resummation model:
\begin{enumerate}
    \item Decide on the overdensity regions to use, e.g.\ depending on the scales on which baryon fractions can be estimated observationally.
    \item Gather DMO halo-matter cross power spectra for a given cosmology for each overdensity region ($\Pmhi(k)$), and calculate the corresponding halo mass fractions ($\fM$) and linear biases ($\bM$, see equation~\ref{eq:bias}).
    \item From the total DMO matter power spectrum and the cross spectra of the largest overdensity region, calculate the cross power of non-halo matter ($\Pmnh$, see equation~\ref{eq:powersum}).
    \item Observe mean halo baryon fractions for each desired overdensity region ($\fbi$), or marginalise over these. Convert these to corrected baryon fractions ($\fbci$) according to equation~\eqref{eq:fbc}, then to retained mass fractions ($\fret$) using equation~\eqref{eq:fretained} with parameters presented in \S\ref{subsec:fretvsfbc}.
    \item Transform the DMO halo cross spectra according to equation~\eqref{eq:halopowerprime}, and use the largest overdensity region to calculate the non-halo cross spectrum with equation~\eqref{eq:nonhalopower_corr}.
    \item If there are multiple overdensity regions, and optionally an inner stellar fraction as well, combine them according to \S\ref{subsec:modelannulus} and \ref{subsec:r2500c_stars}.
    \item Sum all transformed components to obtain $\Pmmp$, and finally square the result relative to the total DMO power spectrum to obtain the suppression signal $\Pmmpp/\Pmm$ (equation~\ref{eq:modelq}).
\end{enumerate}
Note that one key assumption is that matter ejected from haloes clusters roughly as matter already outside DMO haloes, or, alternatively, that the clustering of matter inside haloes is so dominant that the suppression signal is not sensitive to the distribution of ejected matter. We revisit this assumption in \S\ref{subsec:resum_z}. We have also thus far implicitly assumed that the overdensity radius $R_\Delta$ is unchanged by baryonic effects, relative to DMO. As we shall demonstrate in the next section, however, we can choose to calculate retained mass fractions in such a way that $R_\Delta$ is the same for the baryonic haloes and their DMO equivalents by construction.

\begin{figure*}
\includegraphics[width=\columnwidth, trim=17mm 0mm 25mm 12mm]{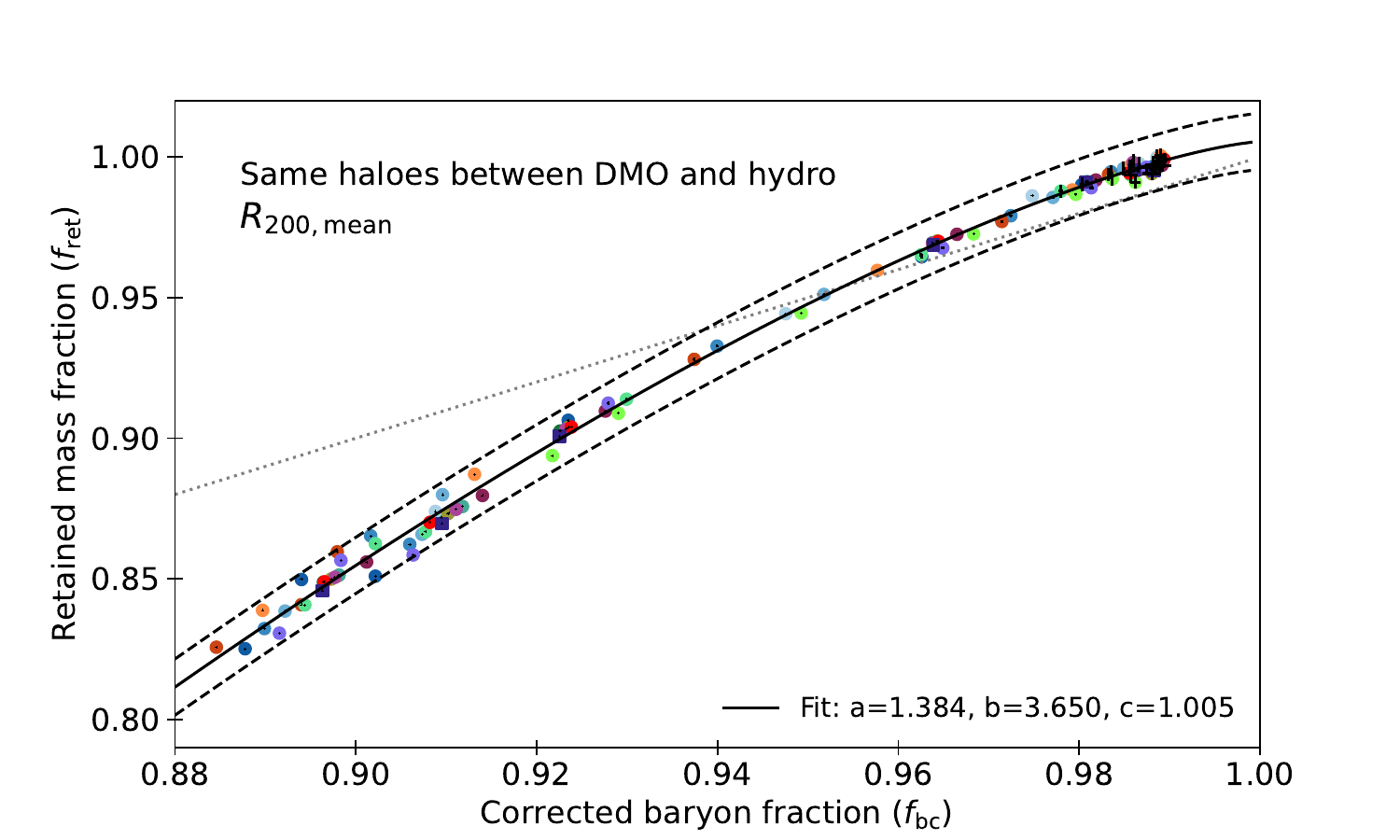}
\includegraphics[width=\columnwidth, trim=8mm 0mm 34mm 12mm]{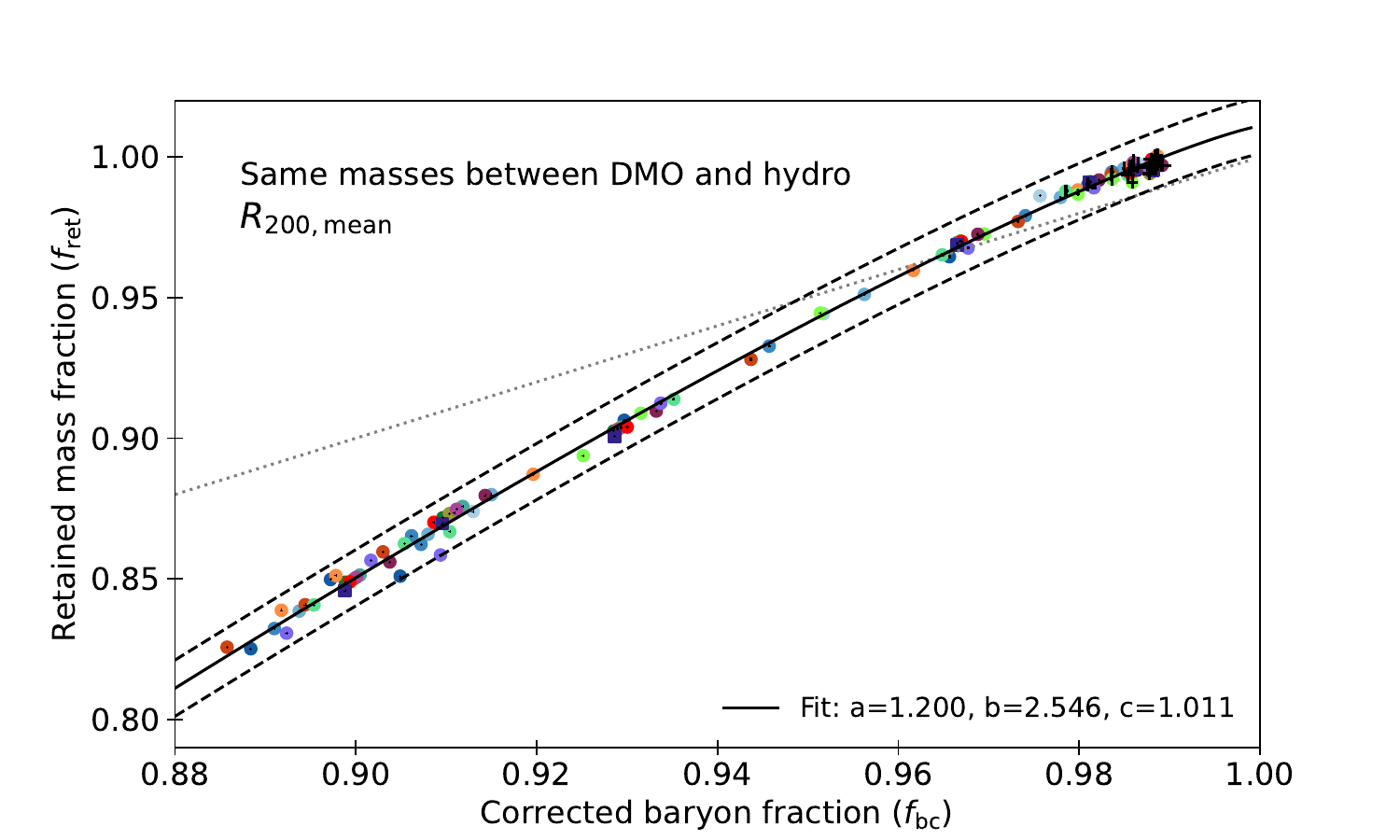}\\
\includegraphics[width=\textwidth, trim=8mm 16mm 0mm 0mm]{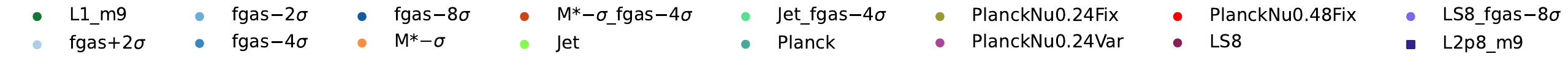}
\caption{The retained mass fraction, $f_\mathrm{ret}=M_\mathrm{tot,bar}/M_\mathrm{tot,DMO}$, as a function of the corrected baryon fraction, $f_\mathrm{bc}=(1-\Omega_\mathrm{b}/\Omega_\mathrm{m})/(1-\bar{f}_\mathrm{b})$, for masses measured within $\radm$ in the \flamingo simulations at $z=0$. Every point represents the average in a $0.5\,\mathrm{dex}$ halo mass bin for all halo masses $\Mdmo\geq\hM{12}$. Different colours represent simulations with different cosmologies and/or feedback. Dotted lines show $f_\mathrm{ret}=f_\mathrm{bc}$, which is the expected retained mass fraction inside a fixed radius if only baryonic mass is removed. The solid lines show the best fit of equation~\eqref{eq:fretained} to all simulations simultaneously. \emph{All} halo average retained mass fractions for \emph{all} simulations lie within $1$ percentage point of this line (indicated by dashed lines).
\textit{Left:} Haloes are matched between the hydro simulation and DMO; both $f_\mathrm{bc}$ and $f_\mathrm{ret}$ are measured at a DMO halo mass $M_i$. \textit{Right:} The measured $f_\mathrm{bc}$ for hydro halo mass $M_i$ translates to an $f_\mathrm{ret}$ for DMO halo mass $M_i$. This version of the relation is most readily applicable to observations.}
\label{fig:fret_fbc_200m_fixedhaloes_fixedmass}
\end{figure*}

A Python package called \texttt{resummation} implementing these steps, along with the necessary DMO cross spectra and other quantities for the \flamingo simulations, is publicly available (see Data Availability). The code calculates power suppression predictions in $<1\,\mathrm{ms}$ for a given cosmology, allowing for fast iterations to marginalise over input halo baryon fractions.

\section{Results}
\label{sec:results}
\subsection{Retained mass vs baryon fraction}
\label{subsec:fretvsfbc}
In order to rescale the dark-matter-only cross spectra to mimic the effects of baryons, we need to know the retained mass fractions $\fret$ -- that is, the fraction of the overdensity mass $M_\Delta$ in halo mass bin $i$ that is still present after feedback and other baryonic processes, relative to DMO. This quantity is not directly accessible in observations, since we do not know the overdensity mass that the halo would have had in the absence of any baryon physics. However, under the assumption that it is the removal (or, when cooling dominates, addition) of baryons from haloes that causes the overdensity mass to change, both directly and through the resulting relaxation processes, we would expect there to be a correlation between the baryon fraction of a halo and its retained mass fraction, which can be calibrated with simulations.

By matching haloes between each hydrodynamical \flamingo simulation and its DMO equivalent \citep[following the procedure described in \S3.2 of][]{Elbers2025}, we can calculate the mean retained mass fraction in each halo mass bin as the ratio of the overdensity mass in the baryonic simulation and in DMO. The mean baryon fractions in the hydrodynamical haloes, $\fbci$, can then be mapped to these retained mass fractions $\fret$. In practice, the baryon fractions are measured at a total \emph{observed} halo mass, but to apply the resummation model we need to know the retained mass fraction at a given \emph{DMO} mass. This is not an issue, since the retained mass fraction itself gives us exactly the ratio between these masses. However, to eliminate this extra step, we derive the mappings between the observed baryon fraction and both the retained mass fraction for those same haloes (i.e.\ at the DMO mass that those haloes would have had in a DMO universe) and the retained mass fraction for haloes that have a total mass in DMO equal to the observed mass (i.e.\ mapping baryon fractions of haloes with mass $M_{\mathrm{obs},i}$ to retained mass fractions of haloes with mass $M_{\mathrm{DMO},j}=M_{\mathrm{obs},i}$). We use the latter relation for all later results. Since for this version of the relation, the overdensity mass of any observed halo and the DMO halo it is mapped to are the same, their overdensity radii are equal as well, and thus we do not need to assume that the ejection of matter leaves the overdensity radius unchanged in our application of the model.

The results of these mappings are shown in Fig.~\ref{fig:fret_fbc_200m_fixedhaloes_fixedmass}. Note that we present these mappings in terms of the corrected baryon fractions, $f_\mathrm{bc}$ (see equation~\ref{eq:fbc}), which takes out some dependence on the cosmic baryon fraction. The corrected baryon fraction is defined such that it is equal to the retained mass fraction when only baryons are removed from haloes, relative to DMO -- that is, if there is no dark matter response or halo relaxation. This one-to-one relation is shown as a dotted line. Every point represents one $0.5\,\mathrm{dex}$ $\M$ halo mass bin, with minimum mass $\Mdmo=\hM{12}$, up to the highest-mass haloes present in the simulation volume. Every colour represents a different hydrodynamical simulation, following the colour scheme and naming convention of \citet{Schaye2023}.

\begin{figure*}
\includegraphics[width=\columnwidth, trim=17mm 0mm 25mm 12mm]{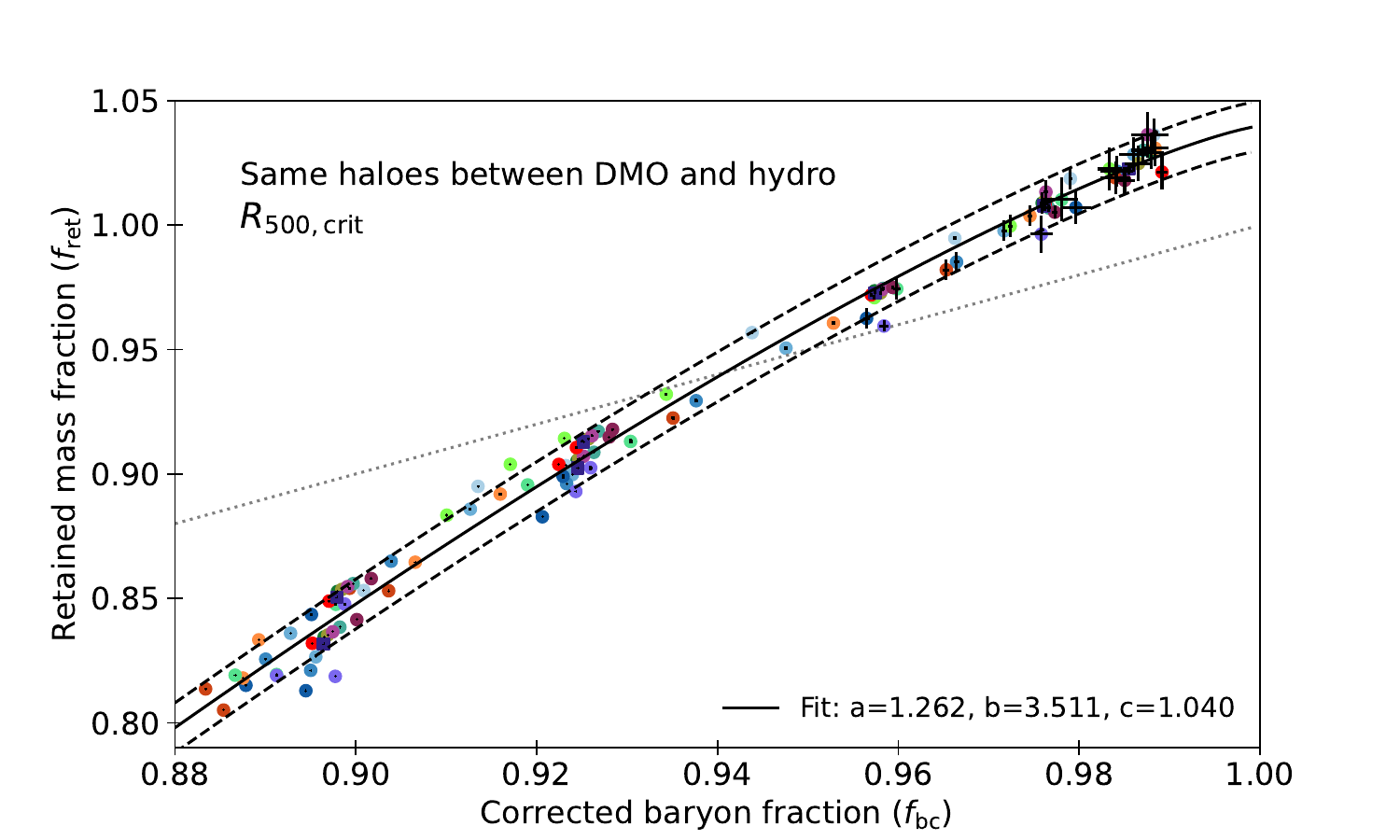}
\includegraphics[width=\columnwidth, trim=8mm 0mm 34mm 12mm]{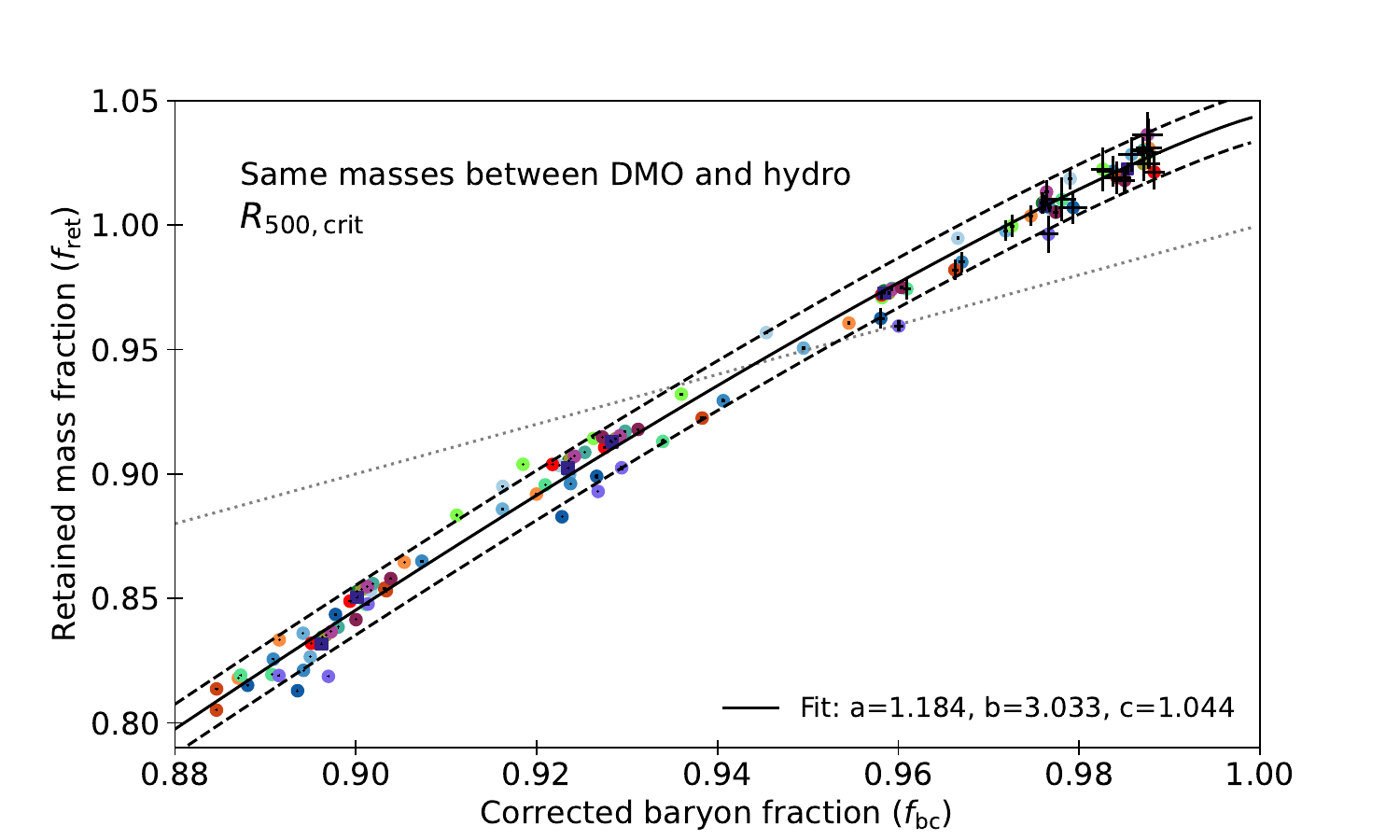}\\
\includegraphics[width=\textwidth, trim=8mm 16mm 0mm 0mm]{figures/fret_vs_fbc_legend.pdf}
\caption{As Fig.~\ref{fig:fret_fbc_200m_fixedhaloes_fixedmass}, but for $\Delta=\mathrm{500c}$. While the scatter increases as we look deeper into the haloes, equation~\eqref{eq:fretained} still describes the retained mass fractions of all relations very well. Note that L2p8\_m9 (square symbols), which has by far the best statistics at equal resolution because it has $(2.8)^3\times$ the volume of the other simulations shown, always falls well within $1$ percentage point of the fit for all halo masses. The most massive haloes have retained mass fractions significantly above unity, meaning that cooling and contraction have increased the mass inside these overdensity regions relative to DMO.}
\label{fig:fret_fbc_500c_fixedhaloes_fixedmass}
\end{figure*}

Strikingly, the correlation between the retained mass fraction and the corrected baryon fraction shown in Fig.~\ref{fig:fret_fbc_200m_fixedhaloes_fixedmass} is very tight. This shows that the retained mass fraction can be predicted accurately from observed baryon fractions, independent of cosmology or feedback strength. This is true regardless of whether we map to the retained mass fractions of the same haloes (left panel) or those of haloes with the same total mass (right panel). The deviations from the one-to-one relation are also clear: as the (corrected) baryon fraction decreases, the retained mass fraction drops further below the dotted line. This is expected: the more mass is removed, the stronger the dark matter response. Additionally, by removing baryons the overdensity radius shrinks, lowering the overdensity mass even without ejecting matter \citep[e.g.][see also \S\ref{subsec:discussion_fret}]{Velliscig2014}. At high baryon fractions, however, the retained mass fractions exceed the dotted line. This implies that for these halo masses, halo contraction due to gas cooling and star formation is more important than mass ejection.

We note that we find that, for the cosmologies explored here, the correlation between the retained mass fraction and the \emph{uncorrected} halo baryon fraction (not shown) is almost as strong as with the corrected one, with simulations LS8\_fgas$-8\sigma$ and Planck$\nu0.48$Fix showing the largest deviation from the other simulations, though still $\leq 1\%$ for $\Delta=200\Omega_\mathrm{m}$. However, this deviation would likely increase for simulations with cosmic baryon fractions, $\Omega_\mathrm{b}/\Omega_\mathrm{m}$, that differ more from those of the fiducial (D3A) cosmology, so transforming to the corrected baryon fraction is expected to yield a tighter relation in general.\footnote{Using $f_\mathrm{bc}$ has the added benefit of allowing one to estimate how much dark matter was lost relative to DMO, by looking at the difference $f_\mathrm{ret}-f_\mathrm{bc}$. However, note that part of this difference is due to the overdensity radius shrinking (see also \S\ref{subsec:discussion_fret}). The latter contribution can be calculated theoretically by assuming a density profile for (the outer part of) the halo.}

The solid black line shows the best fit of equation~\eqref{eq:fretained} to the simulation data, with the fit parameters given in the legend. Dashed lines show the $\pm 1\%$ region around this fit. Notably, \emph{all} of these \flamingo simulations at \emph{all} halo masses $\M\geq\hM{12}$ lie within one percentage point of the fit, regardless of their feedback strength, cosmology, neutrino mass or AGN feedback mode (jet vs thermal), for both mappings.

The highest-mass haloes of each simulation are found in the top right of each panel, where feedback has not been strong enough -- relative to the halo's potential -- to eject much gas beyond $\radm$, or at least not significantly more gas than was gained by inflow \citep[see e.g.][]{MitchellSchaye2022}. As the halo mass decreases, haloes move down the relation to the left. The speed at which they move down the relation depends on the simulation parameters. However, interestingly, when the halo mass becomes low enough for AGN feedback to become ineffective, and supernova feedback begins to dominate ($\M\lesssim\hM{13}$), the points \emph{move back up} the same relation, rather than deviating from it. This shows that this mapping is not specific to AGN feedback, but applies more broadly. We note that we did not cut off the x-axis at an arbitrary value: no halo mass bin in any of the simulations showed $\fbci<0.88$, for any $\Delta$. The scatter between simulations gradually increases with decreasing halo mass, which may be caused by the numerical resolution as low-mass haloes are more poorly resolved (as mentioned previously, the lowest-mass haloes we include here have only $\sim 100$ DM particles).

The mapping between baryon fractions measured within $\radc$ and the corresponding retained mass fraction is shown in Fig.~\ref{fig:fret_fbc_500c_fixedhaloes_fixedmass}. The results are very comparable to those for $\radm$, except that the scatter increases, particularly at lower mass. Additionally, massive haloes with high baryon fractions have retained mass fractions well above unity -- as expected, since gas can cool and gather deeper in the potential well more easily than dark matter.

If we go to even smaller scales, we can still approximately apply the same relation, but the scatter keeps increasing (not shown) -- though we note that this is in part because we select on the $\M$ mass, rather than the mass of the overdensity region itself.

\subsubsection{A universal relation for the retained mass fraction}
\label{subsec:universalfit}
We can fit a relation between $f_\mathrm{ret}$ and $f_\mathrm{bc}$ for any overdensity region $\Delta$ -- however, since all simulations follow the same relation to high precision, we can instead use just the simulation with the largest volume, and therefore the best statistics. The L2p8\_m9 simulation (purple square symbols in Fig.~\ref{fig:fret_fbc_200m_fixedhaloes_fixedmass} and Fig.~\ref{fig:fret_fbc_500c_fixedhaloes_fixedmass}) has the same mass resolution as all other \flamingo simulations shown here, but in a $2.8^3\approx 22$ times larger volume. Using the L2p8\_m9 simulation, we can fit a relation for any overdensity region by applying equation~\eqref{eq:fretained}, but making its parameters dependent on $\Delta$. We find that there is some degeneracy between parameters $a$ and $b$, which we remove by making $b$ a constant. We then replace $a$ and $c$ by the following functions:
\begin{align}
\nonumber
a&=a_0(\Delta_\mathrm{crit}/100)^{a_1},\\
\label{eq:universal_params}
c&=[1+(\Delta_\mathrm{crit}/c_0)^{c_1}]^{c_2},
\end{align}
where $\Delta_\mathrm{crit}$ is the overdensity relative to the critical density of the Universe, e.g.\ $\Delta_\mathrm{crit}=500$ for $\M$ or $\Delta_\mathrm{crit}=200\Omega_\mathrm{m}$ for $\Mm$ at $z=0$.

\begin{figure}
\includegraphics[width=\columnwidth, trim=12mm 16mm 24mm 12mm]{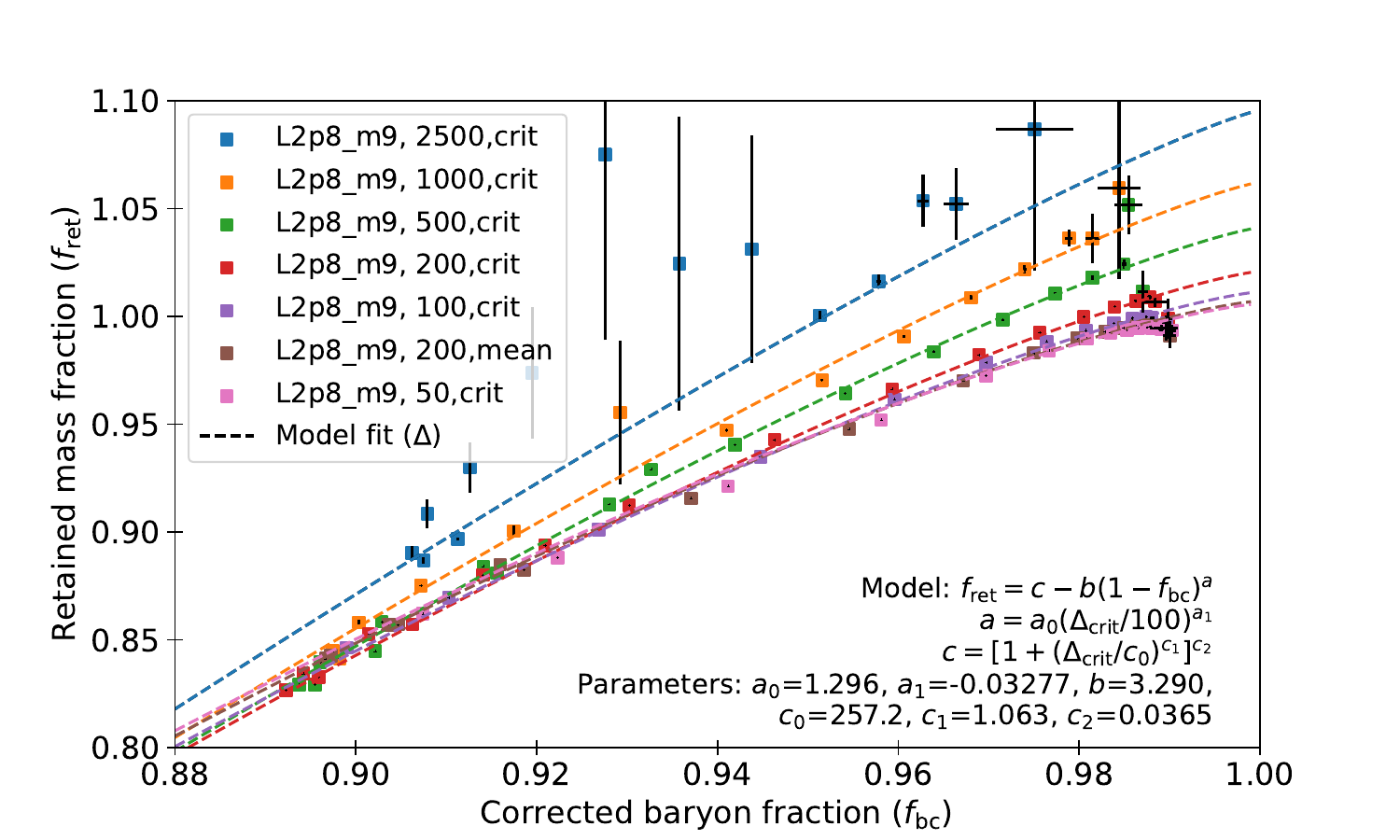}
\caption{Results of fitting the universal function -- equation~\eqref{eq:fretained} combined with equation~\eqref{eq:universal_params} -- to L2p8\_m9 for the corrected baryon fractions and retained mass fractions in seven different overdensity regions, for eighteen $0.2\,\mathrm{dex}$ halo mass bins in the range $\log_{10}(\M/[\mathrm{M}_\odot])=[12.0,15.6]$, at $z=0$. Like in the right-hand panels of Fig.~\ref{fig:fret_fbc_200m_fixedhaloes_fixedmass} and Fig.~\ref{fig:fret_fbc_500c_fixedhaloes_fixedmass}, baryon fractions measured at a hydro halo mass $M$ are mapped to retained fractions at the same DMO halo mass $M$ in the fit. Even though we again see that the scatter increases as we look deeper into haloes, the function reproduces the retained mass fractions on all scales remarkably well for all halo masses probed here.}
\label{fig:fret_fbc_universal}
\end{figure}

The results of fitting the six model parameters $a_0$, $a_1$, $b$, $c_0$, $c_1$ and $c_2$ are shown in Fig.~\ref{fig:fret_fbc_universal}. Here different colours correspond to different overdensity regions, from $R_\mathrm{50c}$ down to $\radcc$. The larger volume allows us to choose smaller mass bins, $0.2\,\mathrm{dex}$, rather than $0.5\,\mathrm{dex}$ as before, in the range $\log_{10}(\M/[\mathrm{M}_\odot])=[12.0,15.6]$ (which includes the highest-mass halo in the simulation). To avoid the large population of poorly-resolved low-mass haloes from dominating the fit, we impose a minimum error on $\fret$ of $0.003$, which are otherwise calculated as the standard deviation on the mean.

Fig.~\ref{fig:fret_fbc_universal} shows that the retained mass fraction changes smoothly with $\Delta$, allowing the same relation to provide an accurate fit to all overdensity regions simultaneously. For large regions, the retained mass seems to converge with decreasing $\Delta$, and the convergence is faster for haloes with low baryon fractions (i.e.\ those where feedback is most effective at removing mass).

Results that use this fit, rather than individual fits for one overdensity region to the collection of all simulations as in Fig.~\ref{fig:fret_fbc_200m_fixedhaloes_fixedmass} and Fig.~\ref{fig:fret_fbc_500c_fixedhaloes_fixedmass}, will be annotated with ``universal fit''.

\subsection{Cross-power model}
\label{subsec:crosspower}
The cross power spectra of the L1\_m9\_DMO simulation, which are used to predict the suppression signal for all \flamingo simulations with the D3A cosmology, are shown in Fig.~\ref{fig:crosspower}. On the left, we show the cross power spectra between matter inside the $\radm$ overdensity regions of different $\M$ mass bins, and all matter; the right panel is similar, but for $\radc$ overdensity regions. Besides the cross spectra for individual halo mass bins, we also show the total cross spectrum of all matter in overdensity regions of (resolved) haloes with all matter (dot-dashed orange line), and of all matter \emph{not} inside these regions with all matter (dot-dashed green line). The sum of these two terms is by construction the total matter power spectrum, shown in black. In the left panel, we also show the resummation model prediction for all components, based on the mean halo baryon fractions of the L1\_m9 simulation, as dashed lines. For clarity, all power spectra have been smoothed by convolving with a Gaussian with a standard deviation equal to one data point (for halo components) or half a data point (for the total spectra).

Comparing the left and right panels, the most important difference is that the cross-power contribution of halo matter is significantly higher when considering $\radm$ regions, compared to $\radc$ regions. As the bias of both regions for the same $\M$ mass is nigh identical (see Appendix~\ref{sec:app_bias}), the difference is simply due to the additional mass located between $\radc$ and $\radm$. Because of this, for $\radm$, the halo power dominates the total power contribution for all $k$, while for $\radc$ it only does so for $\ksc{\gtrsim}{1.5}$. Note that this gives an indication of how sensitive the model outcome is to the assumption of what happens to ejected matter, which in the default version of the model scales the non-halo cross spectrum: for smaller regions, the details of how ejected matter is assumed to contribute to the total spectrum become increasingly important. However, note that when combining multiple overdensity regions, as described in \S\ref{subsec:modelannulus}, it is only the non-halo contribution of the \emph{outer} region that contributes to the final suppression signal. We revisit the non-halo contribution in \S\ref{subsec:resum_z}.

In both panels of Fig.~\ref{fig:crosspower}, we see the same trend with mass: low-mass haloes contribute very little to the total signal, but the contribution grows with mass as both the halo bias and the total mass inside the overdensity do -- until the contribution peaks for $\M\sim\hM{14}$, after which the bias continues to increase but the mass fraction strongly declines. This is in line with the findings of \citet{vanLoonvanDaalen2024} for cosmo-OWLS and BAHAMAS, though we are able to extend this to larger mass scales and with less noise, due to the much larger volume of \flamingo.

Finally, in the left-hand panel we show the outcome of applying the resummation model as dashed lines. The primary effect is that the contribution of almost all halo masses is reduced by the retained mass fraction, derived from the halo baryon fractions measured in L1\_m9, relative to DMO. On large scales, the reduction of the total halo contribution is exactly countered by an increase in the non-halo contribution (dot-dashed green line), by construction. However, the non-halo contribution is subdominant on small and intermediate scales, and therefore the total power there is reduced.

\begin{figure}
\includegraphics[width=\columnwidth, trim=12mm 16mm 24mm 12mm]{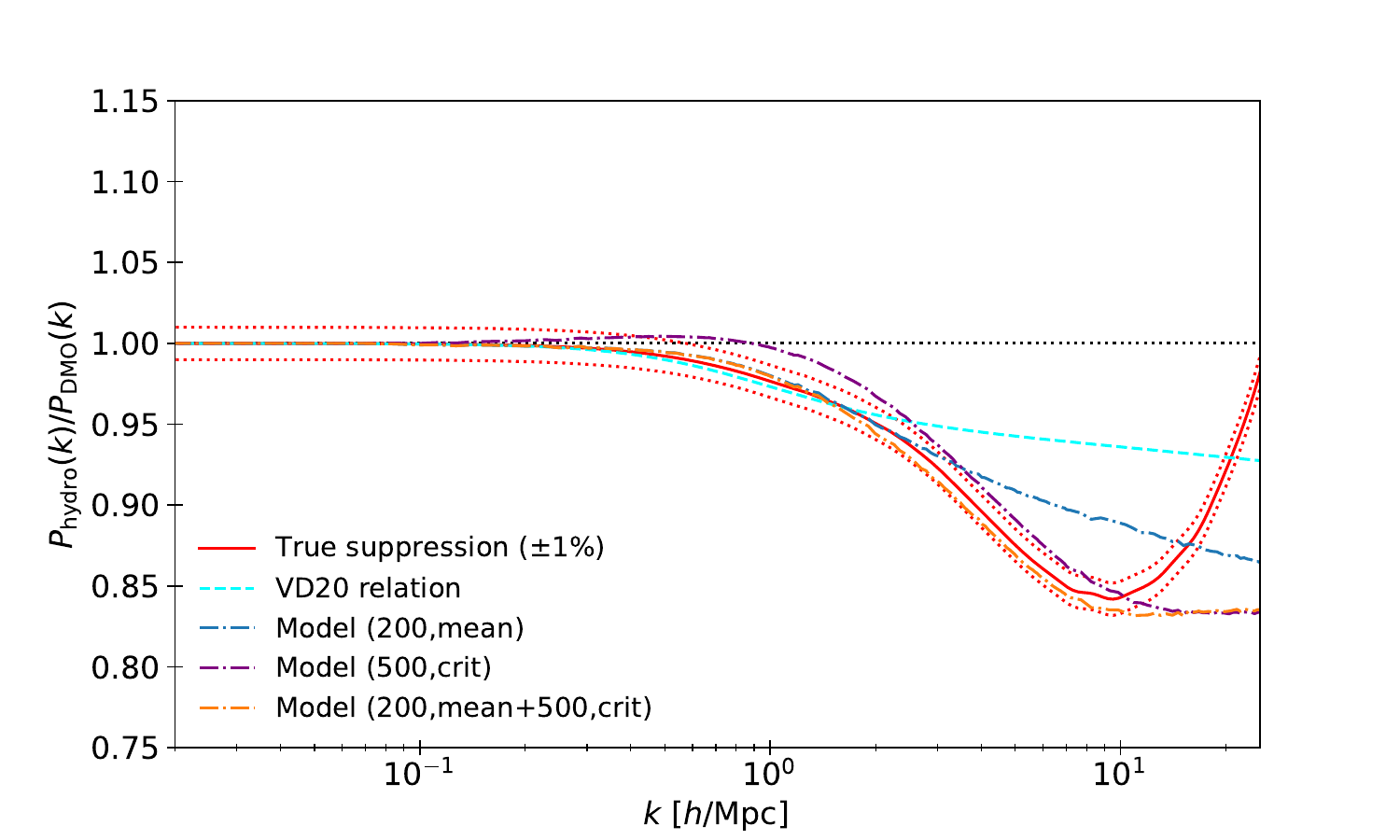}
\caption{Model predictions compared to the true suppression signal for \flamingo simulation L1\_m9, which is shown as a solid red line. Red dotted lines show the $\pm1$ percentage-point band around the true suppression signal. The \citet[][VD20]{vanDaalen2020} relation, based on the mean baryon fraction in haloes of mass $\M\approx 10^{14}\,\mathrm{M_\odot}/h$ and fit to simulation results up to $\ksc{=}{1}$, is shown as a dashed cyan line for comparison. The results from the resummation model are shown as dot-dashed lines. Using only the mean halo baryon fractions within $\radm$ as input reproduces the true suppression signal within $1$ percentage point up to $\ksc{\approx}{3}$, while using only those within $\radc$ follows the true signal down to smaller scale but underpredicts the amount of suppression on large scales. However, combining baryon fractions from both regions, as described in \S\ref{subsec:modelannulus}, allows us to reproduce the true signal to within $1$ percentage point up to $\ksc{\approx}{10}$.}
\label{fig:dpp_fiducial}
\end{figure}

\begin{figure*}
\includegraphics[width=\columnwidth, trim=17mm 6mm 25mm 12mm]{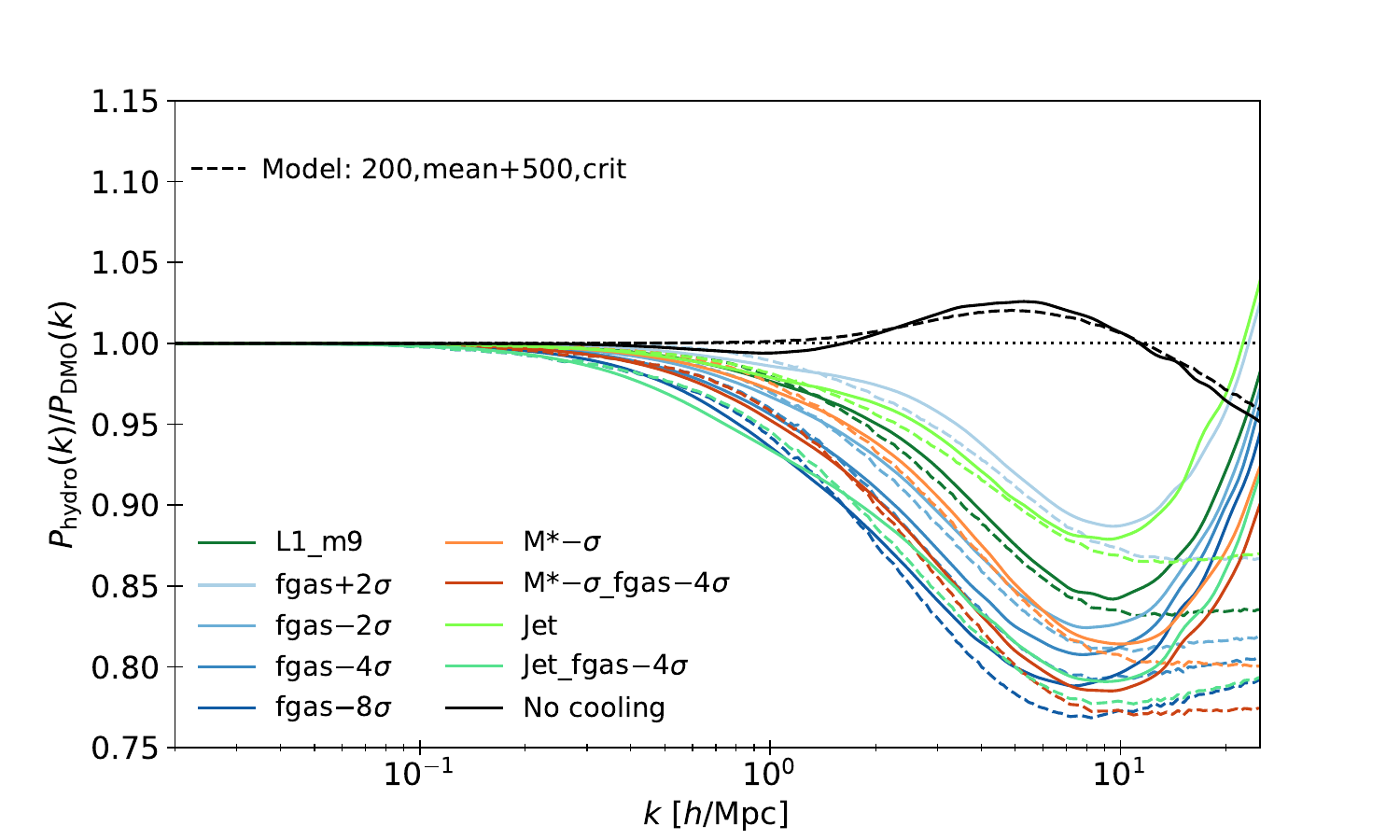}
\includegraphics[width=\columnwidth, trim=8mm 6mm 34mm 12mm]{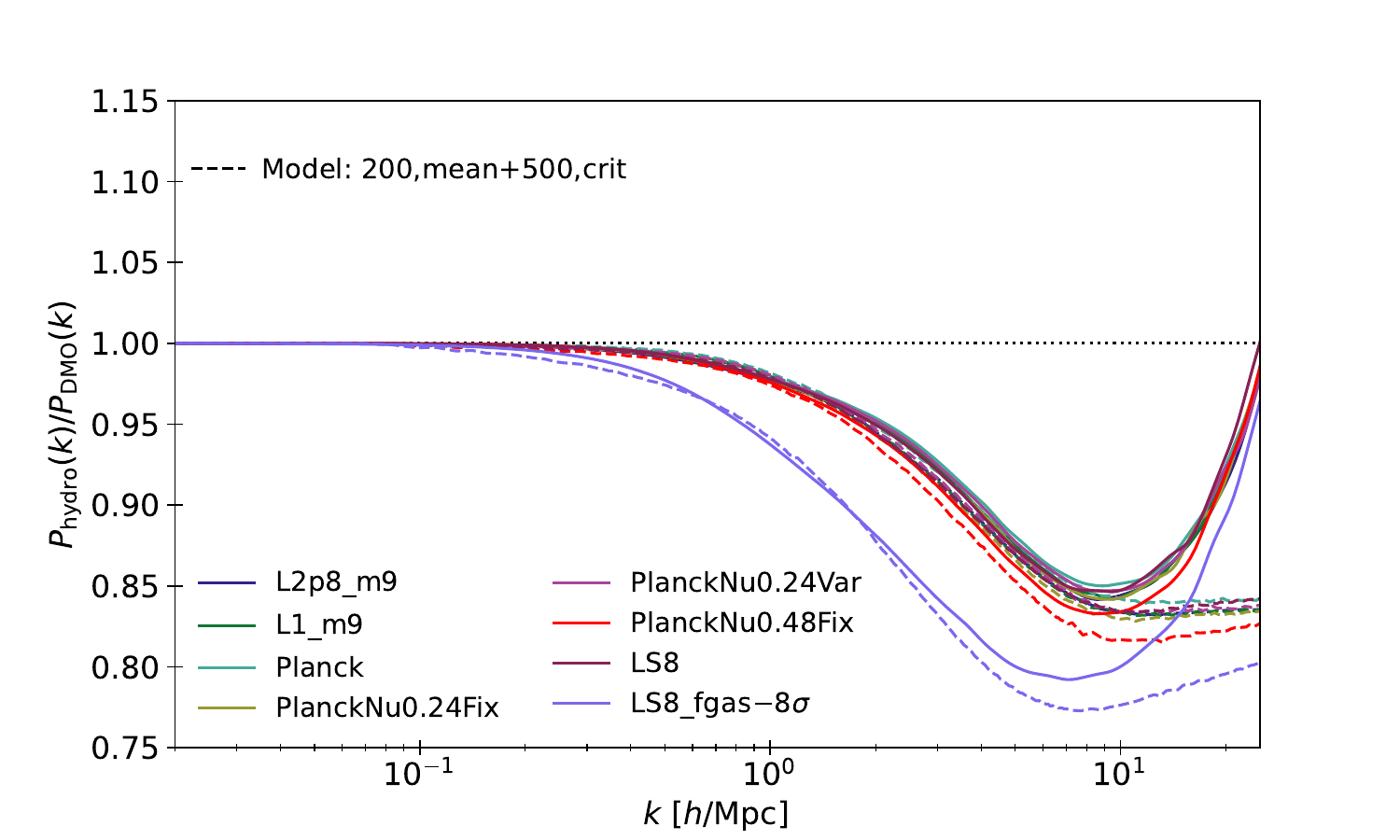}\\
\includegraphics[width=\columnwidth, trim=17mm 6mm 25mm 12mm]{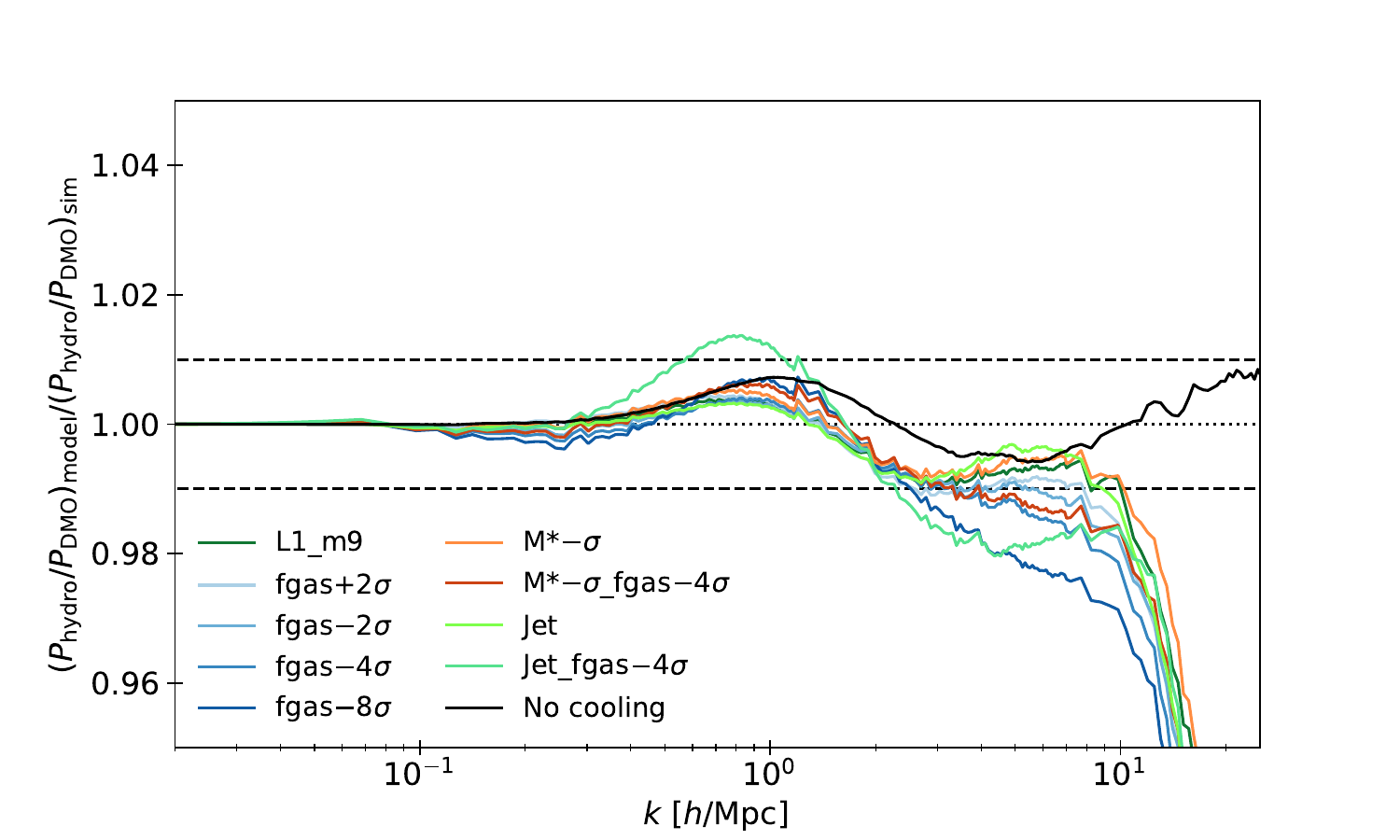}
\includegraphics[width=\columnwidth, trim=8mm 6mm 34mm 12mm]{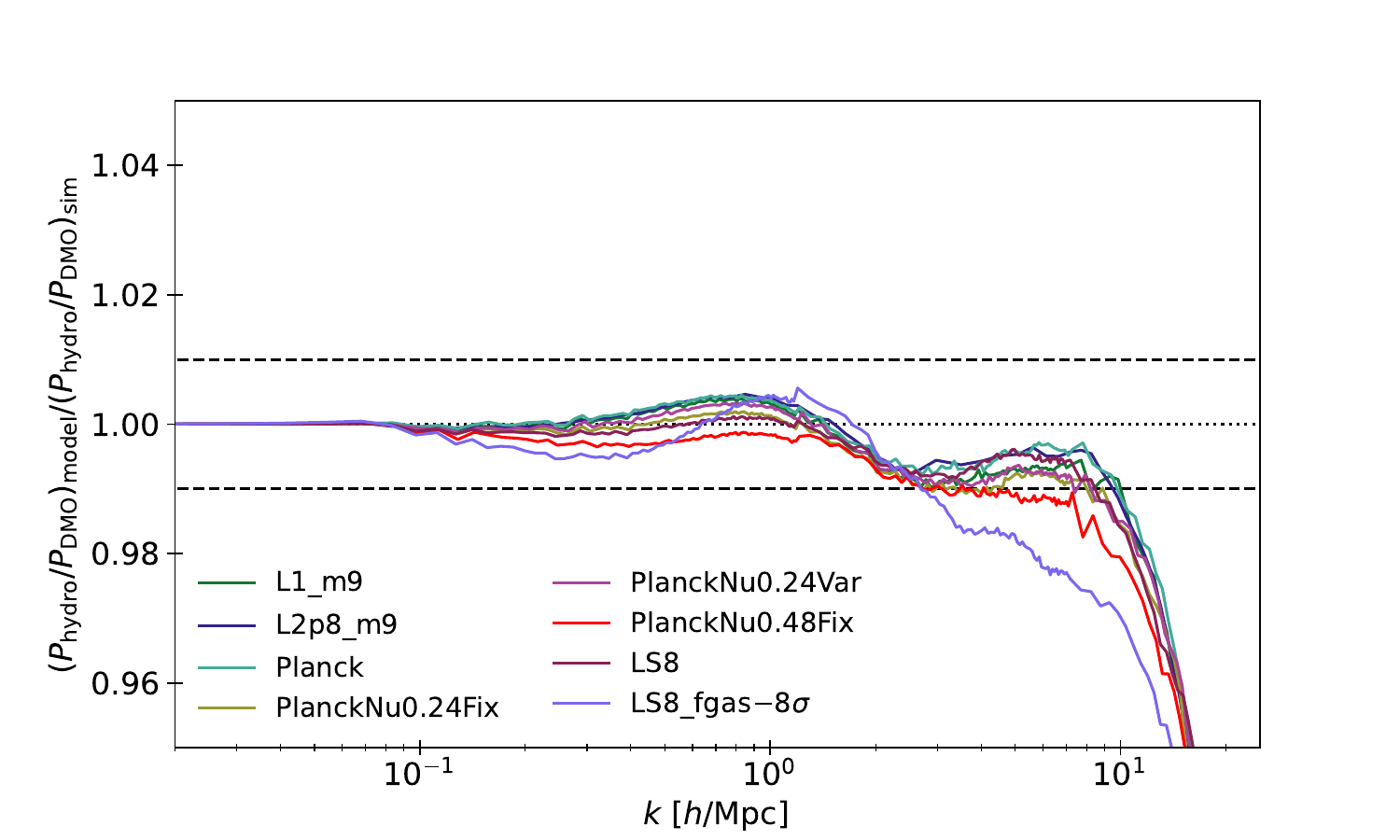}
\caption{The results of applying the resummation model to all baryonic \flamingo variations, using mean halo baryon fractions measured from these simulations within both $\radm$ and $\radc$ for haloes $\M\geq\hM{12.5}$. The top panels show the suppression signals, while the bottom panels show the ratio of the model predictions to the true suppression signals of the simulations. Variations with the default D3A cosmology are shown in the left panels, those with other cosmologies in the right (with L1\_m9 and L2p8\_m9 added as a reference, which are nigh indistinguishable from each other). The true suppression signals of the \flamingo simulations (solid lines) are all accurately reproduced by the resummation model (dashed lines) down to the scale on which the suppression signal starts to turn up, caused by the formation of galaxies in the centre of haloes and the contraction of the inner halo in response, which is not captured by the default model. Strikingly, the model even accurately reproduces the relative power spectrum in a simulation without any cooling or feedback (black line). We emphasise that the model uses zero free parameters to transform given halo baryon fractions into a predicted suppression signal.}
\label{fig:dpp_all}
\end{figure*}

\subsection{Suppression signals from the resummation model}
\label{subsec:resum_results}
We will now consider the predictions of the model in more detail.

In Fig.~\ref{fig:dpp_fiducial}, we compare the true suppression signal for the L1\_m9 simulation, $P_\mathrm{hydro}(k)/P_\mathrm{DMO}(k)$, to different predictions. These suppression signals have been smoothed by convolving with a Gaussian with a standard deviation equal to two data points for clarity. We show the true suppression signal as a solid red line, with dotted lines showing a $\pm 1$ percentage-point band around it -- ideally, our predictions are within this band, at least up to $\ksc{=}{10}$. As a comparison, a dashed cyan line shows the prediction of \citet[][VD20]{vanDaalen2020}, which is based only on the mean baryon fraction of $\M\sim\hM{14}$ haloes. While this accurately predicts the suppression signal on scales where this halo contribution is the dominant one, it fails for scales $\ksc{\gtrsim}{2}$, for reasons explained in Section~\ref{sec:introduction}.

The resummation model instead takes mean halo baryons in multiple halo mass bins as input. Its predictions for varying inputs are shown as dot-dashed lines. Using only baryon fractions measured within $\radm$ regions, the model predicts a suppression signal shown by the blue dot-dashed line. While it, too, accurately reproduces the signal on large scales, it fails for scales $\ksc{\gtrsim}{3}$. This is expected, because for these scales the change in the distribution of matter \emph{within} $\radm$ regions of massive haloes, due to feedback, contributes to the signal as well. If we instead take only mean halo baryon fractions measured within $\radc$ as input, shown by the purple dot-dashed lines, we can accurately follow the shape of the profile up to $\ksc{\approx}{10}$ -- however, since these regions are smaller, we underestimate the suppression on large scales.

But a key strength of the model is that the information contained in baryon fractions measured on different scales can be combined, as explained in \S\ref{subsec:modelannulus}, without assuming a density profile. Using both the baryon fractions measured within $\radm$ and $\radc$ yields the orange dot-dashed line in Fig.~\ref{fig:dpp_fiducial}. This prediction is within $1$ percentage point of the true suppression signal for all $\ksc{\leq}{10}$ -- down to the scale where the contraction of the inner halo starts to become important.

\begin{figure*}
\includegraphics[width=\columnwidth, trim=17mm 6mm 25mm 12mm]{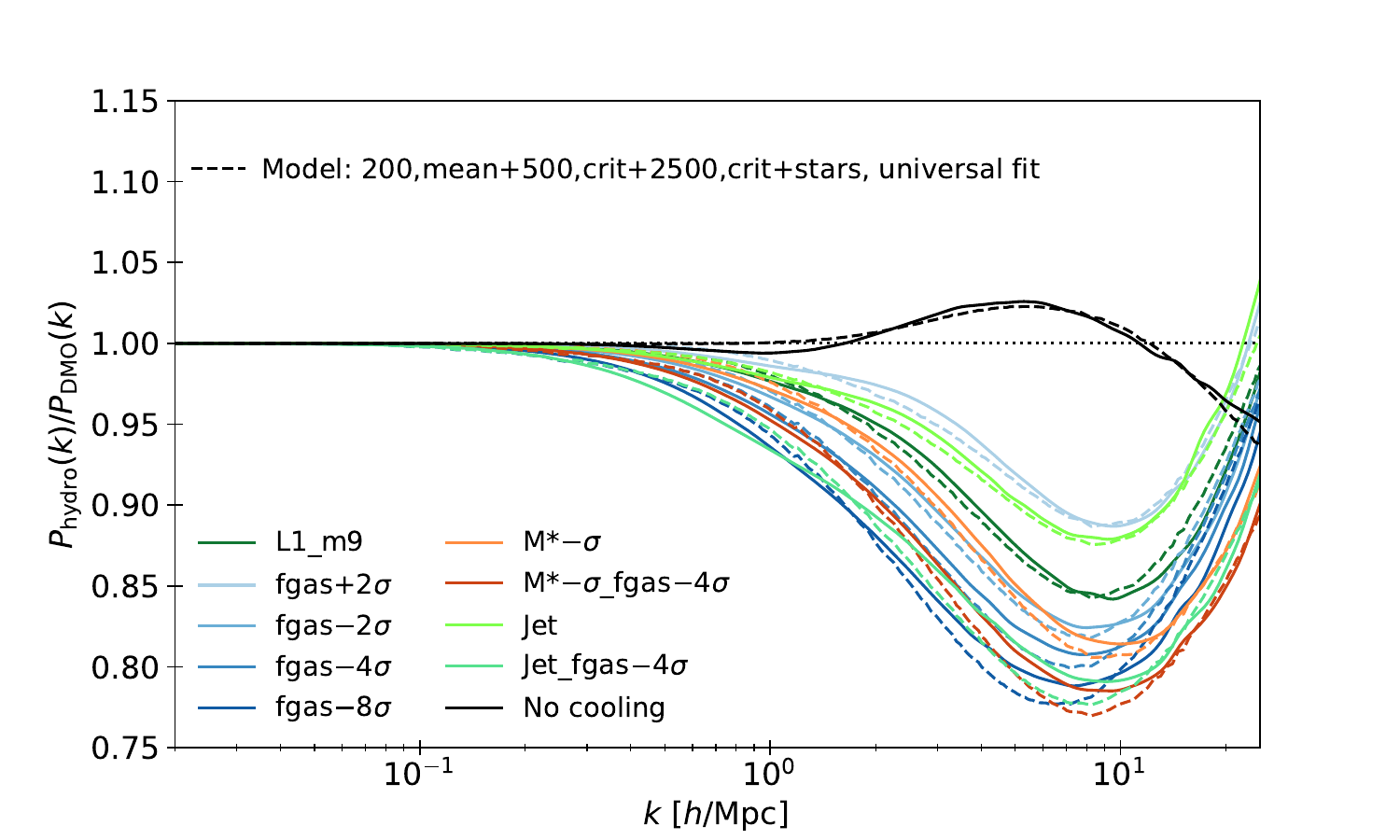}
\includegraphics[width=\columnwidth, trim=8mm 6mm 34mm 12mm]{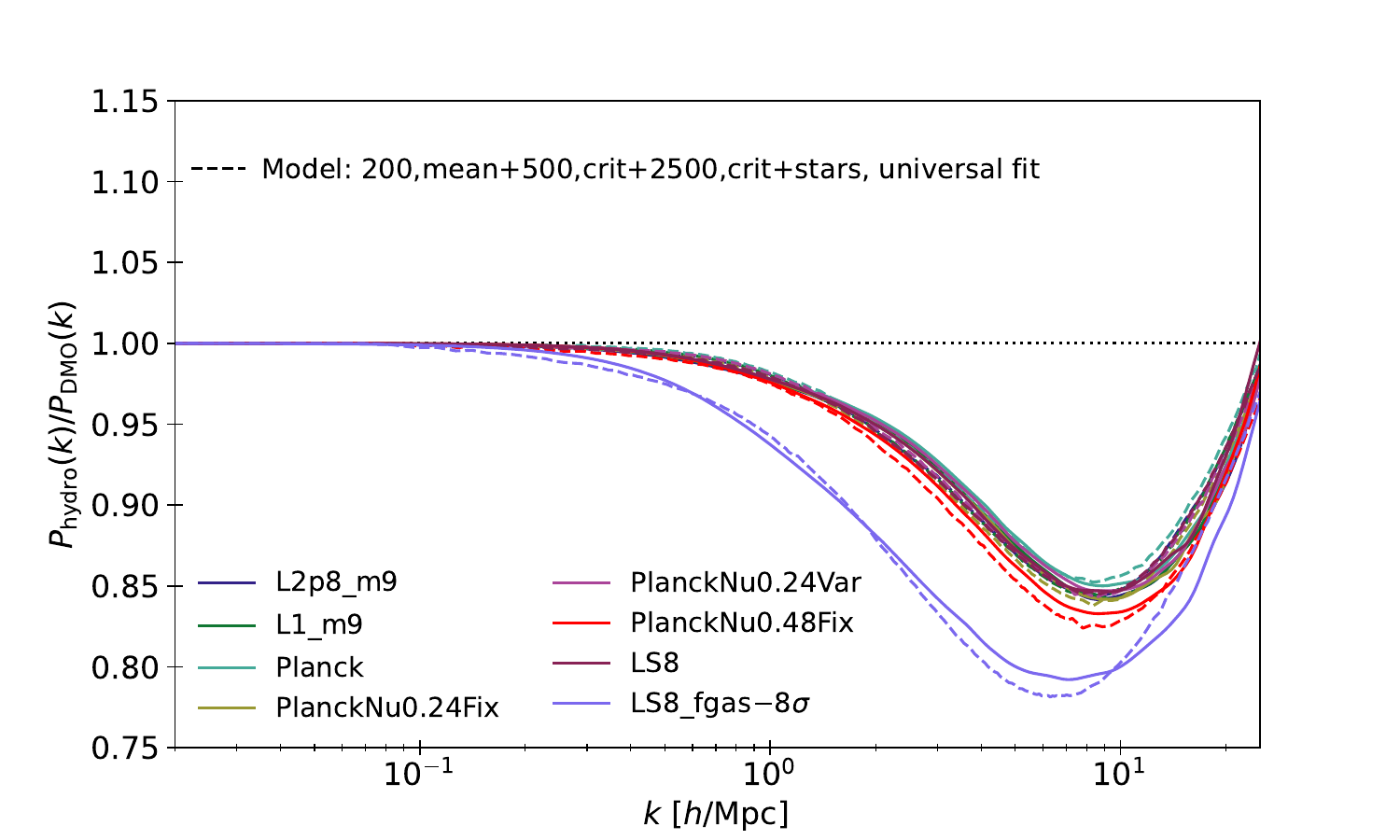}\\
\includegraphics[width=\columnwidth, trim=17mm 6mm 25mm 12mm]{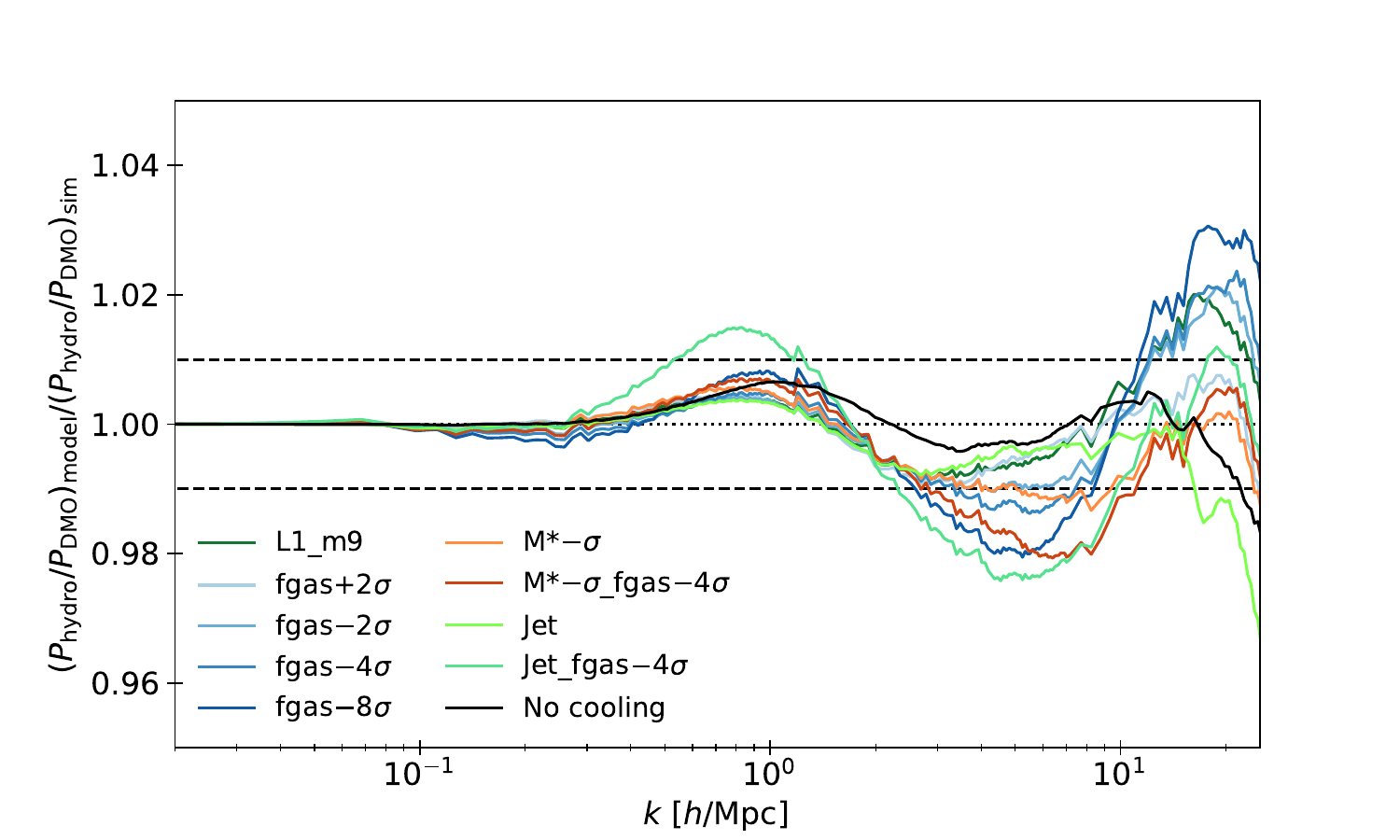}
\includegraphics[width=\columnwidth, trim=8mm 6mm 34mm 12mm]{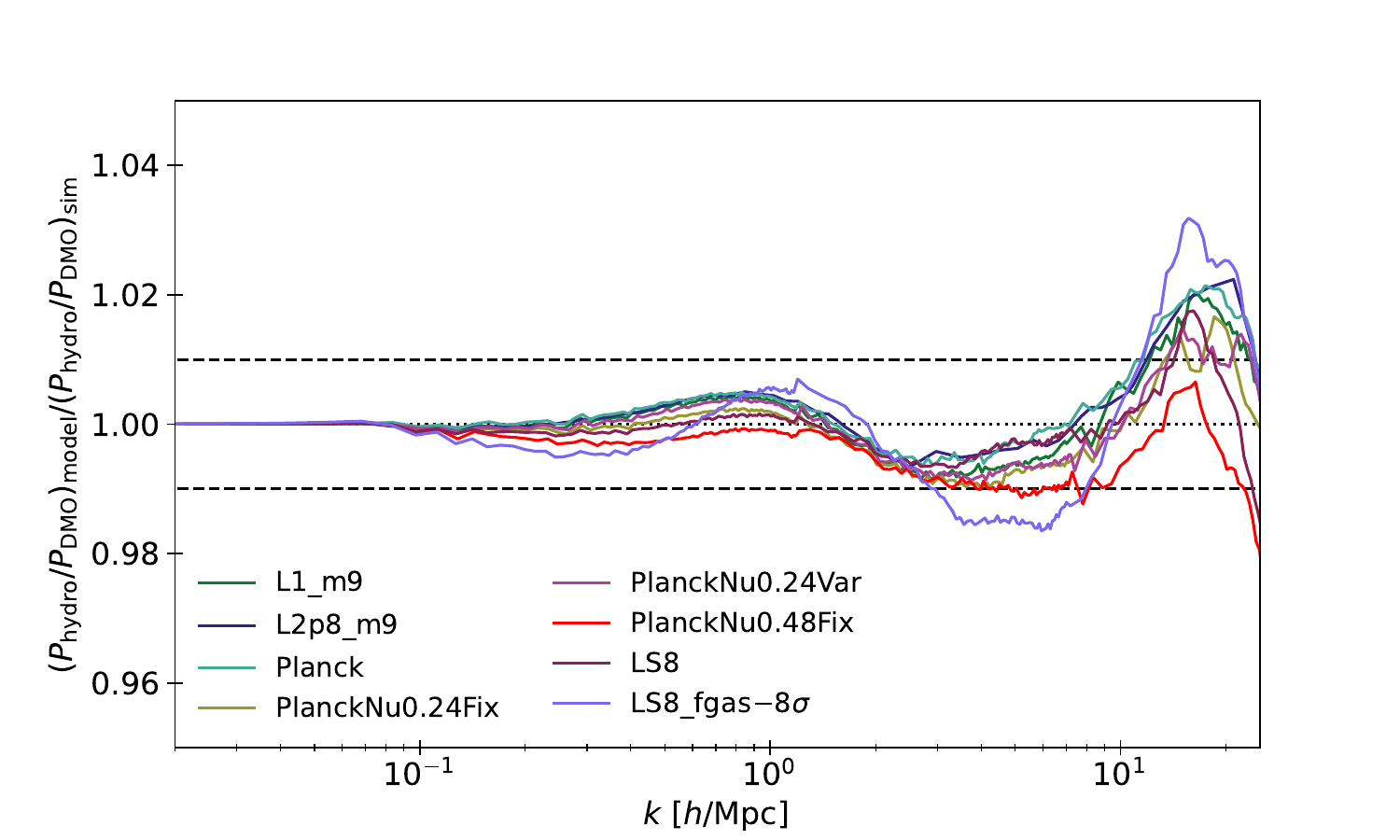}
\caption{As Fig.~\ref{fig:dpp_all}, but additionally using mean gas and stellar fractions within $\radcc$ (see \S\ref{subsec:r2500c_stars}) and the universal fit of equation~\eqref{eq:universal_params}. The universal fit to just the retained mass fractions of L2p8\_m9 is accurate enough to negligibly affect the large-scale agreement between the simulations and the resummation model, while using the small-scale information (including the crude assumption that all stellar mass in the inner region is located at the centre) is enough to capture the upturn at percent level down to the smallest scales for which we expect our simulations to be converged, $\ksc{\approx}{25}$. This version of the model also has zero free parameters, but requires more input fractions than the model presented in Fig.~\ref{fig:dpp_all}.}
\label{fig:dpp_all_stellar_universal}
\end{figure*}

We emphasise that this result is achieved without any free parameters -- therefore, we expect that using mean halo baryon fractions measured from other simulations should also accurately reproduce their suppression signals, without any tweaks to the model itself. We demonstrate that this is indeed the case in Fig.~\ref{fig:dpp_all}, where we apply the resummation model to all baryonic variations of \flamingo. The top panels show the true and modelled suppression signals, while the bottom panels show the ratios of two for each simulation. On the left, we show the variations that use the fiducial D3A cosmology. Using the fits to the $f_\mathrm{ret}-f_\mathrm{bc}$ relation presented in the right-hand panels of Fig.~\ref{fig:fret_fbc_200m_fixedhaloes_fixedmass} and Fig.~\ref{fig:fret_fbc_500c_fixedhaloes_fixedmass}, we convert the baryon fractions measured in each simulation to retained mass fractions, which are used to rescale L1\_m9\_DMO. For all simulations, the predicted suppression signal is well within $1$ percentage point for most $\ksc{<}{10}$, and within $2$ percentage points for all $k$ up to $\ksc{\approx}{10}$, regardless of the strength of supernova or AGN feedback, or the mode of AGN energy injection (i.e.\ jets or not).

In fact, for the \flamingo simulation without gas cooling, which has neither star formation nor feedback but does show the effects of gas pressure relative to DMO, the model reproduces the true signal to within $1$ percentage point on all scales shown, $\ksc{\leq}{25}$ (black lines). This is striking, as the shape of its suppression signal is completely different from that of simulations with strong feedback, and therefore not easily captured by existing fitting functions \citep[e.g.][]{Schaller2025}. The resummation model, which does not rely on any particular functional form as a function of $k$, but instead rescales DMO cross spectra, is much more flexible.

On the right side of Fig.~\ref{fig:dpp_all}, we show the results of applying the resummation model to \flamingo simulations with different cosmologies. Cross spectra from a DMO simulation with a matching cosmology are rescaled to predict the suppression signal. We see that the model performs equally well for these other cosmologies, with no variation of the model performance with cosmological parameters such as $\sigma_8$ or the neutrino mass. However, note that the range in cosmic baryon fractions explored by these simulations is rather limited, with $\Omega_\mathrm{b}/(\Omega_\mathrm{cdm}+\Omega_\mathrm{b})$ spanning approximately $[1.558,1.622]$. We aim to test the resummation model well outside this range in future work. 

Looking at both sides of Fig.~\ref{fig:dpp_all}, we see that in all cases the model accurately reproduces the suppression signal down to the scale where halo contraction due to gas cooling and star formation starts to become important. This is expected, as these processes take place on scales smaller than $\radc$, and are therefore not captured by the mean halo baryon fractions of this or larger regions. However, in \S\ref{subsec:r2500c_stars} we presented an extension of the model to smaller scales, using gas and stellar fractions measured within $\radcc$. The results of adding these fractions are shown in Fig.~\ref{fig:dpp_all_stellar_universal}, where we have additionally switched to using the universal fit to the $f_\mathrm{ret}-f_\mathrm{bc}$ relation as a function of $\Delta$ for all regions used (see equation~\eqref{eq:universal_params} and Fig.~\ref{fig:fret_fbc_universal}).

\begin{figure*}
\includegraphics[width=\columnwidth, trim=17mm 6mm 25mm 12mm]{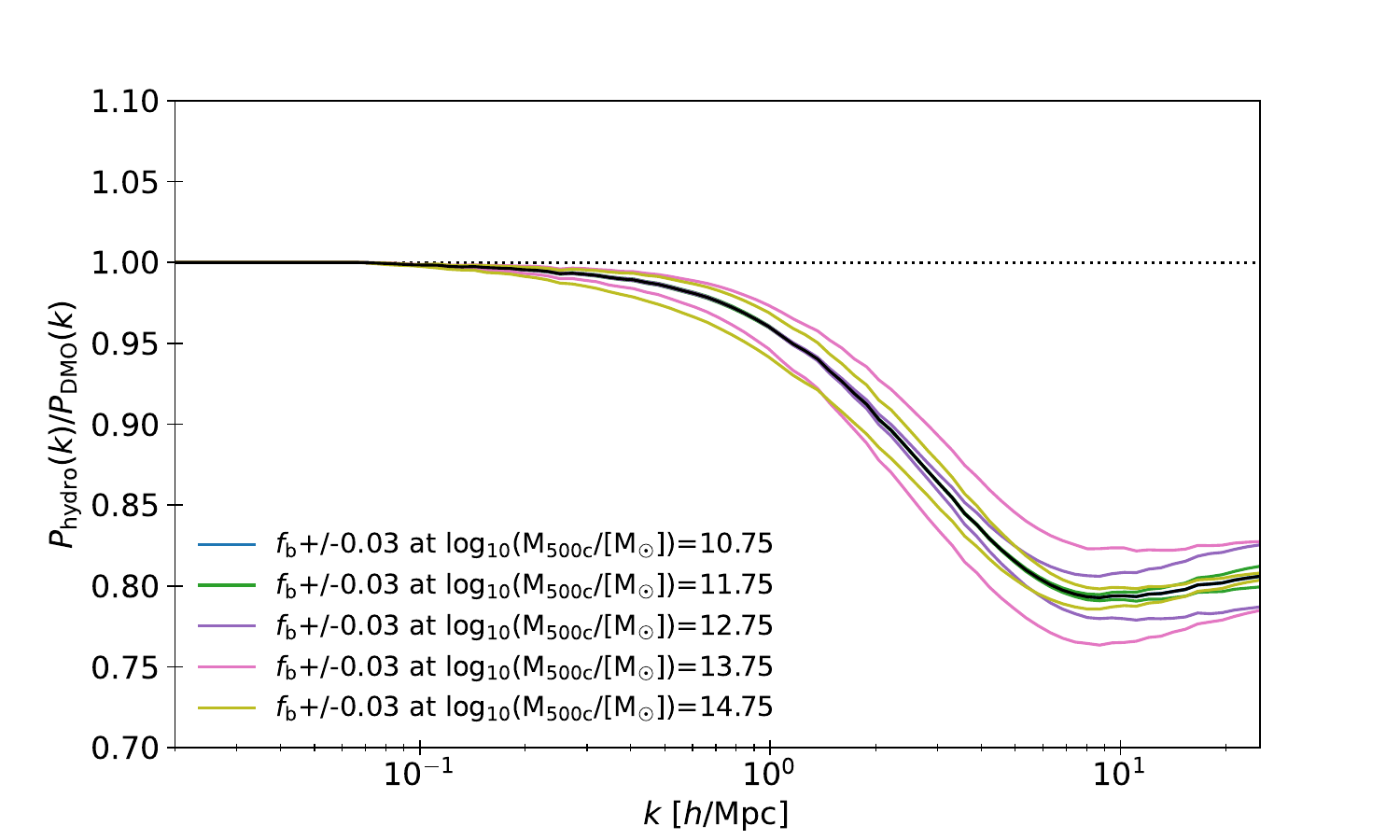}
\includegraphics[width=\columnwidth, trim=8mm 6mm 34mm 12mm]{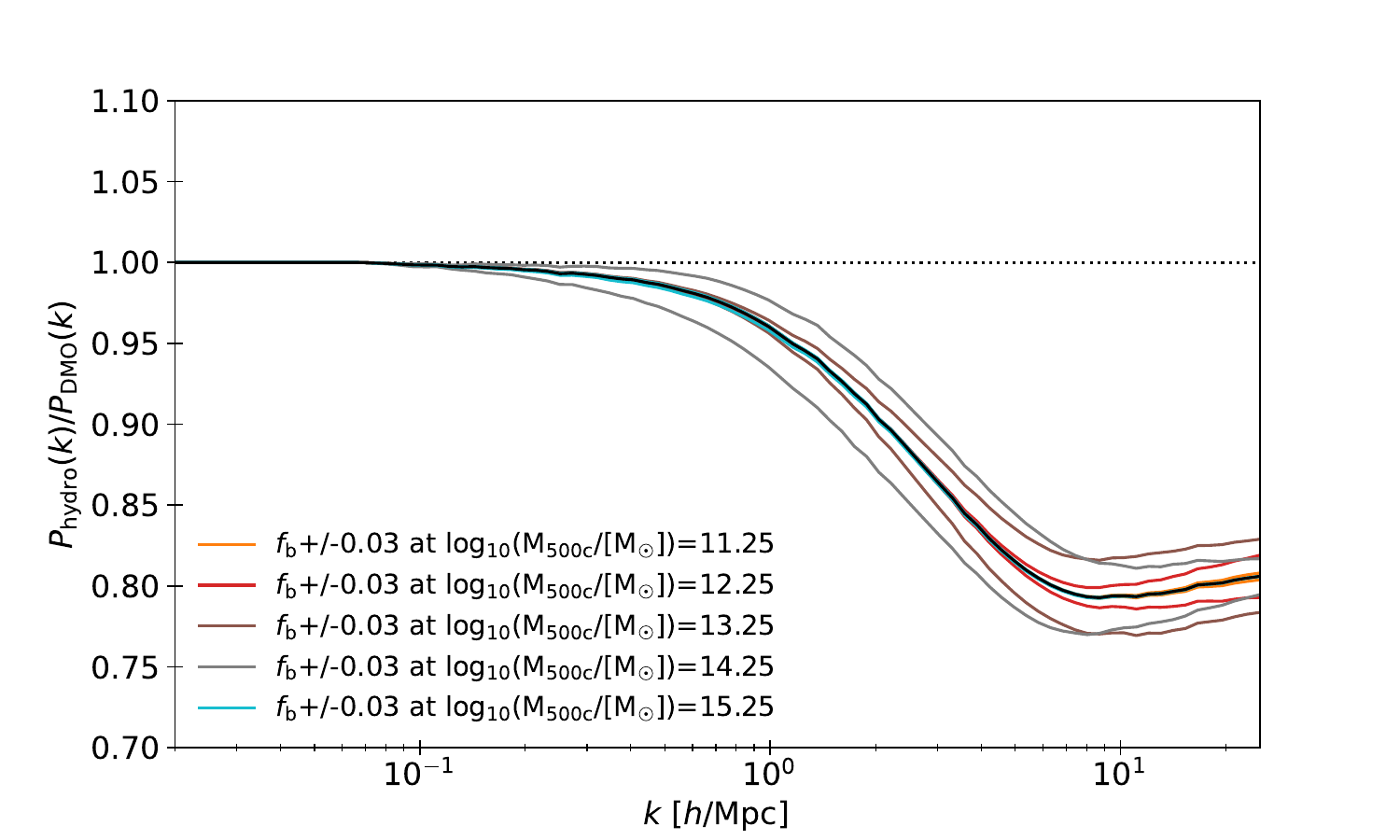}
\caption{The power suppression signal's sensitivity to different halo masses. For visibility, each panel shows every other mass bin. Starting from the (uncorrected) baryon fractions from \flamingo simulation fgas$-4\sigma$, $f_\mathrm{b,200m}$ and $f_\mathrm{b,500c}$, we perturb these by three percentage points in one halo mass bin at a time to investigate its effect on the suppression signal. The masses shown correspond to the central values of logarithmic $0.5\,\mathrm{dex}$ bins. In line with the findings of previous studies, haloes with $\M\sim\hM{14}$ (pink and grey lines) contribute most strongly at all scales $\ksc{\lesssim}{10}$. The baryon fractions of the lowest-mass and highest-mass haloes contribute negligibly to the suppression signal, though the contribution of low-mass haloes grows towards smaller scales.}
\label{fig:dpp_scatter}
\end{figure*}

\begin{figure}
\includegraphics[width=\columnwidth, trim=12mm 16mm 24mm 14mm]{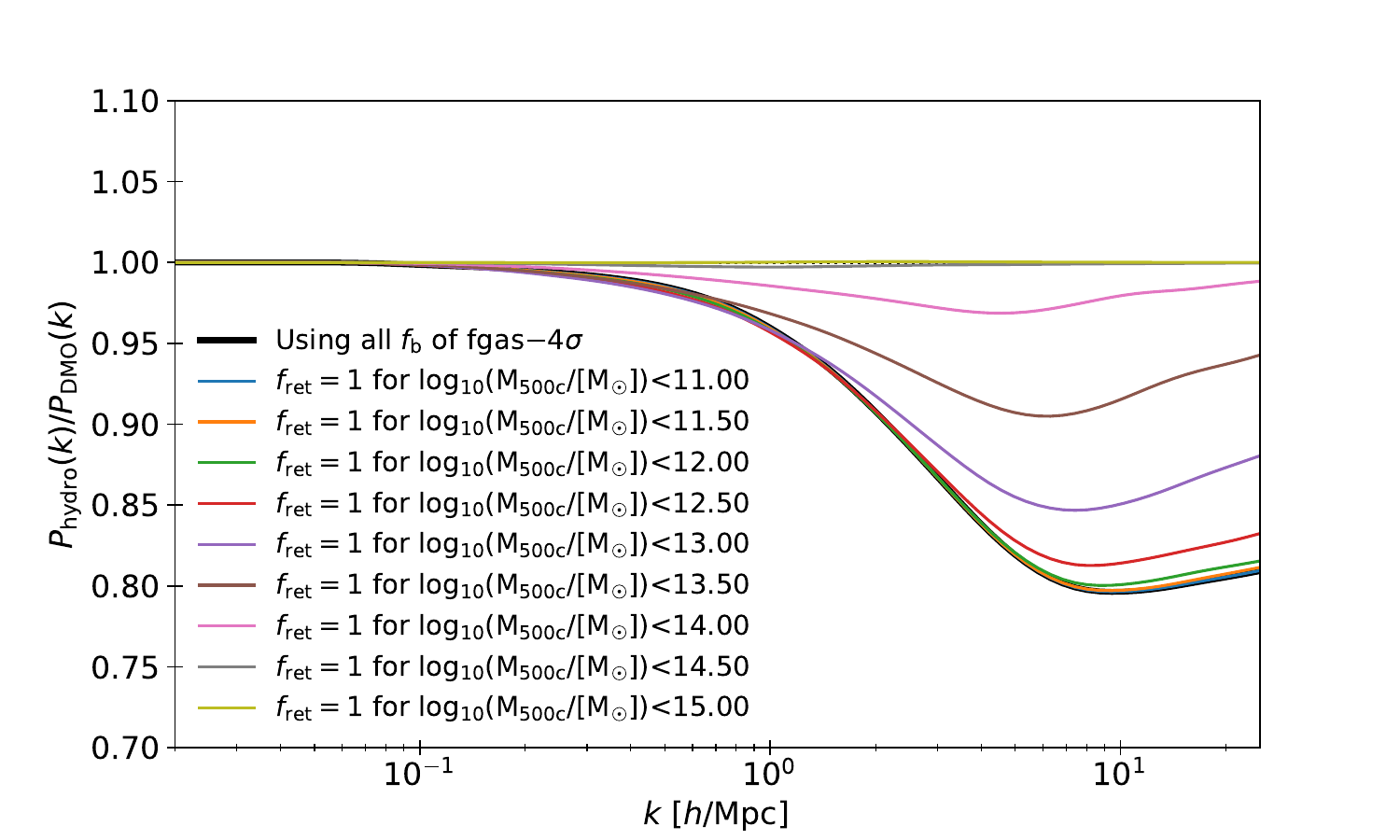}
\caption{Similar to Fig.~\ref{fig:dpp_scatter}, but now we show the effect of including the halo baryon fractions in one additional mass bin at a time. By setting the retained mass fractions of \flamingo simulation fgas$-4\sigma$ to unity below a given halo mass, for both $\frettwo$ and $\fretfive$, we can see the contribution of the baryon fractions of its haloes above this mass to the suppression signal. On large scales, the signal is almost completely determined by haloes in the range $\M=\hM{13.5-14.5}$, while the the haloes that contribute the most to the suppression on scales $1\lesssim \ksc{\lesssim}{10}$ are those in the range $\M=\hM{13.0-14.0}$.}
\label{fig:dpp_builtup}
\end{figure}

The panels shown in the top left and top right of Fig.~\ref{fig:dpp_all_stellar_universal} compare the predicted suppression signal to the true one for the same simulations as before. Now that the upturn in the signal on small scales is captured as well, due to the smaller-scale information included, the model predictions are within $2.5$ percentage points of the true signal for \emph{all} $\ksc{\leq}{10}$, and typically within $1$ percentage point. Additionally, the model deviations are within $3.1$ percentage points down to the smallest scales measured, $\ksc{=}{25}$, for all simulations, showing no signs of divergence. The results are robust to the changes in feedback strength, the type of feedback, and cosmology explored in \flamingo, despite the approximations made, without any free parameters (see \S\ref{subsec:resum_descr}). The model tends to overestimate the suppression signal at $\ksc{\approx}{5}$, where the $\Delta=\mathrm{500c}$ region transitions to the $\Delta=\mathrm{2500c}$ one. Using baryon fractions within e.g.\ $\Delta=\mathrm{1000c}$ would ease the transition and likely lead to even better results.

\vspace{0.2cm}
\noindent
While in all cases we have applied resummation to a DMO simulation matching each hydro simulation in box size, resolution, and initial conditions, this is not necessary to obtain accurate results (neither is it allowed to be if we wish to apply the method to observations, because the real Universe does not have an equivalent simulation). We demonstrate this in Appendix~\ref{sec:app_diff_DMO}.

\subsection{Contribution strength of halo baryon fractions and uncertainty propagation}
\label{subsec:barfrac_response}
One of the strengths of the resummation model is that halo baryon fractions translate straightforwardly into a suppression signal. This makes it possible to marginalise over poorly-constrained baryon fractions for some (or all) halo masses. It is therefore useful to consider the impact of baryon fractions of different halo masses on the suppression signal, as predicted by the model. This tells us both how vital it is to include a particular halo mass scale in the model, and how uncertainties in its baryon fraction propagate into the suppression signal -- and therefore to what precision its mean baryon fraction needs to be known in order to predict the suppression signal to a given accuracy.

In Fig.~\ref{fig:dpp_scatter}, we show the suppression signal for \flamingo simulation fgas$-4\sigma$ as a black line. All other lines show how the signal changes when the mean baryon fraction is varied for a single mass bin. In all cases, we vary both the baryon fraction inside $\radm$ and $\radc$ simultaneously, lowering them by three percentage points (e.g.\ from $12\%$ to $9\%$, strengthening the suppression signal) or raising them by the same amount (e.g.\ from $12\%$ to $15\%$, weakening the signal). The size of the perturbation was chosen arbitrarily. The dependence on halo baryon fraction tends to be very similar for neighbouring halo mass bins, so for clarity, we have split these lines over two panels, alternating between the two panels with every bin. The legends show the central values of the $0.5\,\mathrm{dex}$ halo mass bins that were varied.

Examining Fig.~\ref{fig:dpp_scatter}, we can immediately see that the signal is dominated by halo masses around $\M=\hM{14}$, in line with the findings of e.g.\ \citet{vanDaalen2020}, \citet{Debackere2020}, \citet{Salcido2023} and \citet{vanLoonvanDaalen2024}. The impact of other halo mass bins drops off quickly from there, with haloes in the mass range $\M=\hM{14.5-15}$ still contributing significantly on scales $\ksc{<}{10}$, while haloes with masses $\M>\hM{15}$ contribute negligibly on all scales. On the low-mass end, the large-scale contribution diminishes sharply below $\M\approx\hM{13}$, while that on small scales ($\ksc{\gtrsim}{5}$) drops off much more slowly.

\begin{figure*}
\includegraphics[width=\columnwidth, trim=17mm 6mm 25mm 12mm]{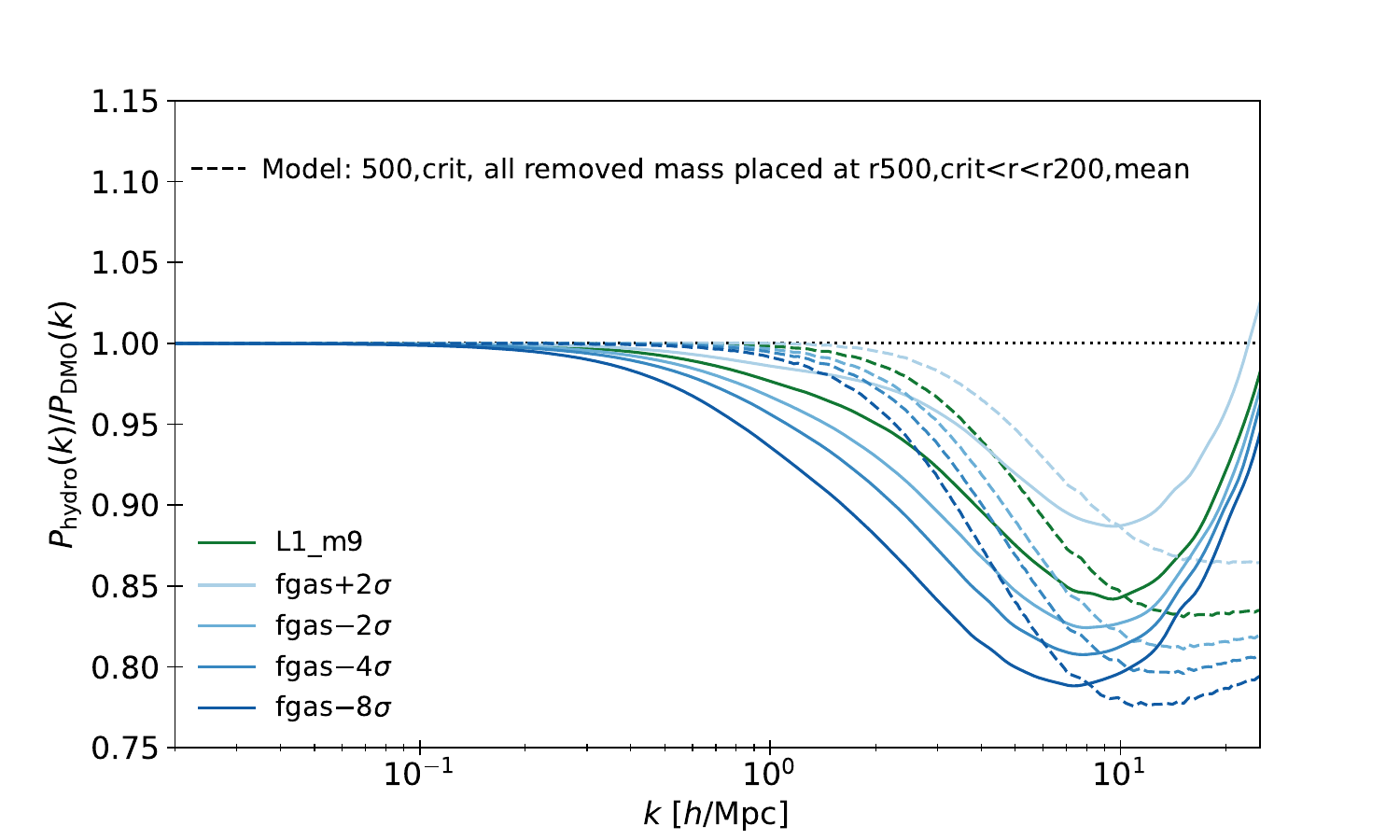}
\includegraphics[width=\columnwidth, trim=8mm 6mm 34mm 12mm]{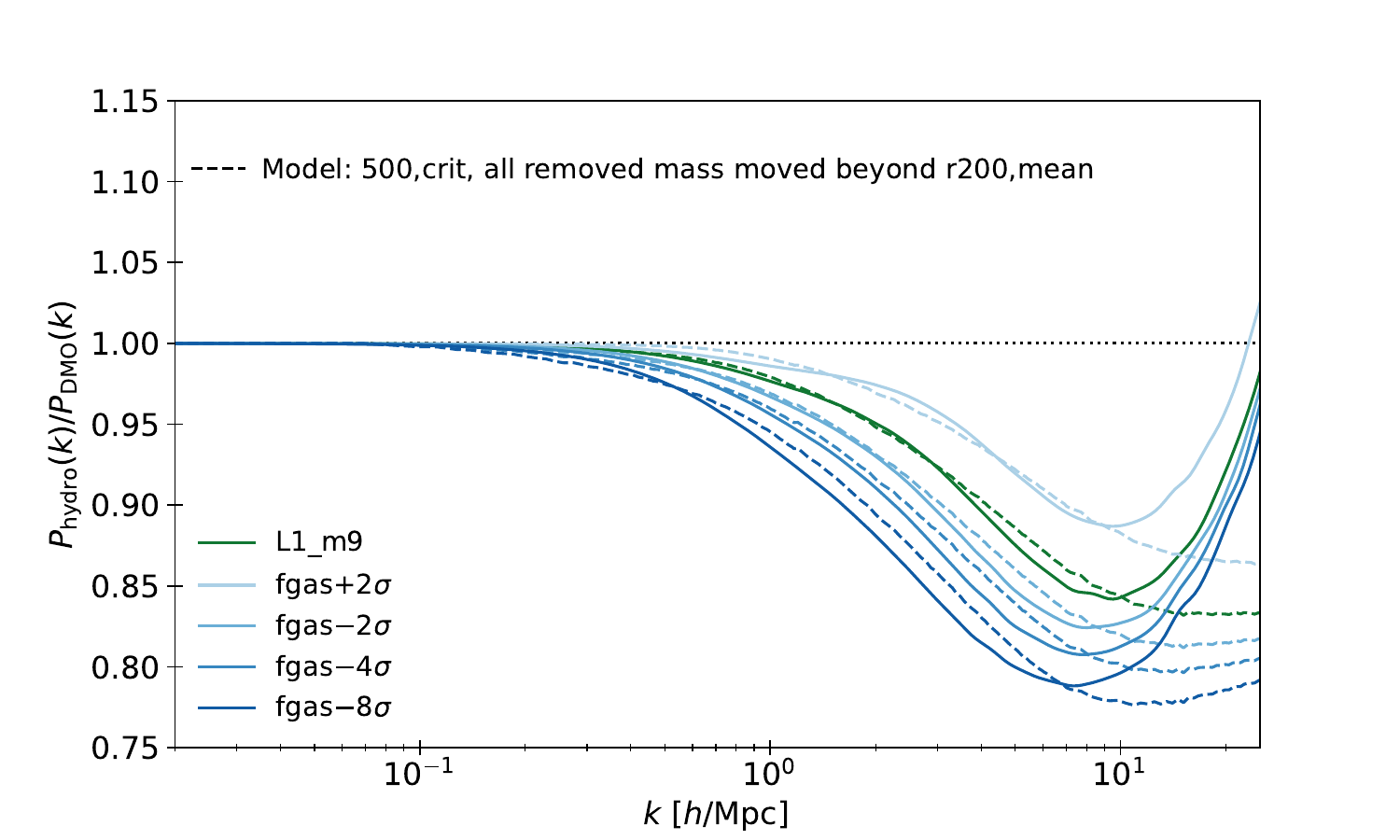}
\caption{The model predictions using only mean baryon fractions measured inside $\radc$ coupled with one of two extreme assumptions. \textit{Left:} Assuming that the mass removed from $\radc$ stays within $\radm$. \textit{Right:} Assuming that any mass removed from within $\radc$ is also removed from within $\radm$. Comparison to the true simulated suppression signal (solid lines) shows that the latter is a much more accurate assumption, but one that slowly becomes less valid as the strength of feedback grows and more matter is actually removed from (or prevented from accreting into) within $\radm$ than within $\radc$.}
\label{fig:dpp_AGN_annulus_outside}
\end{figure*}

Additionally, in Fig.~\ref{fig:dpp_builtup} we show the contribution of the halo baryon fractions above a certain mass scale to the predicted suppression signal, again for fgas$-4\sigma$. Here we set both $\frettwo$ and $\fretfive$ to unity for an increasing number of mass bins, which diminishes the suppression signal. Unlike the results of Fig.~\ref{fig:dpp_scatter}, the contribution of each mass bin is now weighted by the amplitude of each baryon fraction, since no fixed perturbations are used. We see that on large scales, $\ksc{\lesssim}{1}$, the contributions of haloes in the mass range $\M=\hM{13.5-14.5}$ almost entirely set the suppression signal. On smaller scales, $1\lesssim \ksc{\lesssim}{10}$, the largest contributions come from the mass range $\M=\hM{13.0-14.0}$, with significant contributions from mass bins directly adjacent to this range.

In both Fig.~\ref{fig:dpp_scatter} and Fig.~\ref{fig:dpp_builtup}, the baryon fractions in haloes in the lowest two mass bins considered, $\M=\hM{11-11.5}$ and $\hM{10.5-11}$, contribute negligibly on all scales shown. In the case of Fig.~\ref{fig:dpp_builtup}, the contribution of these lowest-mass haloes may be underestimated, as most of them are not well-resolved enough to form stars and therefore expel no gas, which they otherwise might have -- though Fig.~\ref{fig:dpp_scatter} shows that their contribution is small either way. Regardless, the inclusion of these mass scales in the model is still relevant to better measure the non-halo component.

\subsection{Resummation based only on \texorpdfstring{$\boldsymbol{\Delta=}\mathbf{500c}$}{the 500c region}}
\label{subsec:annulus_or_not}
While increasingly accurate measurements of mean halo baryon fractions within $\radc$ are already observationally available for different $\M$ \citep[e.g.][]{Akino2022} or even $M_\mathrm{200c}$ \citep[e.g.][though with some extrapolation]{Dev2024}, measuring these precisely for larger regions is more challenging. Promising techniques that derive large-scale gas fractions from measurements of the kSZ \citep[e.g.][]{Hadzhiyska2025a} or tSZ \citep[e.g.][]{Dalal2025} signal, or those using fast radio bursts to probe the intergalactic medium \citep[e.g.][]{Lemos2025}, may change this in the near future. In the meantime, it is worth investigating how well the suppression signal can be reproduced using only baryon fractions measured within $\radc$.

One of the results of \citet{Debackere2020} was that the suppression signal is sensitive to the density profile of hot gas outside $\radc$. In the terms of the resummation model, the results therefore depend on the retained mass fractions on scales that are larger than the baryon fractions were measured on. We confirm this finding by showing the difference between two extreme assumptions for where the removed mass goes in Fig.~\ref{fig:dpp_AGN_annulus_outside}. On the left-hand side, we show the result of assuming that all mass removed from within $\radc$ stays within $\radm$, by setting $f_{\mathrm{ret},i,\mathrm{200m}}=1$ for all halo masses. Clearly, this is a poor assumption, and the suppression of power on large scales is strongly underestimated as a result. On the right-hand side, we instead assume that all mass removed from within $\radc$ also makes it out of $\radm$ (where it is again assumed to be distributed similar to matter already outside DMO haloes), which means that $f_{\mathrm{ret},i,\mathrm{200m}}=1-(1-f_{\mathrm{ret},i,\mathrm{500c}})\bMifive\fMifive/(\bMitwo\fMitwo)$. While this assumption is closer to the results of the simulations, how well it works is significantly dependent on halo mass and the strength of feedback: when either very little or a lot of mass is removed from within $\radc$, this assumption overestimates the retained mass fraction within $\radm$, i.e.\ \emph{more} mass is actually removed from within $\radm$ (or less of it is accreted) than just that ejected from $\radc$.

\begin{figure*}
\includegraphics[width=\columnwidth, trim=17mm 3mm 25mm 12mm]{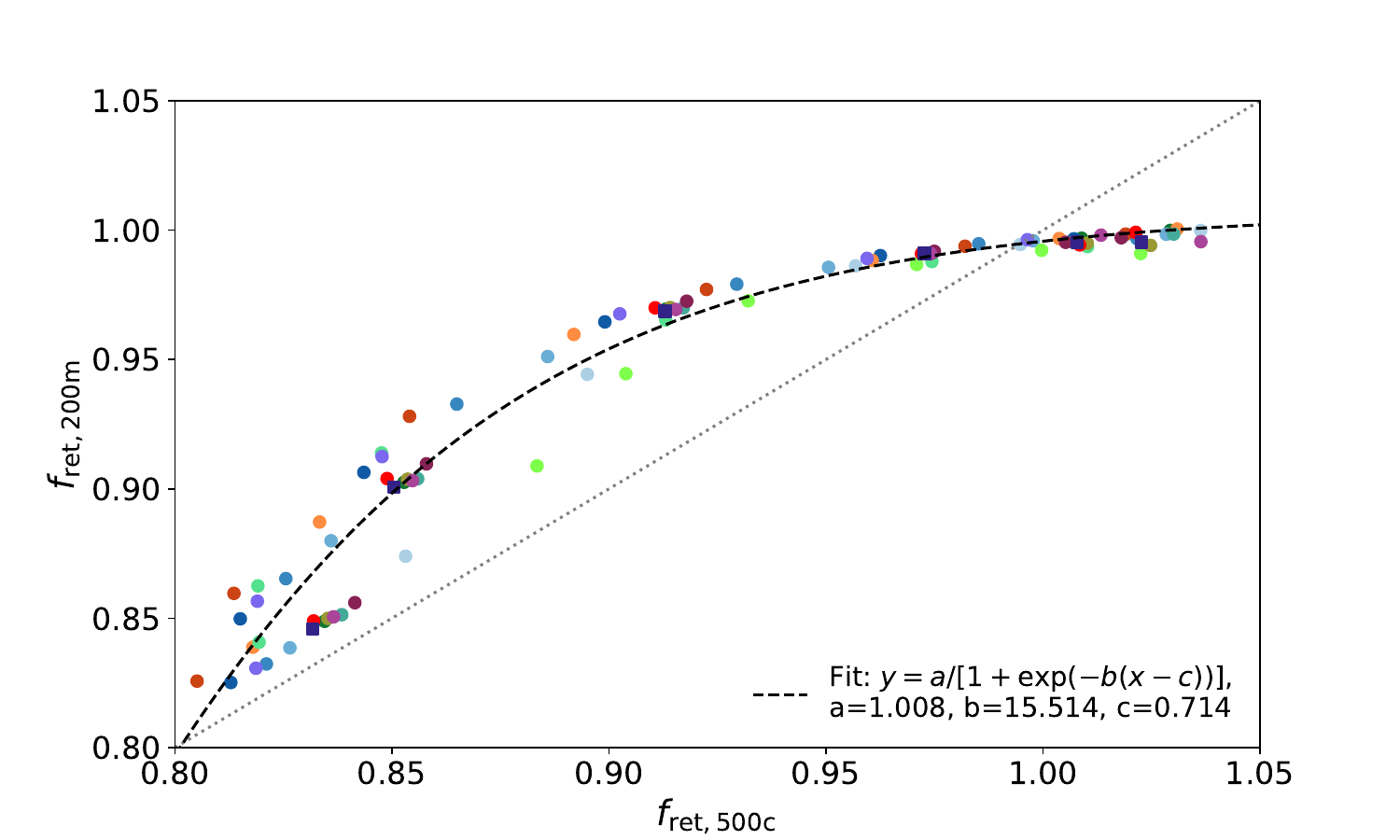}
\includegraphics[width=\columnwidth, trim=8mm 3mm 34mm 12mm]{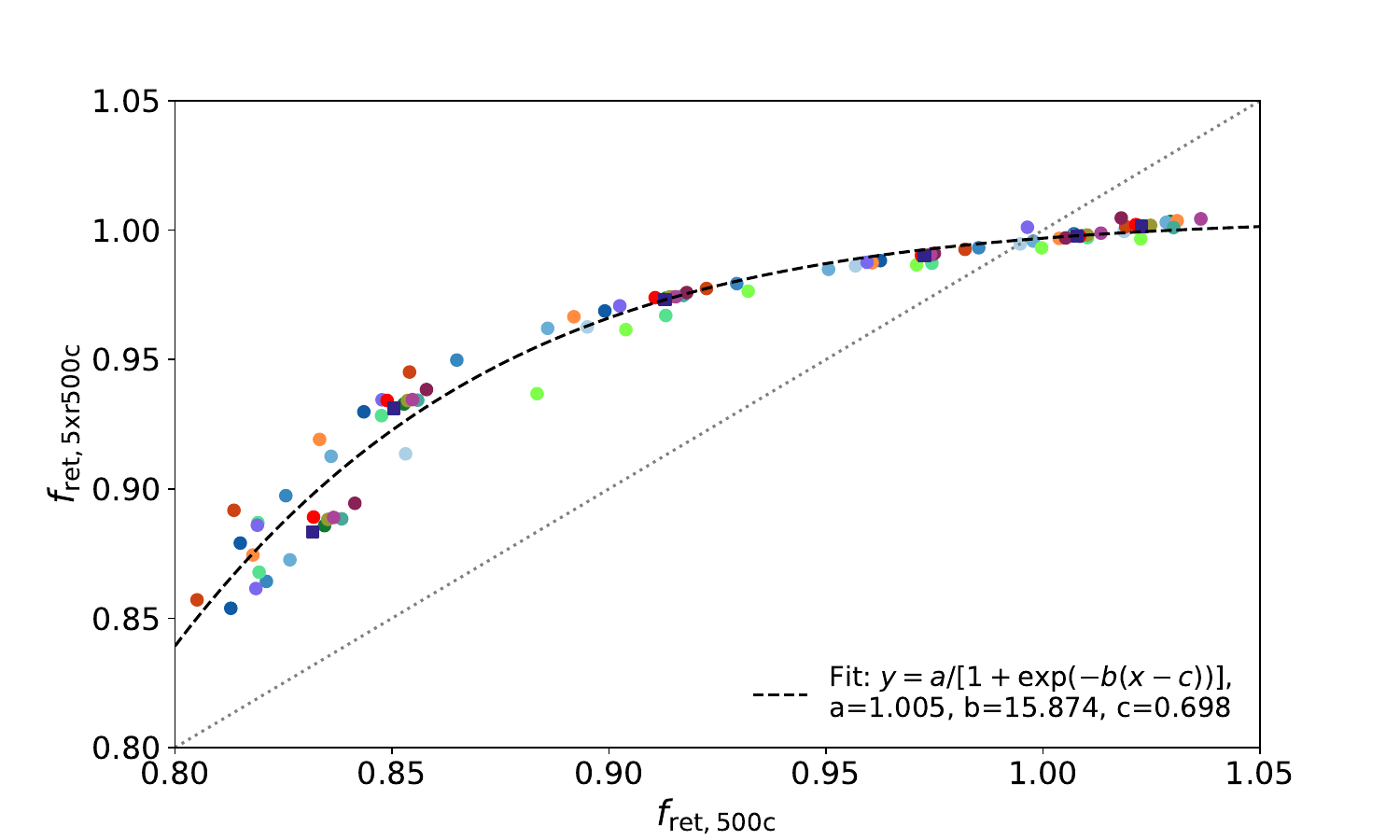}\\
\includegraphics[width=\textwidth, trim=8mm 16mm 0mm 0mm]{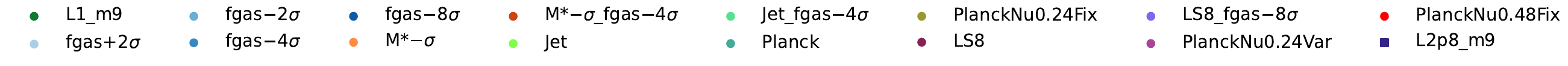}
\caption{The retained mass fraction in a larger region (\textit{left:} $\radm$, \textit{right}: $5\times\radc$) plotted against the retained mass fraction inside $\radc$. As before, every point shows the mean for a $0.5\,\mathrm{dex}$ halo mass bin for a particular simulation, though here for $\Mdmo\geq\hM{12.5}$. Particularly for high halo masses, there is a strong correlation between the inner and outer regions. For lower halo masses, the regions become increasingly decorrelated, meaning it is harder to predict how much mass has been lost from the larger region based on the mass lost in the inner region.}
\label{fig:fret_fret_500c}
\end{figure*}

However, we should expect the retained mass fraction within some larger scale to generally correlate with that on smaller scales. After all, the baryonic feedback processes always take place inside galaxies and are therefore centrally sourced, and stronger feedback or relaxation effects within $\radc$ should generally lead to stronger effects within $\radm$ as well (e.g.\ if we more violently eject mass from within a region, we would expect it to go farther). We therefore investigate how well $f_{\mathrm{ret},i,\mathrm{200m}}$ can be predicted from $f_{\mathrm{ret},i,\mathrm{500c}}$ in Fig.~\ref{fig:fret_fret_500c}. As before, every point corresponds to a single halo mass bin from some \flamingo simulation, although here we take $\M=\hM{12.5}$ as the minimum rather than $\hM{12}$, since the lowest halo mass bin has a relatively large scatter in $\fret$ between simulations while having little weight in the power spectrum. In addition to $f_\mathrm{ret,200m}$, we also show the correlation of $f_\mathrm{ret,500c}$ with the retained mass fraction inside a much larger region, $r<5\times\radc$, given by $f_{\mathrm{ret},i,\mathrm{5xr500c}}$, for comparison. The dotted line shows the one-to-one relation for comparison, while the dashed line shows the result of fitting a logistic function to the points shown, given by:
\begin{equation}
\label{eq:logistic_fret}
f_\mathrm{ret,\Delta}=\frac{a}{1+\exp[-b(f_\mathrm{ret,500c}-c)]}.
\end{equation}
In the legends of both panels we show the best-fitting parameters of this empirical relation.

\begin{figure*}
\includegraphics[width=\columnwidth, trim=17mm 6mm 25mm 12mm]{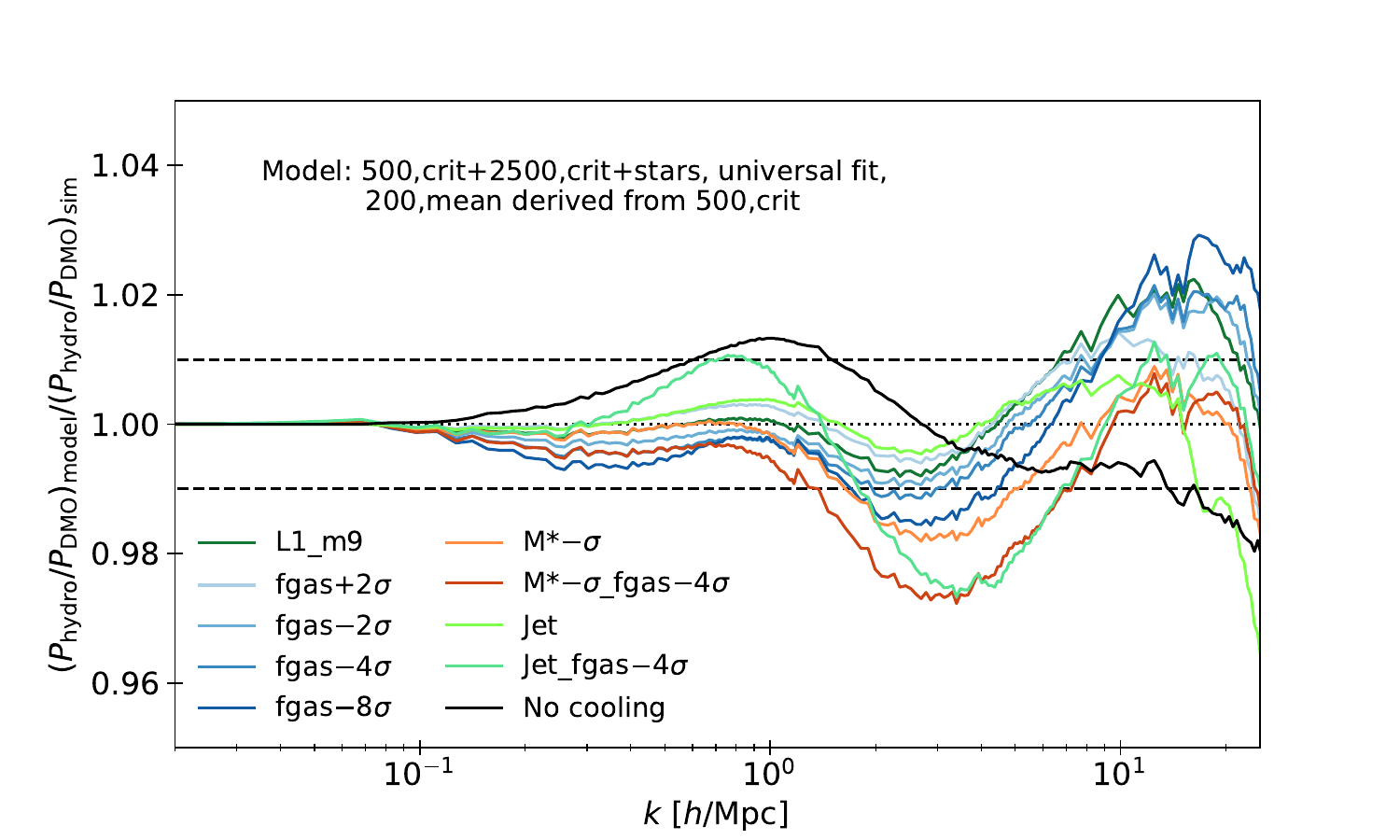}
\includegraphics[width=\columnwidth, trim=8mm 6mm 34mm 12mm]{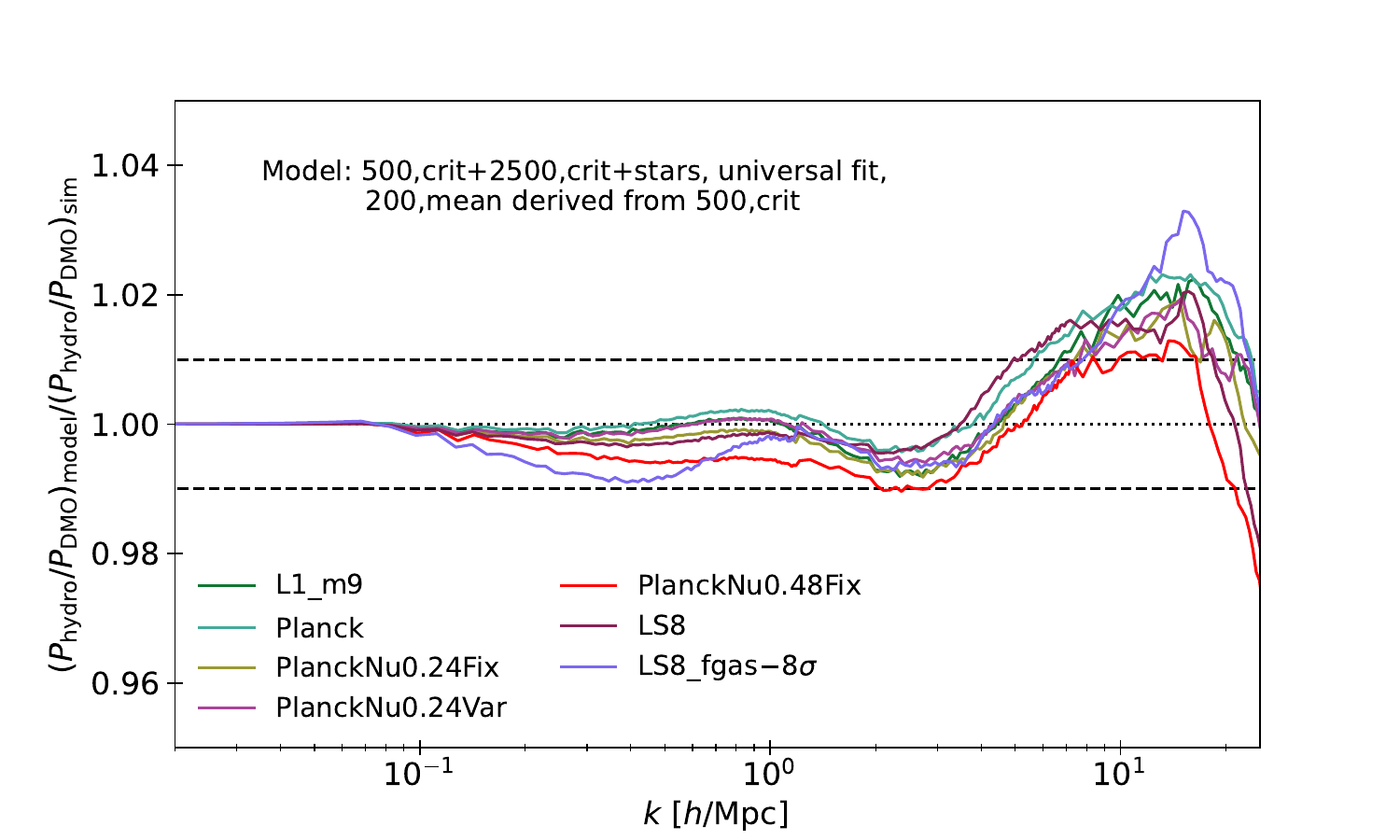}
\caption{Similar to the bottom row of Fig.~\ref{fig:dpp_all_stellar_universal}, and again including stellar fractions within $\radcc$, but now instead of including large-scale baryon fractions, we extrapolate the small-scale information. Using the correlation between the retained fractions within $\radc$ and $\radm$ (see Fig.~\ref{fig:fret_fret_500c} and equation~\ref{eq:logistic_fret}), we predict $f_{\mathrm{ret},i,\mathrm{200m}}$ given $f_{\mathrm{ret},i,\mathrm{500c}}$. This only slightly reduces the accuracy of the model for $\ksc{\lesssim}{10}$, and allows the resummation model to be applied given only mean halo baryon fractions within $\radc$.}
\label{fig:dpp_all_extrapolate}
\end{figure*}

The correlation between both retained mass fractions is particularly strong at high halo masses (large $f_\mathrm{ret,500c}$, see Fig.~\ref{fig:fret_fbc_500c_fixedhaloes_fixedmass}), where the retained mass of the larger region can generally be predicted to within percent accuracy from that of the smaller region. As we go to lower halo mass scales, however, the regions are increasingly decorrelated, and accurately predicting $f_{\mathrm{ret},i,\mathrm{200m}}$ (left) or $f_{\mathrm{ret},i,\mathrm{5xr500c}}$ (right) becomes more sensitive to the strength of feedback and its implementation (thermal or jet AGN feedback). Somewhat surprisingly, the correlation is slightly tighter for the larger region, $r<5\times\radc$ -- likely because any dependence on the halo profile is removed, due to the larger radius being a fixed multiple of the smaller. This hints that we might obtain even better results if we replaced $\radm$ by a fixed multiple of $\radc$ in the resummation model, but we leave this for future work. We note that equation~\eqref{eq:logistic_fret} also works well for predicting $f_{\mathrm{bc},\Delta}$ from $f_\mathrm{bc,500c}$, and translating these into retained mass fractions through equation~\eqref{eq:fretained} to predict $f_{\mathrm{ret},\Delta}$ results in a comparable scatter around the true values as directly using equation~\eqref{eq:logistic_fret} for $f_{\mathrm{ret},\Delta}$.

Since the suppression signal is dominated by massive haloes around $\M\approx\hM{14}$, where the correlation is strong, using only mean baryon fractions within $\radc$ in concert with the best-fitting relation for $f_\mathrm{ret,200m}$ given by equation~\eqref{eq:logistic_fret} should lead to more accurate predictions for the suppression signal than just ignoring the larger region, or making simple assumptions for its retained mass. We show that this is indeed the case in Fig.~\ref{fig:dpp_all_extrapolate}, where we show the predictions of the resummation model relative to the true suppression signal for all \flamingo simulations explored here, using only mean baryon fractions measured inside $\radc$ (and smaller radii, though this only affects the results for $\ksc{\gtrsim}{10}$). The results are very comparable in accuracy to those of the full model, presented in Fig.~\ref{fig:dpp_all_stellar_universal}. This is highly encouraging: if the retained mass fractions within a region as large as $\radm$ in the real Universe are indeed as strongly correlated with the retained mass fraction within $\radc$ as in \flamingo, this means that we only need to measure mean halo baryon fractions within $\radc$ to predict the suppression signal to percent-level accuracy on all scales $\ksc{<}{10}$.

These results indicate that even though the suppression signal is very sensitive to the radius out to which mass is removed, physics apparently significantly constrains the retained mass fraction on large scales when that on smaller scales is known, even in the presence of jets or strong feedback -- at least for large halo masses.

\subsection{Resummation at higher redshifts}
\label{subsec:resum_z}
All results discussed thus far pertain to $z=0$. We will now investigate how the model performs at higher redshifts.

\begin{figure*}
\includegraphics[width=\columnwidth, trim=17mm 6mm 25mm 12mm]{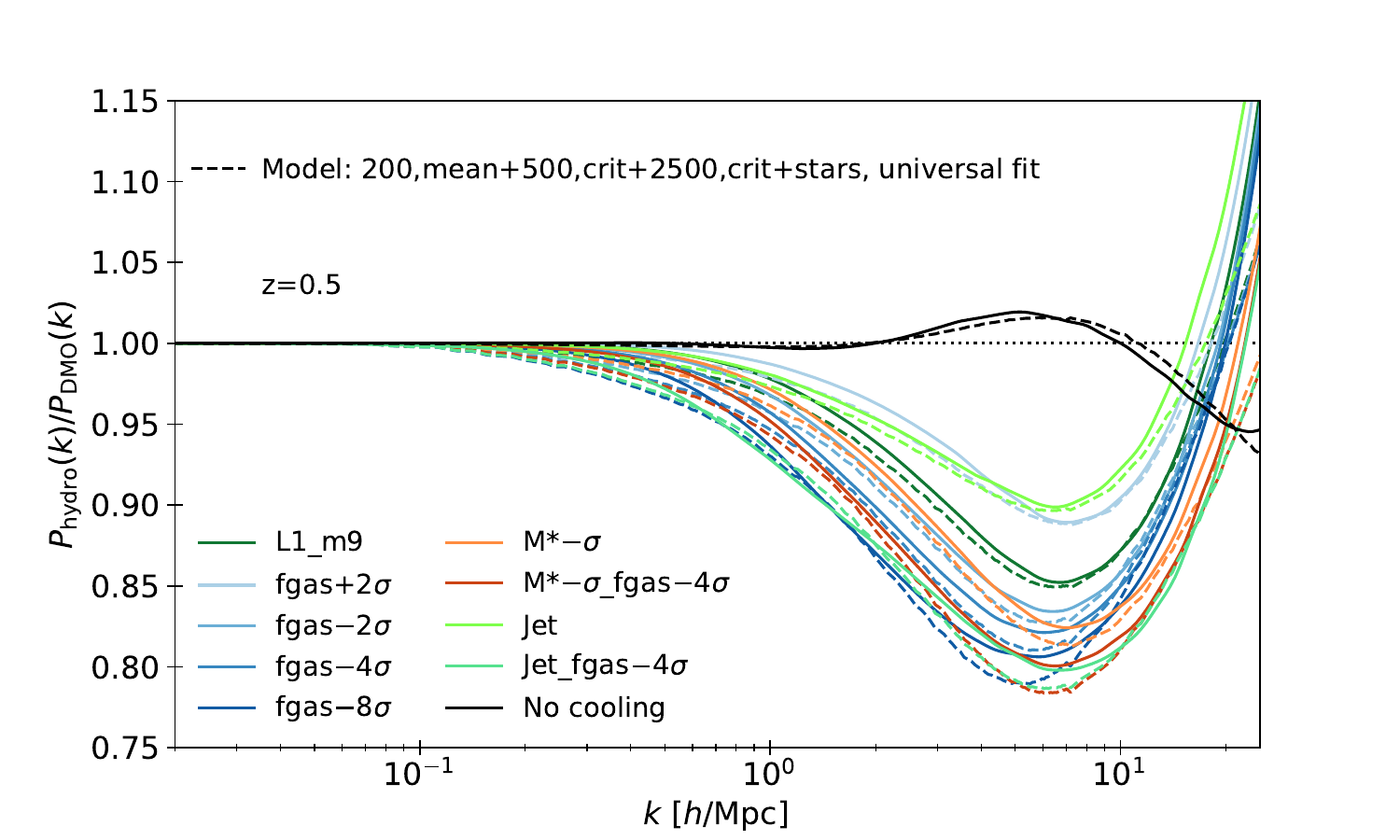}
\includegraphics[width=\columnwidth, trim=8mm 6mm 34mm 12mm]{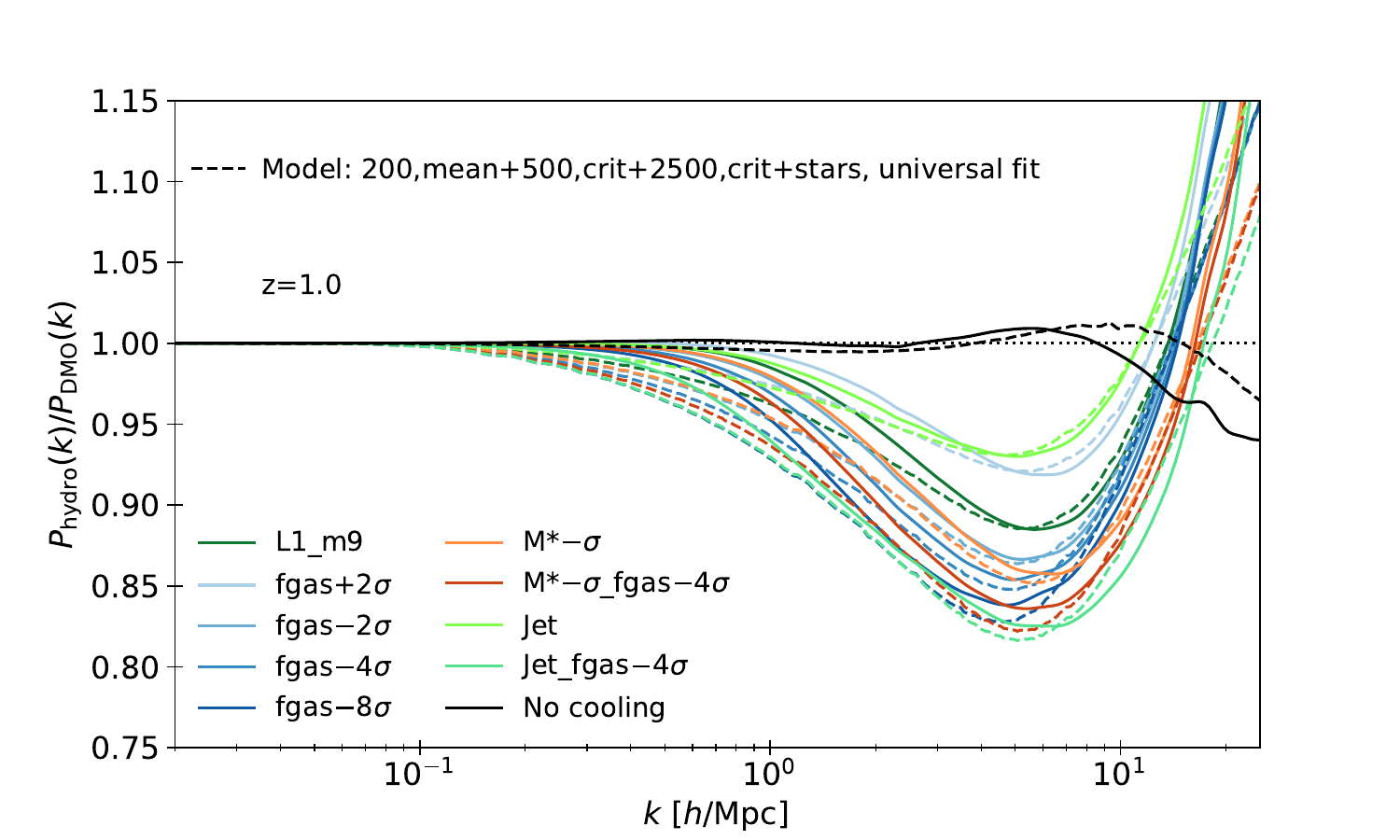}\\
\includegraphics[width=\columnwidth, trim=17mm 6mm 25mm 12mm]{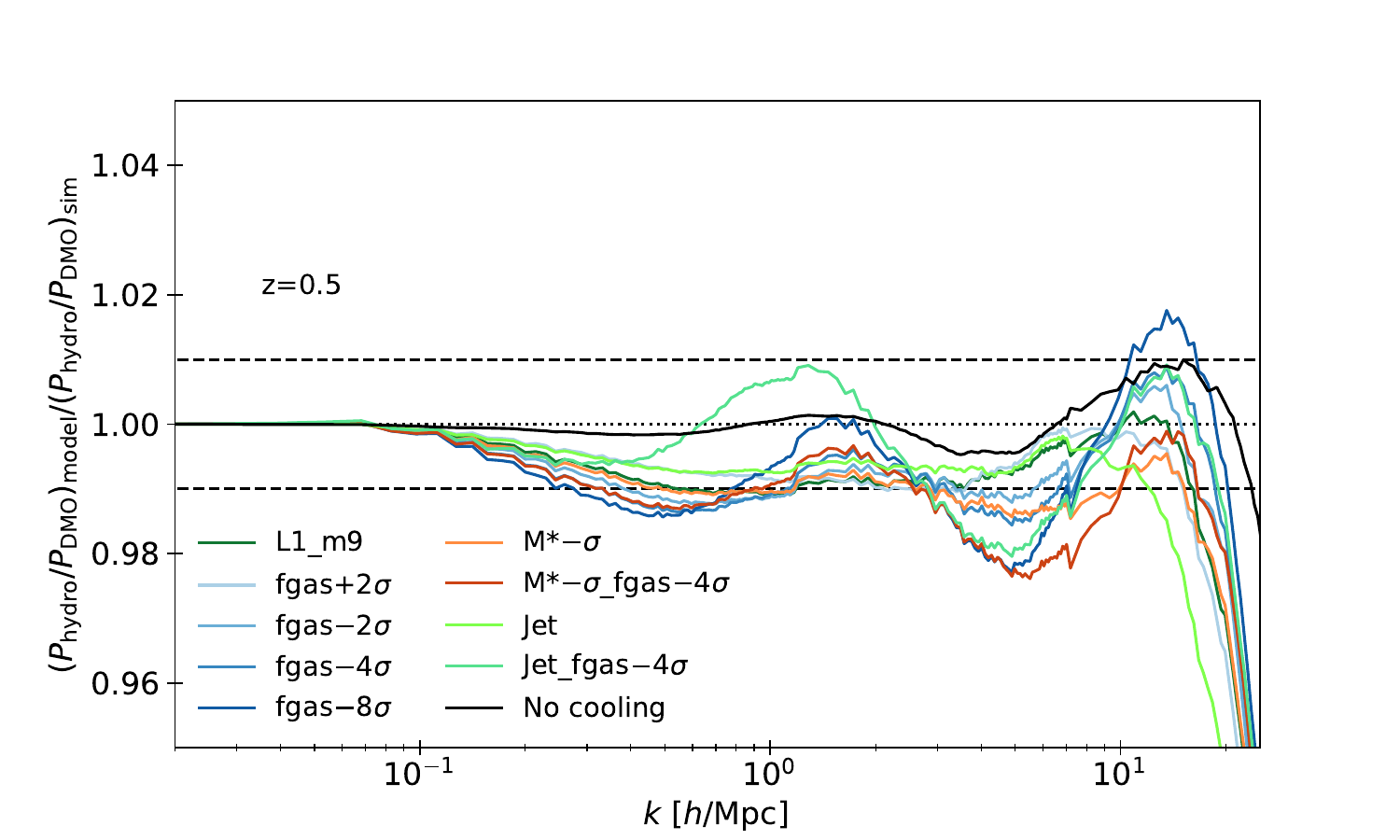}
\includegraphics[width=\columnwidth, trim=8mm 6mm 34mm 12mm]{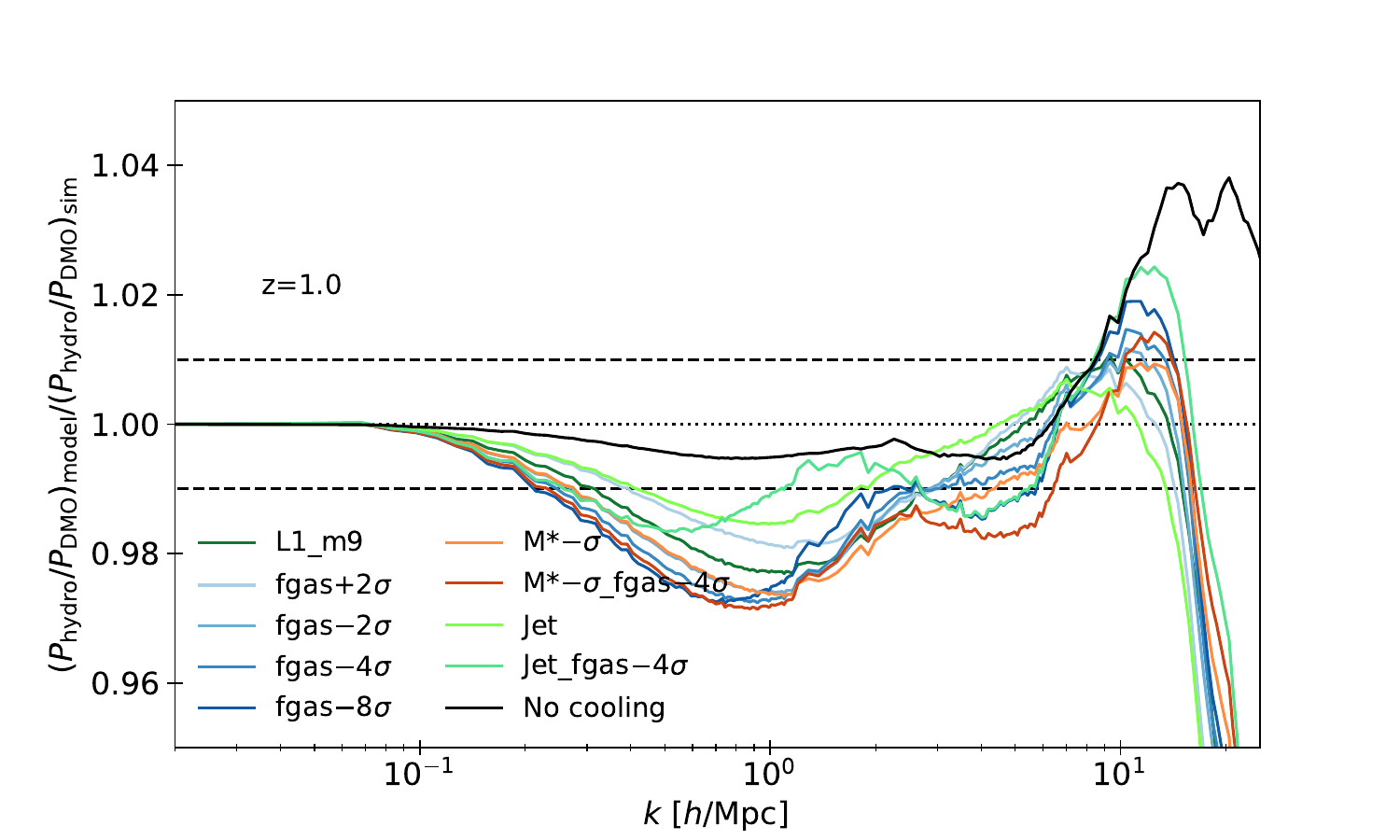}
\caption{Similar to the left-hand side of Fig.~\ref{fig:dpp_all_stellar_universal}, but for $z=0.5$ (left panels) and $z=1$ (right panels). At higher redshifts, the model starts to overpredict the suppression for given baryon fractions on large scales. This is due to less mass being clustered in (resolved) haloes than at $z=0$, which increases the relative importance of the non-halo cross-power component on all scales.}
\label{fig:dpp_z}
\end{figure*}

In Fig.~\ref{fig:dpp_z} we show the suppression signal and its model predictions for the feedback variations once more, but now for redshifts $z=0.5$ (left-hand side) and $z=1$ (right-hand side). The ``universal'' relation between the retained mass fraction and the reduced baryon fraction has a slight redshift dependence, so we refit this relation at each redshift in order to obtain model predictions (see Appendix~\ref{sec:app_z}).

Despite the adjusted retained mass fractions, the model increasingly overpredicts the amount of suppression on intermediate scales with redshift ($0.1\lesssim k\lesssim 3\,
h\,\mathrm{Mpc}^{-1}$), though the overprediction is still at most $3\%$ by $z=1$. By comparing the model-predicted halo-matter cross spectra in individual halo mass bins with those of the hydro simulations (not shown), we find that the cause is the non-halo component. While at $z=0$ the contribution of this component is always below that of the total halo contribution for $\radm$, by a factor of a few on intermediate scales (see the left panel of Fig.~\ref{fig:crosspower}), it becomes more important at higher redshifts, where more mass is not yet in (resolved) haloes. For $z=0.5$, the $\radm$ halo cross-power component only starts to dominate over the non-halo one at $\ksc{\approx}{0.15}$, and at $z=1$ this scale decreases to $\ksc{\approx}{0.45}$. Consequently, at $z>0$ the prediction becomes more sensitive to how accurately matter ejected from overdense regions is modelled.

In the standard resummation model presented in \S\ref{subsec:modelmassloss}, particularly equation~\eqref{eq:nonhalopower_corr}, we rescale the non-halo cross power from DMO uniformly, in such a way that we conserve the product of mass and bias on large scales. However, rather than scaling this component uniformly, it might be more accurate to only scale matter around (massive) haloes, where this material is likely ejected to, rather than scaling all non-halo matter.

We test this by identifying particles inside $r<5\times\radc$ of \emph{any} halo, irrespective of its mass, and calculate the cross power of these particles with all matter. One reason to take the combined region over all halo masses, rather than per halo mass bin, is that for such large radii the overlap between overdensity regions of haloes in different mass bins is significant. For $\radm$ regions with $0.5\,\mathrm{dex}$ mass bins, the percentage of the mass that is counted multiple times between mass bins is only $0.16\%$ (corresponding to $0.0065\%$ of all mass), but for $5\times\radc$ this would increase to $5.8\%$ (corresponding to $3.1\%$ of all mass).\footnote{Note that particles in the same halo mass bin are never counted more than once, since we flag particles that are in overdensity regions of a given halo mass bin rather than copying them into a catalogue.} In L1\_m9\_DMO at redshift zero, $40.5\%$ of the mass resides inside $\radm$ of some resolved halo, while $45.7\%$ is outside all $5\times\radc$, leaving $13.8\%$ of the mass in between these radii.

\begin{figure}
\includegraphics[width=\columnwidth, trim=12mm 34mm 24mm 12mm]{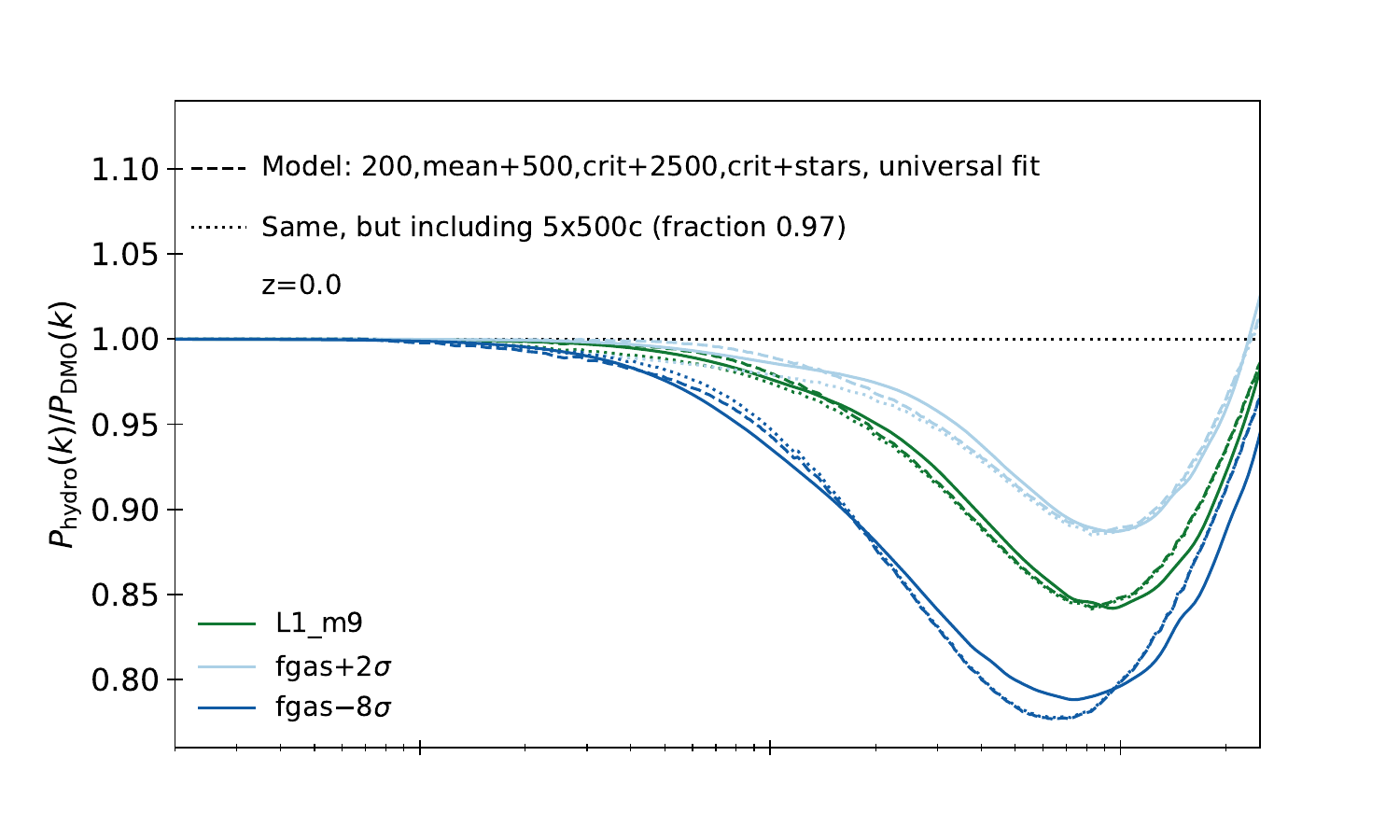}\\
\includegraphics[width=\columnwidth, trim=12mm 34mm 24mm 2mm]{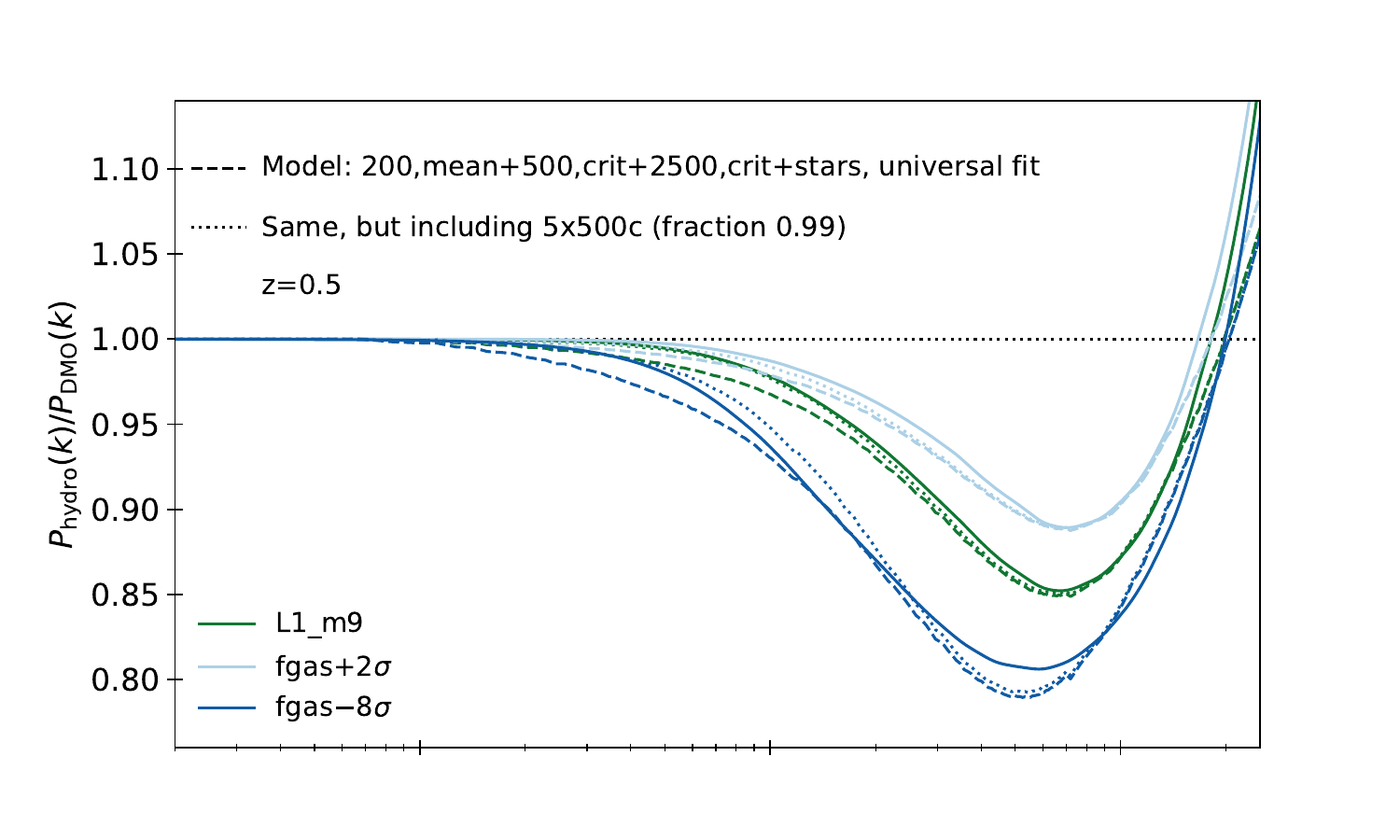}\\
\includegraphics[width=\columnwidth, trim=12mm 16mm 24mm 2mm]{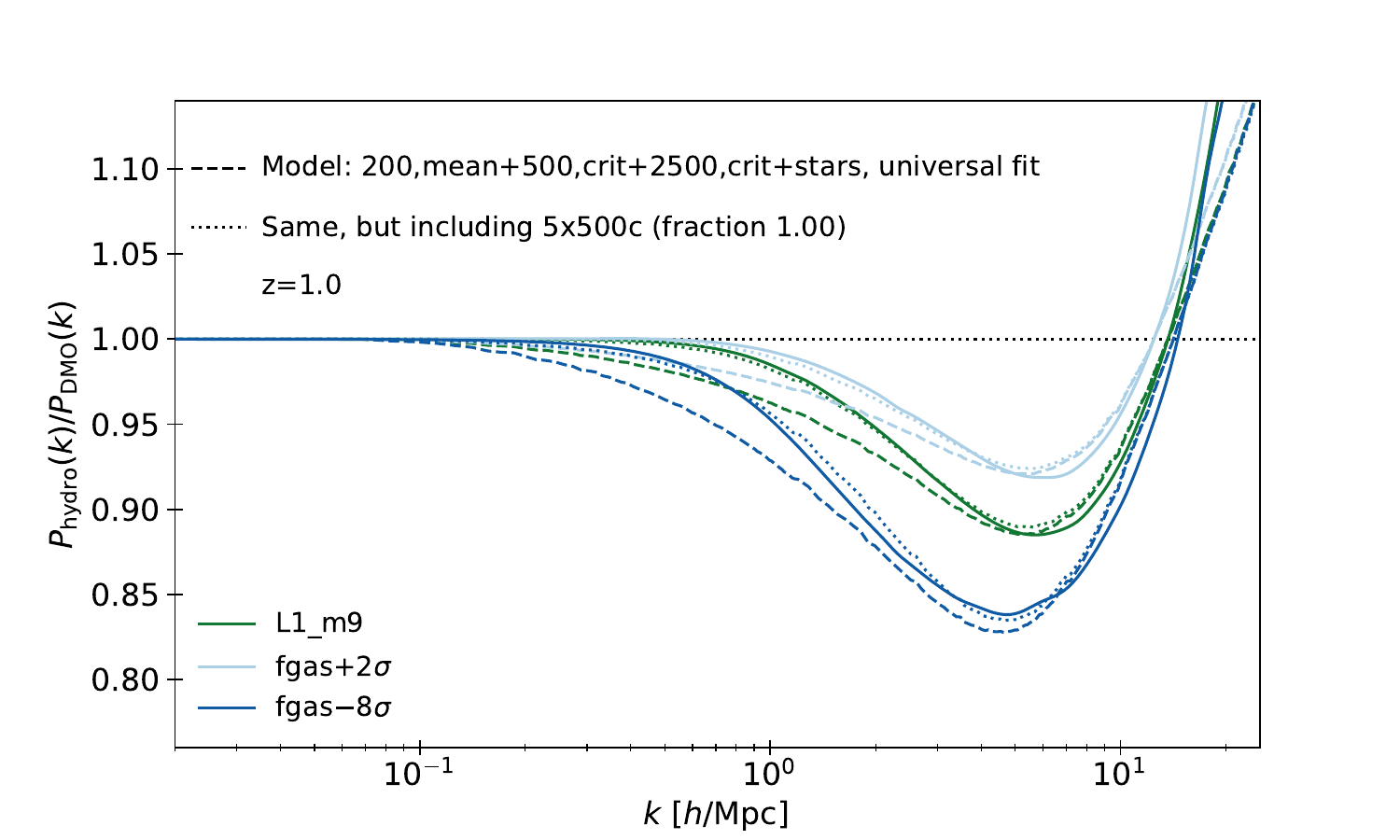}\\
\includegraphics[width=\columnwidth, trim=12mm 16mm 24mm 2mm]{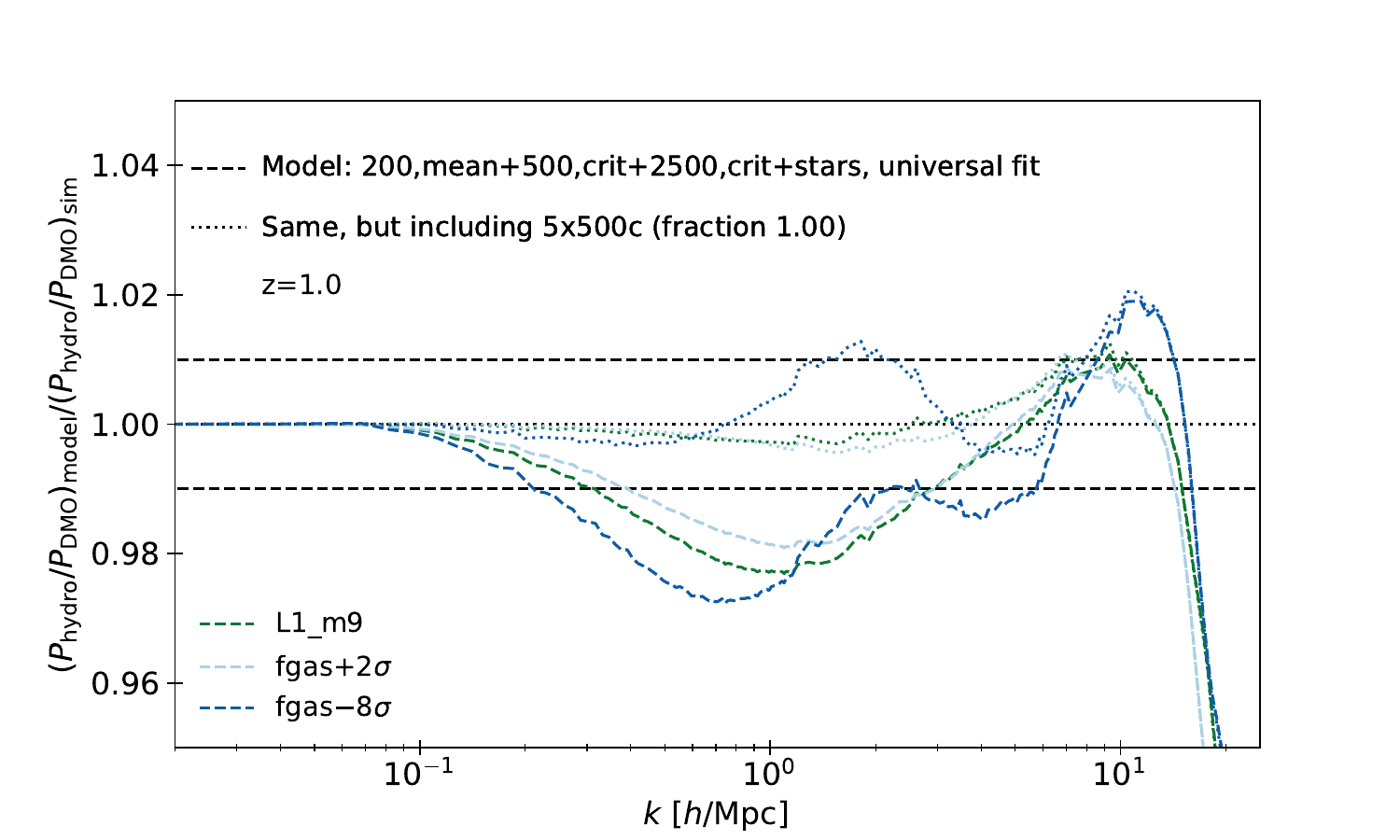}
\caption{Model results for the default model (dashed lines) and a 1-parameter extension that splits the non-halo matter into matter at distances $\radm<r<5\times\radc$ from any halo, and matter outside all $5\times\radc$ regions (dotted lines), applied to three \flamingo simulations. The additional parameter, $f_\mathrm{ret,5x500c}$, sets the total fraction of the DMO mass still retained within $5\times\radc$. Here we set $f_\mathrm{ret,5x500c}=0.97$ for $z=0$ (top panel), $0.99$ for $z=0.5$ (second panel), and $1.00$ for $z=1$ (third panel). The bottom panel shows the ratio of the suppression signal in the model relative to that of the simulations for $z=1$, showing that the additional parameter yields more accurate suppression signals at high redshift.}
\label{fig:dpp_z_farout}
\end{figure}

After obtaining the cross power of the combined $5\times\radc$ regions with all matter, $P_\mathrm{mh,5x500c}$, we subtract its product with its mass fraction, $f_\mathrm{M,5x500c}$, from the total matter power spectrum to get a new estimate for the non-halo contribution, which now only includes matter for which $r>5\times\radc$ for all haloes:
\begin{equation}
\label{eq:Pmnh5x500c}
P_\mathrm{mnh,5x500c}(k)=\Pmm(k)-f_\mathrm{M,5x500c}P_\mathrm{mh,5x500c}(k).
\end{equation}
We then subtract from $P_\mathrm{mh,5x500c}$ all previously calculated $r<\radm$ cross power components, yielding the contribution of matter at $\radm<r<5\times\radc$:
\begin{align}
\nonumber
P_\mathrm{mh,5x500c-200m}(k)&=f_\mathrm{M,5x500c}P_\mathrm{mh,5x500c}(k)-\\
\label{eq:Pmh5x500c-200m}
&\phantom{=}\,\,\sum_i\fMitwo\Pmhitwo(k).
\end{align}
Finally, we replace step (v) of the model (see \S\ref{subsec:model_summary}) such that a fraction $\fretcomb$ of the mass removed from $\radm$ regions is uniformly added to the $\radm<r<5\times\radc$ component, and the remaining fraction, $1-\fretcomb$, is added to the new non-halo component, again in such a way that the product of mass and bias is conserved. Setting $\fMcomb'\equiv\fretcomb\fMcomb$, equation~\eqref{eq:modelP} is then adapted as follows (once again omitting the explicit dependence on $k$ for conciseness):
\begin{align}
\nonumber
\Pmmp&=P_\mathrm{mnh,5x500c}'+P_\mathrm{mh,5x500c-200m}'+\Pmhtwo'\\
\nonumber
&=\frac{1-\fMcomb'\bMcomb}{1-\fMcomb\bMcomb}P_\mathrm{mnh,5x500c}+\\
\nonumber
&\phantom{=}\,\,\frac{\fMcomb'\bMcomb-\sum_i\fMitwop\bMitwo}{\fMcomb\bMcomb-\sum_i\fMitwo\bMitwo}P_\mathrm{mh,5x500c-200m}+\\
\label{eq:modelP_5x500c}
&\phantom{=}\,\,\sum_i \fMitwop\Pmhitwo,
\end{align}
where the last term can be further split into smaller regions as in \S\ref{subsec:r2500c_stars}.

What the value of the parameter $\fretcomb$ should be is not clear \emph{a priori}, though one would expect it to be close to unity. This is especially true at high redshift, where ejected matter has not had time to spread out over large scales. In Fig.~\ref{fig:dpp_z_farout}, we show the results of taking $\fretcomb=0.97$ at $z=0$, $\fretcomb=0.99$ at $z=0.5$, and $\fretcomb=1.0$ at $z=1$. In all cases, this new model accurately reproduces the simulation results, though at the cost of a single redshift-dependent parameter. In future work, we will explore ways to set its value based on observed baryon fractions or theory as well, and investigate other redshifts.

\section{Discussion}
\label{sec:discussion}
The improved resummation model presented in this work is able to reproduce the $z=0$ matter power suppression signal for a wide range of \flamingo simulations, without any free parameters. Based only on their halo spherical overdensity masses and mean halo baryon fractions, for example $\radc$ and $\radm$ or similar, the model yields percent-level accuracy up to $\ksc{\approx}{10}$, and even to $\ksc{\gtrsim}{20}$ if gas and stellar fractions are known inside $\radcc$ as well. Similar accuracy can be obtained at redshifts up to at least $z=1$ by adding a single parameter, although our results imply that this parameter may be independent of baryonic feedback and could therefore be a fixed function of redshift (see \S\ref{subsec:resum_z}). We will explore this further in future work. Even without this parameter, the resummation model achieves $\lesssim 3\%$ accuracy up to at least $z=1$ for all $\ksc{\lesssim}{10}$.

Below, we discuss several insights, strengths and limitations of the model.

\begin{figure*}
\includegraphics[width=\columnwidth, trim=17mm 6mm 25mm 12mm]{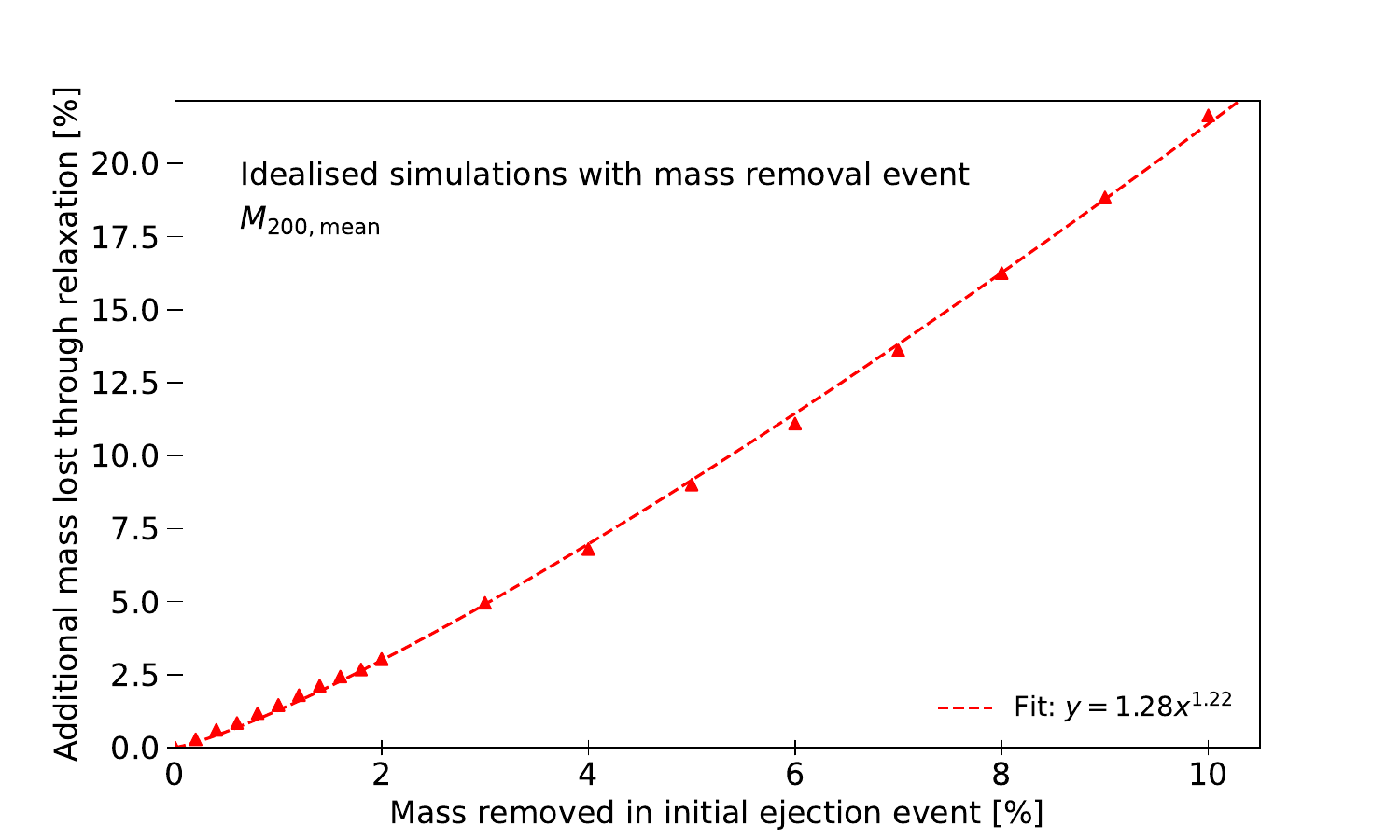}
\includegraphics[width=\columnwidth, trim=8mm 6mm 34mm 12mm]{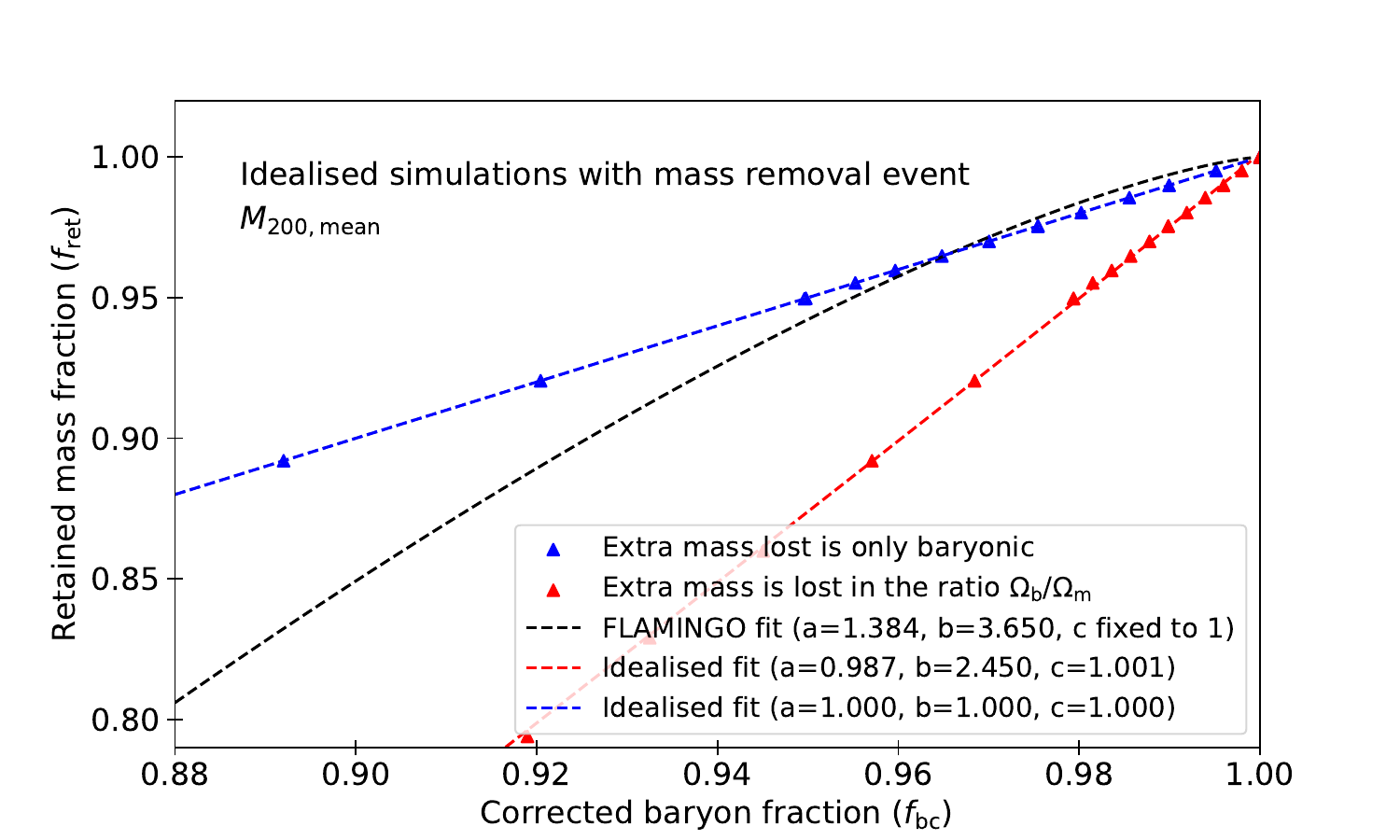}
\caption{Results from an idealised \swift DMO simulation where we remove some percentage of the halo mass from the centre of a $\Mm=\hM{14}$ halo, then let the halo relax. Mainly through relaxation, the halo loses additional mass after ejection. \textit{Left:} The percentage of extra mass lost, relative to the original halo mass of $\hM{14}$, as a function of the initial mass ejected. The extra mass lost is always at least as large as the ejected mass, and can exceed twice the ejected mass for large ejections. \textit{Right:} The retained halo mass fraction as a function of the final baryon fraction of the halo, which is assumed to have the cosmic baryon fraction before ejection. The initially removed mass is always assumed to be baryonic, but to calculate the final baryon fraction of these haloes, we assume either that the mass lost through relaxation has the cosmic baryon fraction (red points), or is only baryonic (blue points, by definition $f_\mathrm{ret}=f_\mathrm{bc}$). Also shown is the universal \flamingo fit (black dashed line), shifted down slightly so it converges to $f_\mathrm{ret}=1$, to allow more direct comparison to these results.}
\label{fig:idealized_sims}
\end{figure*}

\subsection{The origin of the retained mass fraction relation}
\label{subsec:discussion_fret}
Central to the resummation model is the $f_\mathrm{ret}-f_\mathrm{bc}$ relation, which, for a given halo overdensity region and redshift, accurately returns the mass ratio between a halo in a hydrodynamic simulation or in the real Universe and its DMO equivalent, as a function of its halo baryon fraction (see Fig.~\ref{fig:fret_fbc_200m_fixedhaloes_fixedmass}, \ref{fig:fret_fbc_500c_fixedhaloes_fixedmass} and Fig.~\ref{fig:fret_fbc_universal}, as well as Appendix~\ref{sec:app_z}). This relation is seemingly universal, at least within this family of simulations, showing no significant dependence on cosmology, feedback strength, or halo mass (at least for $\M\gtrsim \hM{12}$). The existence of this relation, and its slope exceeding unity for all but the most massive haloes, implies that when mass is lost due to feedback or other baryonic processes, a predictable amount of cold dark matter mass is removed from overdense regions as well. As mentioned early in \S\ref{subsec:fretvsfbc}, this is expected based on purely gravitational processes: after mass is removed from the centre, the density enclosed in the ejection radius will decrease, and so overdensity radii will also decrease, which means mass between the new and original overdensity radii will be ``lost''. Additionally, the halo potential will be lessened, reducing the binding energy of the particles, allowing some to stream out of the halo. Afterwards, a gravitational exchange of energy between halo particles will take place (halo relaxation), which will cause additional particles to move outward, or even become unbound, while for others the binding energy will increase. It is however not clear \emph{a priori} whether this should lead to a precisely predictable amount of (relative) mass loss.

To test whether this picture is accurate and how much freedom it allows, we ran a series of idealised DMO simulations using the \swift simulation code \citep{Schaller2024}. Using ICICLE \citep{Drakos2017}, we generate initial conditions for a $\Mm=\hM{14}$ exponentially-truncated NFW halo with concentration $c\approx 4.5$ using $138\,500$ dark matter particles, such that $10^5$ of these reside within $\radm$. We note that the choice of halo mass is unimportant, as the mass can be rescaled without affecting the experiment. Similarly, the exponential truncation helps to define a stable ``edge'' to the halo, but does not influence the outcome. After letting the halo settle into a steady state, we instantaneously remove some fraction $f_\mathrm{ej}$ of its $\Mm$ mass, crudely mimicking AGN feedback ejecting gas. We do so starting from the particle at the halo centre and moving radially outward until the desired fraction is reached, then remove these particles from the simulation.\footnote{We also tried choosing some inner radius, and removing a random subset of the particles inside it, corresponding to the mass fraction to be ``ejected''. The results are indistinguishable.} We then continue the simulation. Immediately after this idealised ejection event, relaxation occurs, and the halo overdensity mass behaves as a damped oscillator as a function of time before settling on a new equilibrium within several Gyr. We then measure the new overdensity mass and compare it to the mass just before ejection. Finally, we repeat the simulation a number of times with different values of $f_\mathrm{ej}$.

The results of this test are shown in Fig.~\ref{fig:idealized_sims}. On the left, we show the percentage of additional mass lost during relaxation as a function of the initial percentage ejected, $100\times f_\mathrm{ej}$. The dashed line shows a power-law fit, indicating that the fraction of extra mass lost is a slightly superlinear function of the fraction of mass ejected initially. Part of the former is due to the overdensity radius shrinking instantaneously after the ejection event, which lowers the mass by definition rather than by ejection, but the majority is lost during relaxation and truly ejected dynamically. For example, one can derive mathematically that for an NFW profile with $c=4.5$, after removing $2\%$ of the initial halo mass from the centre, $\radm$ becomes about $0.9\%$ smaller. This decreases $\Mm$ by less than $0.6\%$, in addition to the mass ejected. Meanwhile, as we can read from the left-hand panel of the figure, the total additional mass lost after relaxation in this case is about $3\%$, or more than five times the mass that the shrinking of the virial radius alone accounts for. For the largest ejection fraction probed here, $10\%$, this factor goes up to more than $7$.

On the right, we show what these idealised simulations predict for the $f_\mathrm{ret}-f_\mathrm{bc}$ relation. We calculate $f_\mathrm{ret}$ as the ratio of the final halo mass to the mass just before ejection. To calculate a final baryon fraction, we assume that the halo initially had the cosmic baryon fraction, and that all mass removed in the ejection event is baryonic. We then need to make an assumption about the mass removed. For the red points, we assume that the extra mass lost during relaxation has the cosmic baryon fraction as well, while for the blue points, we assume that it is instead only baryonic matter (note that this affects only $f_\mathrm{bc}$, not $f_\mathrm{ret}$). By construction, the latter assumption results in $f_\mathrm{ret}=f_\mathrm{bc}$ (see equation~\ref{eq:fbc}). We fit equation~\eqref{eq:fretained} to these results, shown as red and blue dashed lines. The black dashed line is the fit to \flamingo shown in the left-hand panel of Fig.~\ref{fig:fret_fbc_200m_fixedhaloes_fixedmass}, shifted vertically to converge to $(1,1)$ as well. The more dark matter would be lost, relative to baryonic matter, the steeper the relation found.

We see that in this idealised setup, the $f_\mathrm{ret}-f_\mathrm{bc}$ relation is highly dependent on the ratio of baryonic to dark matter ejected during relaxation. Despite this, we can conclude that relaxation following ejection certainly has the potential to explain why the slope of the relation is significantly steeper than unity in all \flamingo simulations. Looking at the slope of the relation explicitly, the results imply that haloes with high baryon fractions remove mostly baryonic mass through relaxation, while haloes with relatively low baryon fractions may remove mass in a baryons-to-dark matter ratio of approximately $\Omega_\mathrm{b}/\Omega_\mathrm{m}$.

\begin{figure*}
\includegraphics[width=\columnwidth, trim=17mm 0mm 25mm 12mm]{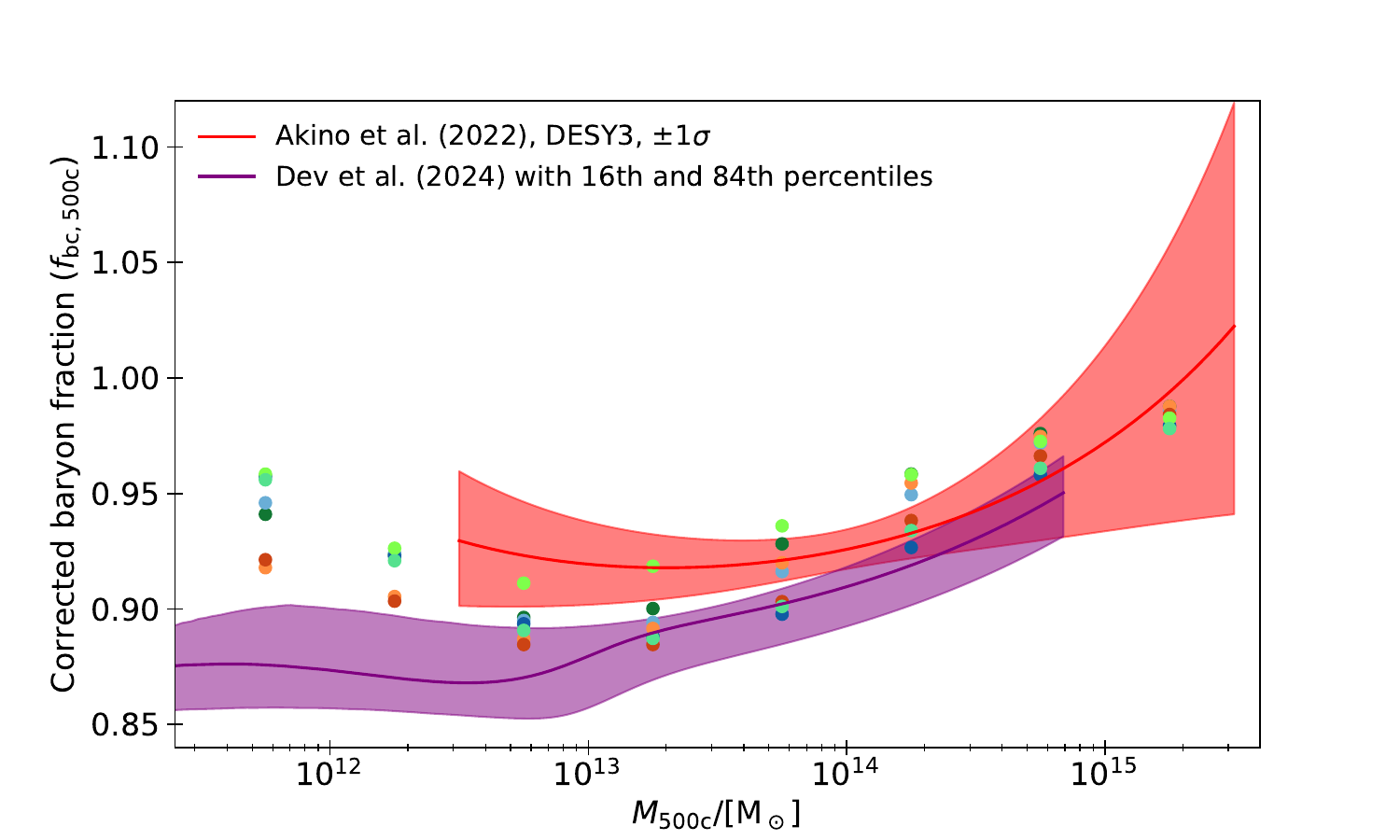}
\includegraphics[width=\columnwidth, trim=8mm 0mm 34mm 12mm]{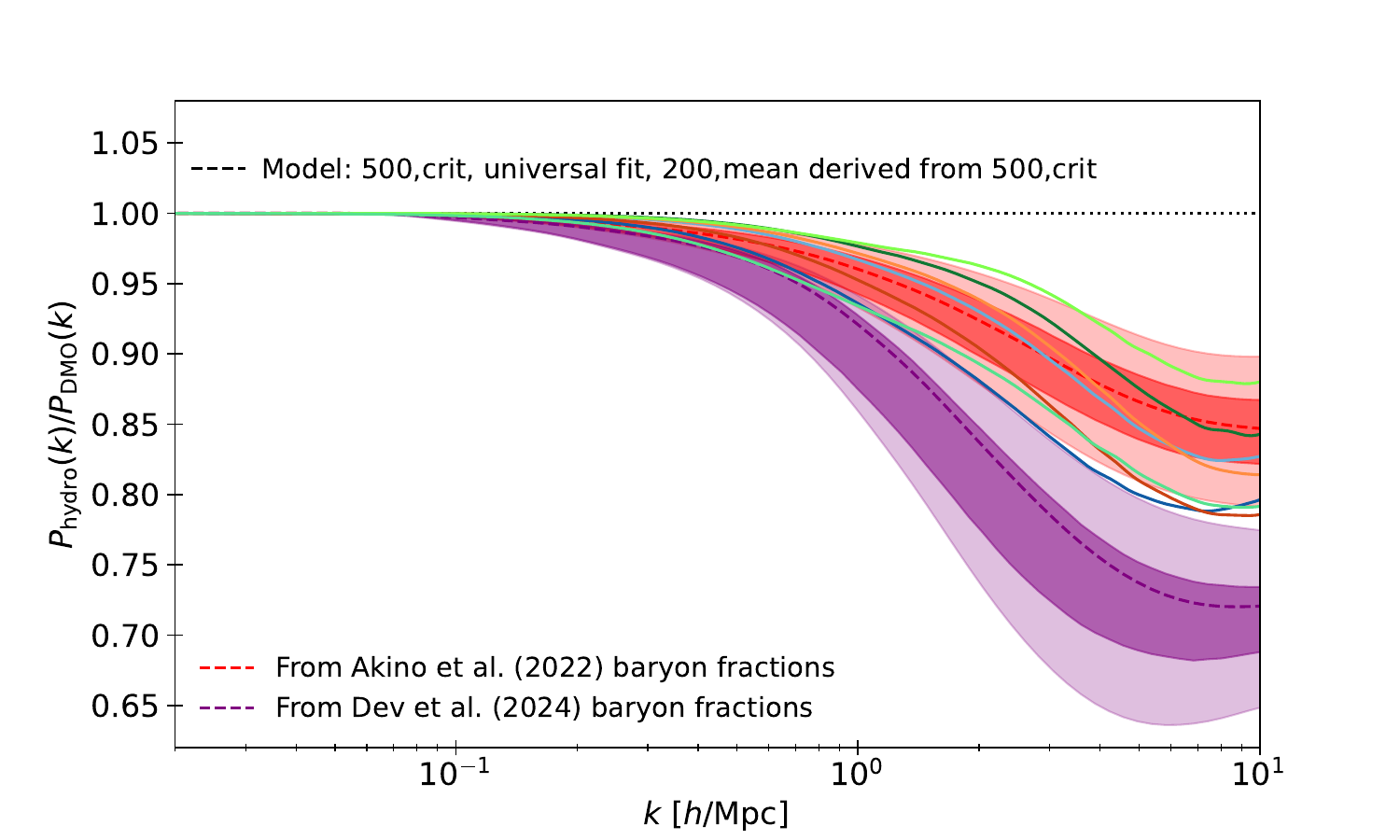}\\
\includegraphics[width=\columnwidth, trim=8mm 8mm -2mm 0mm]{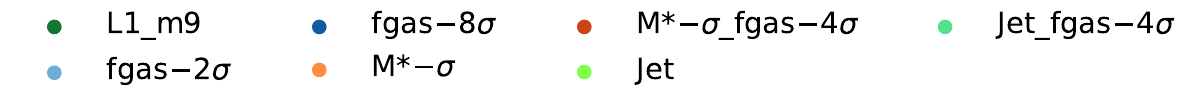}
\includegraphics[width=\columnwidth, trim=-2mm 8mm 8mm 0mm]{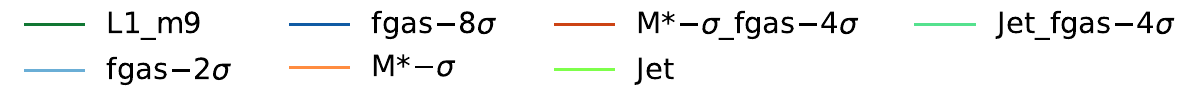}
\caption{A comparison with observations. \textit{Left:} The corrected baryon fraction within $\radc$ as a function of $\M$, derived from the baryon fractions presented in \citet{Akino2022} (red) and \citet{Dev2024} (purple), as well as for several \flamingo simulations. Shaded regions show the confidence intervals around the observational results ($\pm 1\sigma$ and 16th and 84th percentiles, respectively). \textit{Right:} Using these baryon fractions and extrapolating them to $\radm$ (see \S\ref{subsec:annulus_or_not}), we calculate their expected matter power spectrum suppression signals. Note the different axis ranges compared to previous figures. Dark shaded regions around the model prediction show the expected range in suppression if we draw randomly from the observed distribution in $f_\mathrm{b}$, assuming independent Gaussian errors. Light shaded regions show the expected range if the errors are fully correlated. Of the \flamingo simulations shown here, only Jet\_fgas$-4\sigma$ and fgas$-8\sigma$ are consistent with the \citet{Dev2024} result for $\ksc{\lesssim}{1}$, and only fgas$-8\sigma$ is at least marginally consistent with it for all $\ksc{\lesssim}{10}$. Meanwhile, all simulations shown are at least marginally consistent with \citet{Akino2022}.}
\label{fig:fbc_dpp_obs}
\end{figure*}

Given the dependence on the baryon-to-CDM mass ratio of additional mass removed, the results of these idealised simulations do not explain the universality of the $f_\mathrm{ret}-f_\mathrm{bc}$ relation. Processes that were ignored here may play an important part there. For example, in reality haloes experience many small ejection events throughout their lifetime, rather than a single large event, and additionally reaccretion and gas cooling flows would change how the halo profile is restored following ejection.

Finally, some dependence on the halo profile is expected, as its shape should affect the outcome of relaxation. Repeating this experiment for haloes with different masses and concentrations (not shown), we find only a weak dependence on concentration, where haloes with higher concentrations lose slightly less additional mass. This is in line, qualitatively, with the results of \citet{Elbers2025}. The \flamingo results presented in this work, including those with different cosmologies, imply that any profile dependence (along with any other dependencies) of the $f_\mathrm{ret}-f_\mathrm{bc}$ relation must indeed be small.

\subsection{Predictions based on observed baryon fractions}
\label{subsec:discussion_obs}
Similar to \texttt{SP(k)} \citep{Salcido2023}, the resummation model directly maps observed baryon fractions and their uncertainties to the suppression of the matter power spectrum, though without assuming a functional form for the $f_\mathrm{b}-\M$ relation. While baryon fractions measured within any overdensity radius can be used with our model, fractions within $\radc$ ($f_\mathrm{b,500c}$) are currently most readily available. Lacking precise observations at larger radii, like $\radm$, we can use the results of \S\ref{subsec:annulus_or_not} to predict $f_\mathrm{ret,200m}$ from $f_\mathrm{b,500c}$.

The left-hand panel of Fig.~\ref{fig:fbc_dpp_obs} shows corrected baryon fractions within $\radc$ for several \flamingo simulations. Also shown are observed baryon fractions for two compilations: \citet{Akino2022}, in red, and \citet{Dev2024}, in purple. The \citet{Akino2022} data was derived for an X-ray-selected sample of 136 galaxy groups and clusters (HSC-XXL). The version of the data shown here was scaled to the fiducial \flamingo cosmology \citep[DESY3, see][]{Schaye2023}, and a $\pm 1\sigma$ confidence interval was derived using Monte Carlo simulation \citep[][private communication]{Salcido2025}. The \citet{Dev2024} data was compiled by combining observations across a wide range of wavelengths, and includes cold gas, hot gas (X-ray) and stars. These authors used eROSITA X-ray data that was not X-ray selected, which yield substantially lower gas fractions than pre-eROSITA X-ray-selected data. The halo masses for this data set were converted from $M_\mathrm{200c}$ to $\M$ by undoing the same conversion factor employed by the authors, $\M/M_\mathrm{200c}\approx 0.69$. Quantiles for these baryon fractions were provided by the authors. For both data sets, we assume the DESY3 cosmological baryon fraction to convert the baryon fractions to corrected baryon fractions. We see that the \citet{Dev2024} baryon fractions are lower than those of \citet{Akino2022}, implying stronger feedback, as the authors themselves have noted. Indeed, above $\M\approx\hM{12.5}$ only the most extreme feedback models in \flamingo are consistent with the baryon fractions of \citet{Dev2024} within $\approx 1\sigma$, and below this halo mass scale none are.

We use these observed baryon fractions as input to the resummation model, linearly extrapolating them in $(\log_{10}\M,f_\mathrm{bc,500c})$ where needed, although we note that the most important mass range for predicting the suppression signal is covered by both data sets (see \S\ref{subsec:barfrac_response}). In addition, we use the results of \S\ref{subsec:annulus_or_not} to predict $f_\mathrm{ret,200m}$ from $f_\mathrm{b,500c}$. Finally, we probe the uncertainty regions around the mean observed baryon fractions in two ways: once by assuming the uncertainties are fully correlated, meaning we use the upper or lower edge of the envelope around the data as input to the model; and once by assuming no correlations, by drawing random mean halo baryon fractions in each halo mass bin independently, interpreting the uncertainty regions of the \citet{Dev2024} data as $1\sigma$ Gaussian uncertainties as well. In the latter case, we take into account the asymmetry of the uncertainties by using the 84th and 16th percentiles to estimate $1\sigma$ uncertainties independently above and below the data. The results are shown in the right-hand panel of Fig.~\ref{fig:fbc_dpp_obs}. Since we did not use small-scale information here, we have decreased the range in $k$ compared to previous figures. We also expanded the vertical range to fully show the strong suppression predicted by the \citet{Dev2024} baryon fractions.

As the results of \S\ref{subsec:barfrac_response} implied, the contributions of different halo masses can partially counteract each other, e.g.\ a higher baryon fraction in one halo mass bin can compensate for a lower baryon fraction in another. Because of this, the uncertainty regions assuming uncorrelated errors (dark shaded regions) can be much tighter than those assuming fully correlated errors (light shaded regions), at least on scales small enough that one or two halo mass bins do not dominate the signal (see Fig.~\ref{fig:dpp_scatter}). The \flamingo simulations shown here are all at least marginally consistent (within $1\sigma$) with the \citet{Akino2022} results assuming the latter case, with only fgas$-2\sigma$ being consistent for all $k$ shown if the errors are uncorrelated, and M$^*$$-\sigma$ almost so. The suppression predicted by the \citet{Dev2024} baryon fractions, however, is so strong that of the simulations shown in Fig.~\ref{fig:fbc_dpp_obs}, only Jet\_fgas$-4\sigma$ and fgas$-8\sigma$ are consistent with it within $1\sigma$ for $\ksc{\lesssim}{1}$. On smaller scales, just fgas$-8\sigma$ is marginally consistent with this result within $1\sigma$, and that only if we can assume fully correlated errors. This is in line with recent results from kSZ studies \citep[e.g.][]{Bigwood2024,McCarthy2025,Hadzhiyska2025b,Kovac2025} as well as more up-to-date X-ray studies \citep[e.g.][]{Ferreira2024}, particularly those using eROSITA data \citep[e.g.][]{Grandis2024,Kovac2025}, which all imply significantly lower baryon fractions than previous X-ray studies like \citet{Akino2022}. Very recent ACT tSZ \citep{Dalal2025} and FRB \citep{ReischkeHagstotz2025} studies also indicate stronger feedback than the \citet{Akino2022} data implies, though the uncertainties on both data sets are still large. This suggests that simulations with even stronger feedback should be explored in the near future.

\vspace{0.2cm}
\noindent
One may wonder whether the resummation model could be inverted to yield baryon fractions, given a suppression signal, similar to what \citet{Salcido2025} did for \texttt{SP(k)}. These could then be directly compared to observations. However, our model treats the baryon fraction of every halo mass bin as independent, which gives it great flexibility, but does mean that there is also a strong amount of degeneracy when trying to invert the model. Therefore, we would need to assume a functional form for $f_\mathrm{b}$ as a function of $\M$; we find that a sigmoid function, specifically the hyperbolic tangent function, provides a great fit to the simulation results for $\M\gtrsim\hM{12.5}$. Even in this case, there is still some degeneracy between halo mass bins, and between the baryon fractions of different overdensity regions. If we additionally assume the mapping between $f_\mathrm{ret,500c}$ and $f_\mathrm{b,200m}$ from \S\ref{subsec:barfrac_response}, we can break most of the remaining degeneracy and recover the baryon fractions of several \flamingo simulations. However, when minimising the difference between the simulated suppression signal and the model-predicted one as a function of the $f_\mathrm{b,500c}$ sigmoid's parameters, the high-mass baryon fractions are sometimes driven to unrealistically large values, as the $f_\mathrm{ret,200m}$ values saturate at high $f_\mathrm{ret,500c}$ (see the left-hand panel of Fig.~\ref{fig:fret_fret_500c}). We leave improving the model inversion to future work.

\subsection{Other applications: halo mass function}
\label{subsec:discussion_hmf}
In future work, we plan to explore other observables that can be predicted with the resummation framework. However, there is one quantity that can be straightforwardly predicted from just the $f_\mathrm{ret}-f_\mathrm{bc}$ relation: the DMO halo mass function (HMF), given observed halo masses and their baryon fractions. By simply dividing the observed masses, which include dark matter and baryons, by the predicted retained mass fraction, we recover the average DMO halo mass those haloes would have had in a dark-matter-only universe.

\begin{figure}
\includegraphics[width=\columnwidth, trim=12mm 16mm 24mm 14mm]{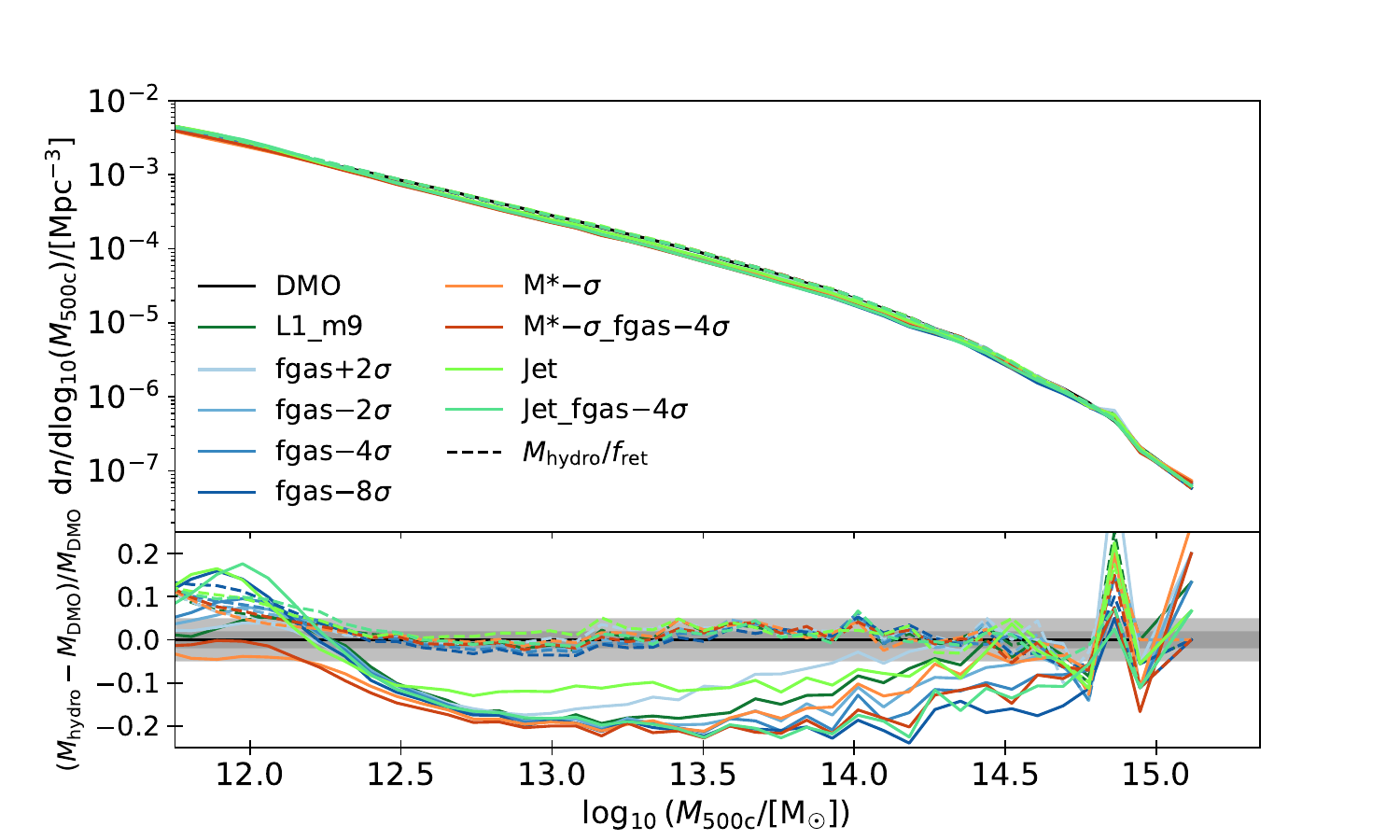}
\caption{The $\M$ halo mass function (HMF) for L1\_m9\_DMO and various hydrodynamical variations. Shown as dashed lines are the HMFs we obtain when we divide the halo masses in the hydrodynamical simulations by their expected retained mass fractions, obtained from their baryon fractions. The bottom panel shows that without this correction (solid lines), the HMF shifts downward by up to $\approx 20\%$ depending on the strength of feedback, over a wide range of mass scales. Grey shaded regions show deviations of $2\%$ and $5\%$. After correcting the hydrodynamical masses with the retained mass fractions (dashed lines), the DMO HMF is recovered to percent-level accuracy over more than two decades of mass for all simulations.}
\label{fig:hmf_fret}
\end{figure}

We show the result of applying the mass-independent, ``universal'' $f_\mathrm{ret}-f_\mathrm{bc}$ relation of Fig.~\ref{fig:fret_fbc_universal} to the halo masses of several \flamingo simulations in Fig.~\ref{fig:hmf_fret}, binning the resulting masses to obtain the HMF. We show the HMF down to the mass scale corresponding to $100$ DMO particles, and the three most massive bins are combined into one to ensure that every bin contains at least $10$ DMO haloes. In the top part of the figure we show the HMF itself, and in the bottom part we examine the relative difference between all HMFs and the DMO HMF. Grey shaded regions show deviations of $2\%$ and $5\%$. The results for the unscaled hydro masses are similar to the results presented in \citet[][their Fig.~20]{Schaye2023}, but now presented for $\M$: the different \flamingo simulations shift the HMF downward by up to $\approx 20\%$ due to feedback (solid lines), with the effect extending to higher halo masses for simulations with stronger AGN feedback. However, when we divide the masses by the predicted retained mass fraction for their mean halo baryon fractions (dashed lines), the effects of feedback are compensated over more than two decades of halo mass, and we recover the DMO HMF to percent-level accuracy. While the $f_\mathrm{ret}-f_\mathrm{bc}$ relation used here was fit only to haloes from the L2p8\_m9 simulation with $\M\geq\hM{12}$, it applies to all \flamingo simulations, and divergence at lower halo masses is slow.

We note that \citet{Euclid2024} recently also explored a relation to map halo masses to their DMO equivalent to recover the (cumulative) HMF, and they too start from what we call the corrected baryon fraction. However, their equivalent of the retained mass fraction is defined by adding a parameter $\delta_f$ to the measured baryon fraction instead, which they calibrate to simulations. The accuracy they obtain is similar to ours, however, the relation of the current work is applicable over a much wider range in halo mass scales and feedback strengths, without requiring recalibration.

The retained mass fraction given a halo baryon fraction is one of the data products returned by the \texttt{resummation} Python package, and the code can thus easily be used to calculate rescaled halo mass functions as well.

\subsection{Strengths and limitations}
\label{subsec:discussion_strlim}
 The main strength of the resummation model is that it allows halo baryon fractions to be mapped directly to a matter power suppression signal with zero free parameters, meaning that observed baryon fractions could be straightforwardly marginalised over to give a confidence interval around the predicted suppression. In this way it is similar to \texttt{SP(k)} \citep{Salcido2023} -- however, the resummation model does not assume any scaling law between baryon fractions and halo masses, and is much more flexible in the suppression signals it can generate, having not been trained on any simulated suppression signals.
 
 Compared to the baryonification/baryon-correction models \citep[e.g.][]{SchneiderTeyssier2015,Schneider2019,Arico2020,Arico2021,Schneider2025}, which can also predict quantities beyond the suppression signal and can operate at the field level, the applicability of the resummation model is more narrow. However, given appropriate halo baryon fractions, the resummation model again has zero free parameters, at least at low redshift, and does not assume any functional forms for density profile shapes, whereas the default baryonification models have seven free parameters and do require these assumptions. For the same reasons, the resummation model compares favourably to halo-model based approaches \citep[e.g.][]{Semboloni2011,Zentner2013,Debackere2020,Pandey2025}, including the popular \textsc{hmcode} \citep{Mead2015,Mead2021}, which contain many parameters that are trained on a limited set of DMO and hydrodynamical simulations, typically leaving at least one parameter modelling the strength of AGN feedback free. The resummation model, then, offers a great alternative to less flexible models for the suppression signal, and could be complementary to baryonification-based approaches: using its predictions for the retained mass fractions and suppression signals may help constrain the parameter space of baryonification models, allowing for more precise predictions of other quantities like halo pressure and temperature profiles.

The main underlying assumption of our model is that the $f_\mathrm{ret}-f_\mathrm{bc}$ relation -- linking the mean halo baryon fraction to the fraction of mass lost due to baryonic effects, relative to DMO -- which applies to all well-resolved \flamingo haloes regardless of their baryonic physics, is truly universal, i.e.\ that it also applies to haloes in the real Universe. We have checked that haloes from the cosmo-OWLS \citep{LeBrun2014} and BAHAMAS \citep{McCarthy2017} simulations indeed lie on top of the same relation we present here -- despite their cosmic baryon fractions lying outside the range probed by \flamingo -- and encourage other simulators to do the same. It would be particularly interesting to see if this relation holds for simulations with different gravity and/or hydrodynamics solvers or subgrid implementations than those used here, as these may impact the result of relaxation. The results of Appendix~\ref{sec:app_resolution} do indicate that the relation may shift by $\sim 1$ percentage point for lower-mass haloes when the simulation's mass resolution is improved by a factor of 8, which we will investigate further in future work.

The second-most important assumption of our model is that the non-resolved-halo matter component is sub-dominant on non-linear scales, and that we therefore do not need to accurately model the distribution of ejected matter. As we have shown in \S\ref{subsec:resum_z}, this assumption breaks down at high redshifts, although our results there indicate that adding a possibly also universal but redshift-dependent parameter ($\fretcomb$) could solve this issue. Some less important assumptions taken include that the linear halo bias is preserved as haloes lose mass (though we find that the relative change in bias is typically $\lesssim 1\%$), and that the halo positions do not change due to feedback in a way that significantly impacts their clustering \citep[justified by the results of][]{vanDaalen2014,Zennaro2025}.

We note that, through the retained mass fraction relation, the resummation model links the baryon fractions measured within some radius $R_\Delta$ to the mass contained inside a radius $R_{\Delta,\mathrm{DMO}}$ for some DMO halo with the same total mass. Since these masses are equal, the radii are equal as well, and this contributes to the rescaling of the halo-matter cross spectra yielding such accurate results. As shown in the left-hand panels of Fig.~\ref{fig:fret_fbc_200m_fixedhaloes_fixedmass} and Fig.\ref{fig:fret_fbc_500c_fixedhaloes_fixedmass}, an almost equally tight relation exists between the mean baryon fraction of a halo and its retained mass fraction -- compared to its equivalent halo in a DMO universe -- which could in principle be used for the same purposes (possibly including some interpolation in mass). However, in this case the radii $R_\Delta$ and $R_{\Delta,\mathrm{DMO}}$ linked together are \emph{not} equal, and the resulting resummation-type model would yield less accurate results.

For applications to observations, a limitation is that it is difficult to measure the spherical overdensity masses and gas fractions. These generally require the assumption of spherical symmetry and a density profile. In future work, we plan to explore whether the model can be adapted to work with more readily observable halo masses.

Finally, a limitation of the model is that it requires DMO halo-matter cross spectra for the cosmology that matches the one assumed when calculating the (corrected) baryon fractions, and the variation in the cosmologies explored in \flamingo is limited. In future work, we therefore aim to present an emulator for DMO halo-matter cross spectra as a function of cosmology.

\section{Summary}
\label{sec:summary}
In order to derive unbiased cosmological parameters from Stage-IV weak lensing surveys like Euclid, we need models that can predict the matter power spectrum up to at least $\ksc{\approx}{10}$ with percent-level accuracy. The main challenge in this endeavour, as first put forward in \citet{vanDaalen2011}, is that baryonic feedback significantly redistributes matter on large scales, relative to dark matter alone, the details of which are still not understood to the required level of precision. Many modelling efforts and mitigation strategies based on the results of hydrodynamical simulations exist
, but these all assume functional forms for halo density profiles or the suppression signal itself, have free parameters that need to be marginalised over, or both. Here, we presented an improved version of the resummation model first described in \citet{vanLoonvanDaalen2024}, which maps observed halo baryon fractions of groups and clusters to a flexible suppression signal without any free parameters, with a typical accuracy of $\lesssim 1\%$.

Conceptually, the model takes observed halo masses and baryon fractions within one or more spherical overdensity radii $R_\Delta$, translates these into retained mass fractions $\fret\equiv M_{\mathrm{hydro},i,\Delta}/M_{\mathrm{DMO},i,\Delta}$ for different halo mass bins $i$ through the seemingly mass-independent and feedback-independent $f_\mathrm{ret}-f_\mathrm{bc}$ relation presented in \S\ref{subsec:modelmassloss}, and uses these to rescale DMO halo-matter cross spectra to account for the mass lost due to baryonic feedback. The mass removed from haloes is assumed to cluster similarly to matter already outside haloes, allowing us to rescale the DMO non-resolved-halo cross spectrum as well. All these cross spectra are then \emph{resummed} to obtain a new matter power spectrum including baryons. Extensions to larger or smaller scales are straightforward (see \S\ref{subsec:modelannulus}-\ref{subsec:r2500c_stars}), and allow for $\sim 1\%$ accuracy up to $\ksc{\approx}{25}$ given stellar and gas fractions within the inner halo. The steps of the model are summarised in \S\ref{subsec:model_summary}, and a Python package called \texttt{resummation} is made publicly available. For a given \flamingo cosmology, the model maps halo baryon fractions to power suppression predictions in $<1\,\mathrm{ms}$, allowing for fast iterations to marginalise over the uncertainties of observed baryon fractions of each mass bin independently.

We have quantified the contributions to the suppression signal of haloes of different masses (\S\ref{subsec:barfrac_response}), explaining and confirming the findings of e.g.\ \citet{vanDaalen2020} on the dominant role played by haloes around $\M=\hM{14}$, in line with \citet{vanLoonvanDaalen2024}. We have shown that the baryon fraction within $\radc$ is also a strong predictor of the retained mass fractions on larger scales as well (\S\ref{subsec:annulus_or_not}), allowing the resummation model to yield accurate predictions even when only $f_\mathrm{b,500c}(\M)$ is available. We have also shown that the model's assumption about the minor role of non-resolved-halo matter starts to break down at high redshift, but that this can be accounted for as well (\S\ref{subsec:resum_z}) -- though whether this too is model-independent remains to be seen. Finally, in Section~\ref{sec:discussion}, we have shown that halo relaxation following ejection likely shapes the $f_\mathrm{ret}-f_\mathrm{bc}$ relation (\S\ref{subsec:discussion_fret}), applied our model to observed baryon fractions within $\radc$ to show the suppression signals they predict (\S\ref{subsec:discussion_obs}), shown that the $f_\mathrm{ret}-f_\mathrm{bc}$ relation can also be used to accurately recover the DMO HMF from the observed halo mass function (\S\ref{subsec:discussion_hmf}), and discussed the strengths and limitations of the model (\S\ref{subsec:discussion_strlim}).

Future work on the resummation model could include: improved modelling of the non-resolved-halo component, particularly for high $z$; emulation of DMO halo-matter cross spectra; investigating the dependence of the $f_\mathrm{ret}-f_\mathrm{bc}$ relation on resolution, numerical solvers and subgrid implementation, as well as its dependence on $\Omega_\mathrm{b}/\Omega_\mathrm{m}$ outside the limited range probed by \flamingo; reformulating the model in terms of multiples of $\radc$ and/or more readily observable halo mass definitions; inverting the model to yield baryon fractions given a suppression signal, without degeneracies; and improving the model in the presence of asymmetric redistributions of matter, such as through jets.

\section*{Acknowledgements}
We thank Jaime Salcido for providing the baryon fractions and Monte Carlo errors derived from \citet{Akino2022}, converted to the D3A cosmology. We thank Ajay Dev for providing the baryon fractions and percentiles derived from the combined dataset of \citet{Dev2024} and a useful discussion on their cosmological and mass definition dependencies. We also thank the anonymous referee for their useful comments, which have led to the improvement of this manuscript.

This work used the DiRAC@Durham facility managed by the Institute for Computational Cosmology on behalf of the STFC DiRAC HPC Facility (\url{www.dirac.ac.uk}). The equipment was funded by BEIS capital funding via STFC capital grants ST/K00042X/1, ST/P002293/1, ST/R002371/1 and ST/S002502/1, Durham University and STFC operations grant ST/R000832/1. DiRAC is part of the
National e-Infrastructure.

\section*{Data Availability}
The halo-matter DMO cross spectra, as well as baryon fractions and stellar fractions for the \flamingo simulations used in this work, are publicly available as part of the Python package \texttt{resummation}. This can be installed via the \texttt{PyPI} package index (\texttt{pip install resummation}) or downloaded from GitHub\footnote{\url{https://github.com/FLAMINGOSIM/Resummation}}, which includes examples. The matter power spectra of all \flamingo simulations are available on the \flamingo project's web page.\footnote{\url{https://flamingo.strw.leidenuniv.nl/}}

The full FLAMINGO simulation data will eventually be made publicly available, though we note that the data volume (several petabytes) may prohibit us from simply placing the raw data on a server. In the meantime, people interested in using the simulations are encouraged to contact the corresponding author.



\bibliographystyle{mnras}
\bibliography{resummation.bib} 




\appendix
\section{Bias measurements}
\label{sec:app_bias}
\begin{figure*}
\includegraphics[width=\columnwidth, trim=17mm 0mm 25mm 12mm]{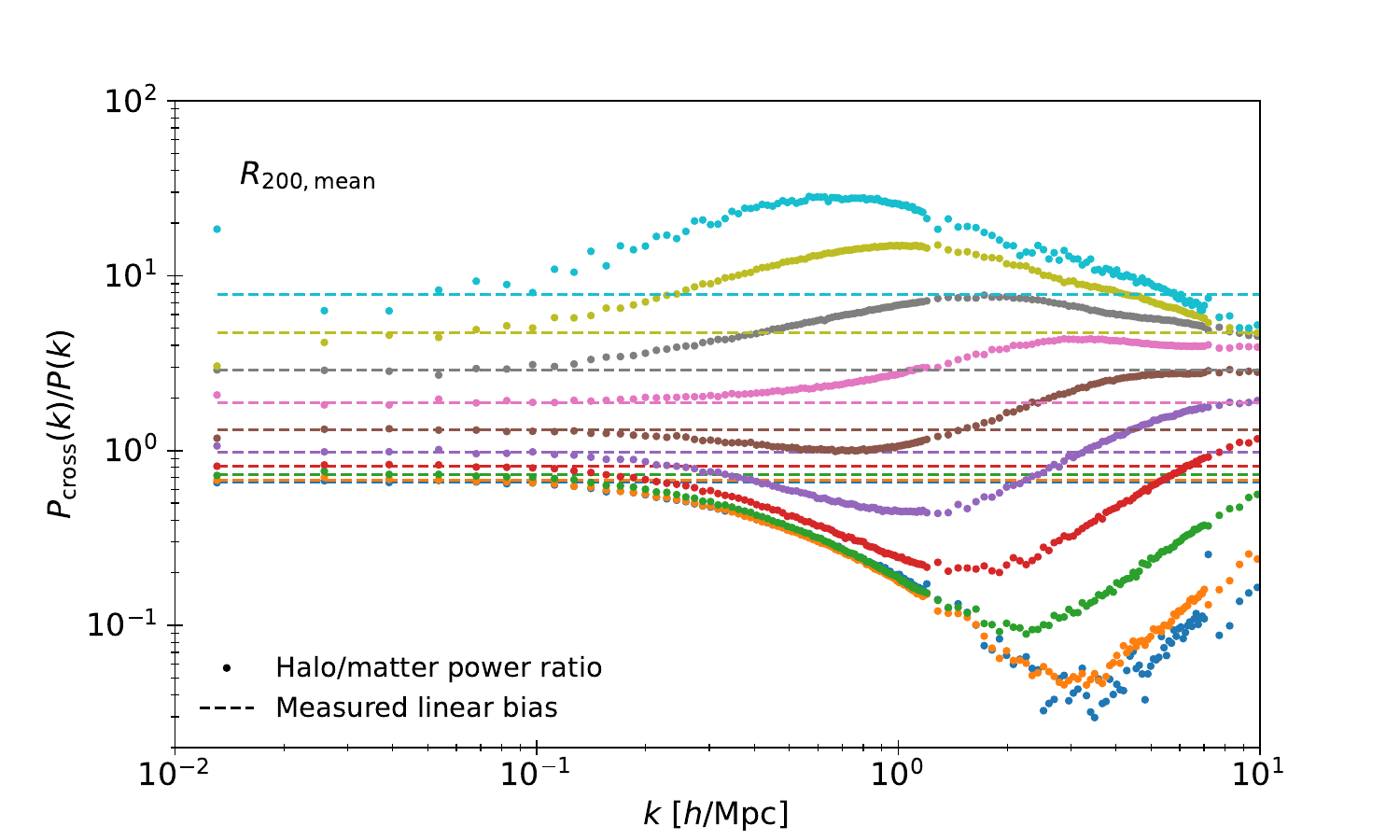}
\includegraphics[width=\columnwidth, trim=8mm 0mm 34mm 12mm]{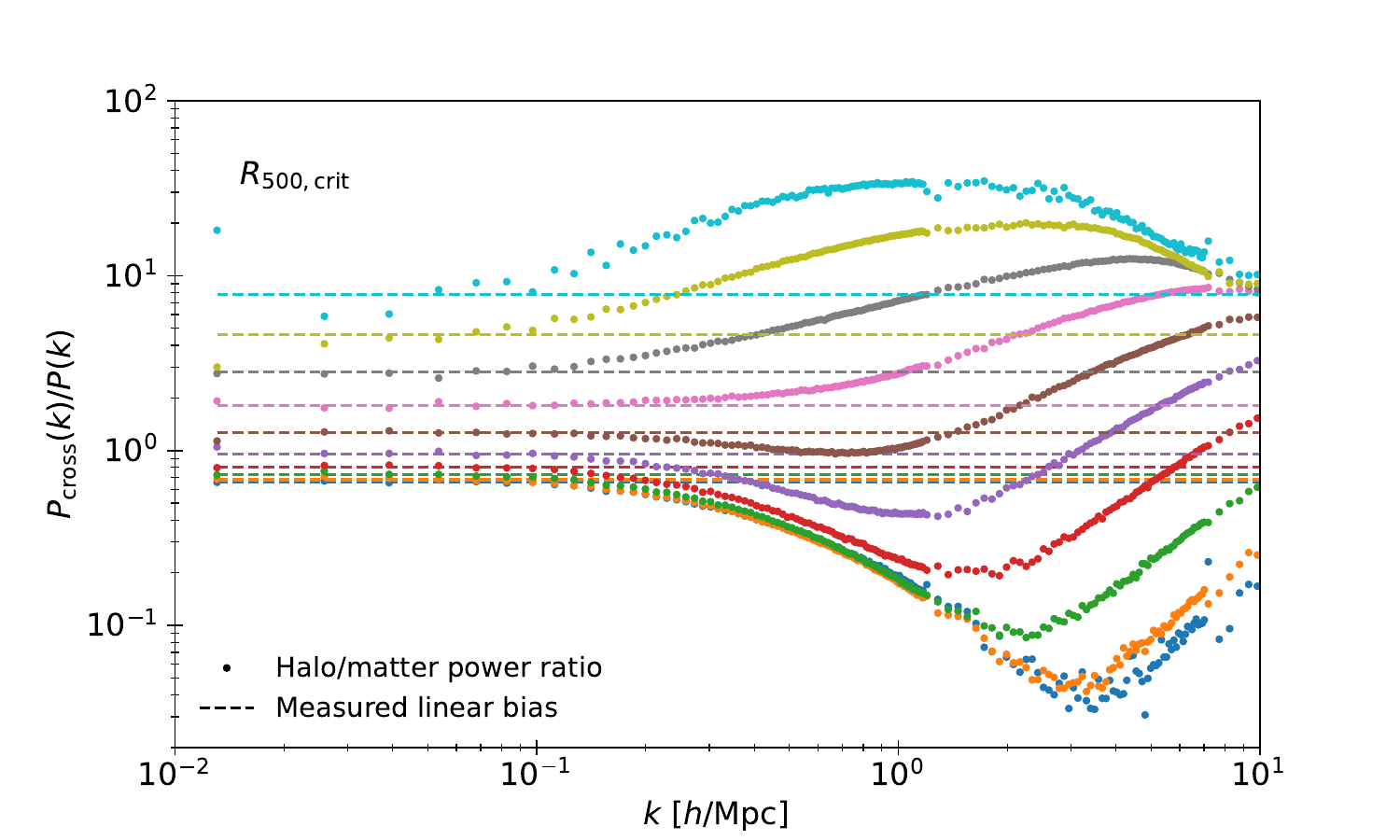}
\\
\includegraphics[width=0.8\textwidth, trim=0mm 16mm 0mm 0mm]{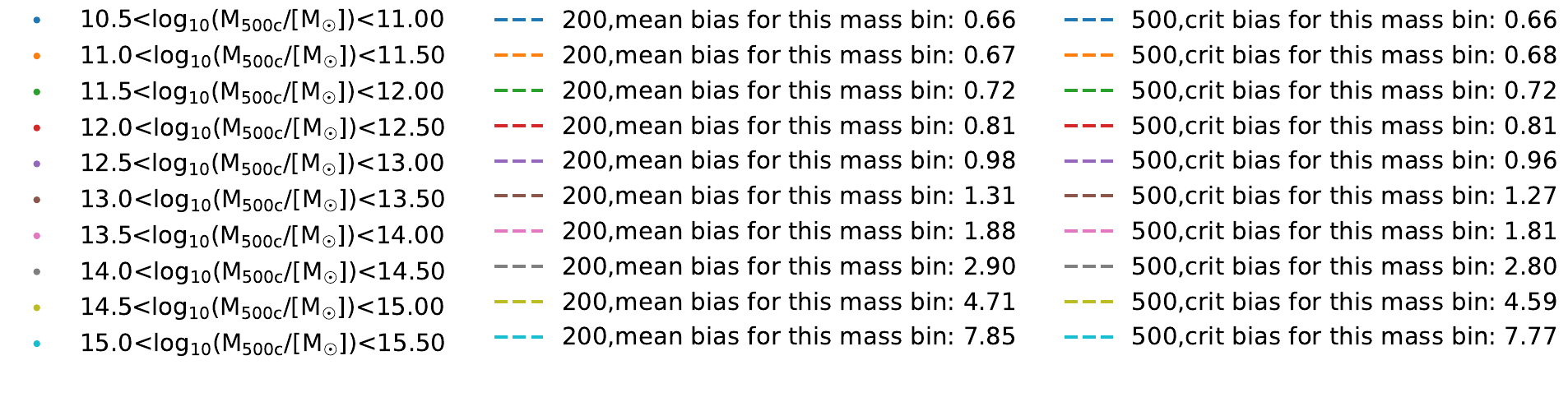}
\caption{The cross power of matter inside haloes of a certain mass and all matter, divided by the total matter power spectrum, for the default L1\_m9\_DMO simulations. On large scales (low $k$), this ratio converges to a constant, the linear halo bias. The bias measured in each halo mass bin is indicated with a dashed line, with the value given in the legend. \textit{Left:} Defining halo matter as the total mass inside its $\mathrm{200m}$ overdensity region, though mass selection is always performed on the $\M$ mass. \textit{Right:} Defining halo matter as the total mass inside its $\mathrm{500c}$ overdensity region.}
\label{fig:bias}
\end{figure*}
We show the ratios of the halo-matter cross spectra of the different $0.5\,\mathrm{dex}$ mass bins with the matter power spectrum of L1\_m9\_DMO, as well as the biases measured from them, in Fig.~\ref{fig:bias}. On the left we use cross spectra of all matter inside $\radm$ of some halo in the mass bin, while on the right we use matter inside $\radc$. As one would expect, the bias is approximately constant on large scales. We record the average value of the ratio for modes $\ksc{<}{0.08}$ -- excluding the first mode -- as the linear halo bias, and indicate these in the figure as dashed lines. The biases are nigh identical for both regions.

\section{Retained mass fraction when only baryonic mass is lost}
\label{sec:app_deriv}
Let us indicate the total mass of a halo in a scenario without mass ejections as $M_\mathrm{tot}$, and the total mass of that same halo in a scenario with mass ejections as $M_\mathrm{tot}'$. We will refer to the halo in the former scenario as the DMO halo, and to the halo in the latter as the baryonic halo. We will assume that the baryon fraction of the DMO halo is $M_\mathrm{bar}/M_\mathrm{tot}=\Omega_\mathrm{b}/\Omega_\mathrm{m}$. That of the baryonic halo is some observable fraction $f_\mathrm{b}=M_\mathrm{bar}'/M_\mathrm{tot}'$. The retained mass fraction of the baryonic halo relative to the DMO halo is $f_\mathrm{ret}=M_\mathrm{tot}'/M_\mathrm{tot}$. We can split the numerator into a CDM and a baryonic component, yielding $f_\mathrm{ret}=(M_\mathrm{cdm}'+M_\mathrm{bar}')/M_\mathrm{tot}$.

If we now assume that only baryonic mass is lost in ejection events, then $M_\mathrm{cdm}'=M_\mathrm{cdm}$. We can then rewrite the retained mass fraction as:
\begin{align}
\nonumber
f_\mathrm{ret}&=\frac{M_\mathrm{cdm}+M_\mathrm{bar}'}{M_\mathrm{tot}}\\
&=1-\frac{M_\mathrm{bar}}{M_\mathrm{tot}}+\frac{M_\mathrm{bar}'}{M_\mathrm{tot}}\\
\nonumber
&=1-\frac{\Omega_\mathrm{b}}{\Omega_\mathrm{m}}+f_\mathrm{b}f_\mathrm{ret}.
\end{align}
Taking the last term on the right-hand side to the left-hand side, we find:
\begin{equation}
(1-f_\mathrm{b})f_\mathrm{ret}=1-\frac{\Omega_\mathrm{b}}{\Omega_\mathrm{m}} \,\,\Rightarrow\,\, f_\mathrm{ret}=\frac{1-\Omega_\mathrm{b}/\Omega_\mathrm{m}}{1-f_\mathrm{b}},
\end{equation}
matching our definition of the corrected baryon fraction in equation~\eqref{eq:fbc}.

\section{Fitted results at higher redshifts}
\label{sec:app_z}
\begin{figure*}
\includegraphics[width=\columnwidth, trim=17mm 6mm 25mm 12mm]{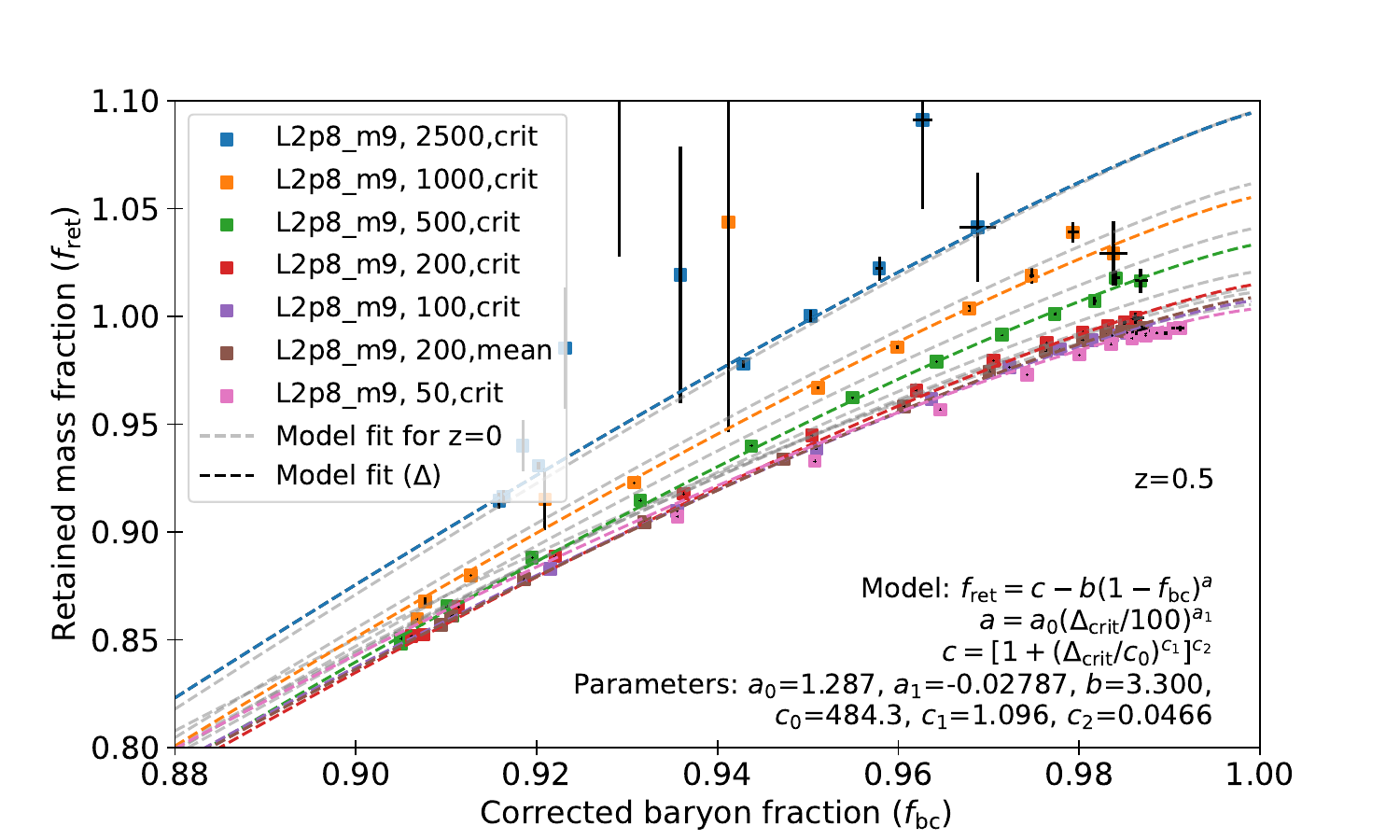}
\includegraphics[width=\columnwidth, trim=8mm 6mm 34mm 12mm]{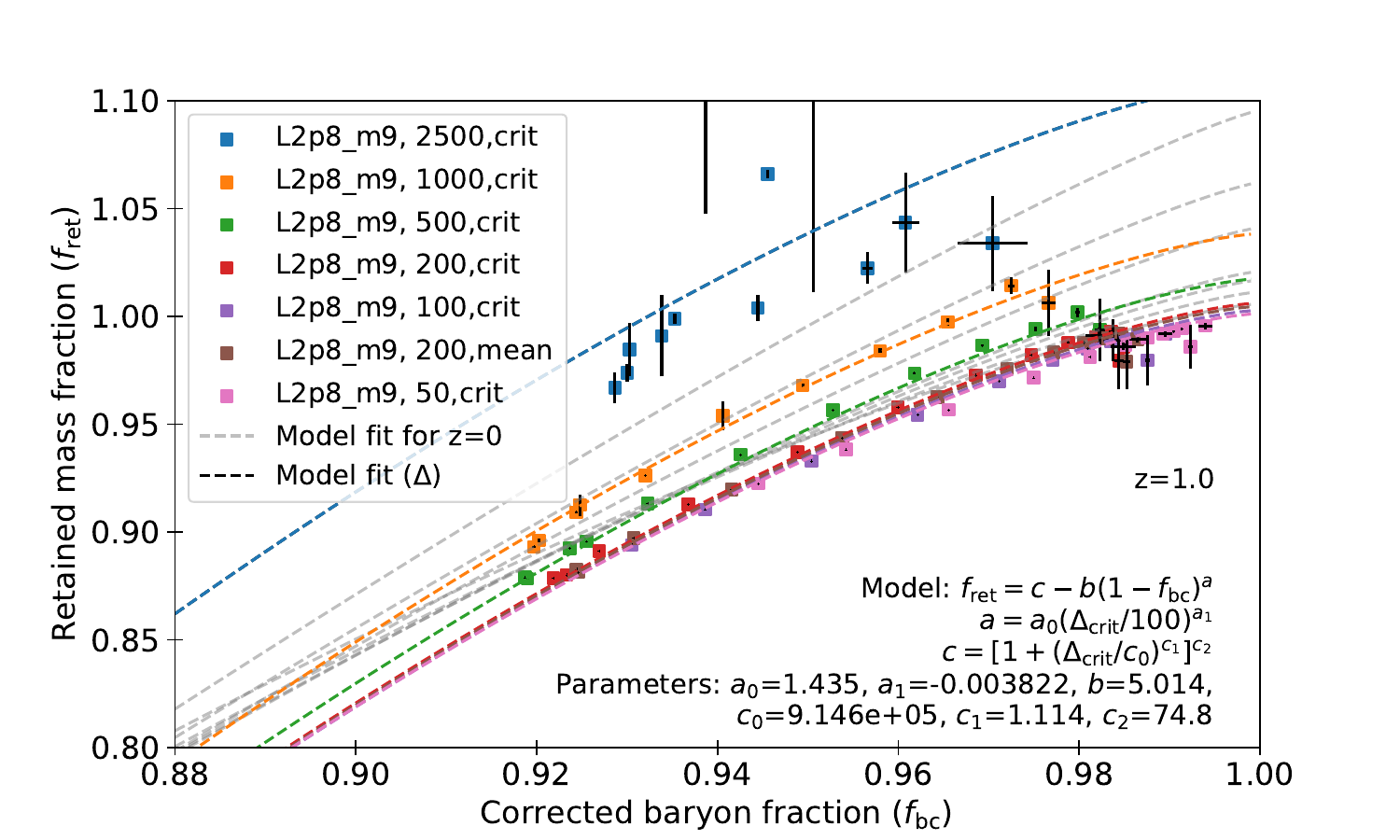}
\caption{As Fig.~\ref{fig:fret_fbc_universal}, but showing the results of fitting the retained mass fraction at $z=0.5$ (left panel) and $z=1$ (right panel). The $z=0$ fits are shown in grey. The retained mass fraction shows only a slight evolution with redshift, generally going down with redshift at fixed corrected baryon fraction, except in the innermost region ($\Delta=\mathrm{2500c}$) where the opposite is true. The universal fit is able to capture this behaviour well.}
\label{fig:fret_fbc_universal_z}
\end{figure*}
In Fig.~\ref{fig:fret_fbc_universal_z} we plot the retained mass fraction against the corrected baryon fractions measured in L2p8\_m9 matched against L2p8\_m9\_DMO for different overdensity regions, as in Fig.~\ref{fig:fret_fbc_universal}, but now for redshifts $z=0.5$ (left) and $z=1$ (panel). While at redshift zero the critical overdensity corresponding to $\radm$ is simply $\Delta=200\Omega_\mathrm{m}$, in general this is $\Delta=200\Omega_\mathrm{m}(1+z)^3/E(z)$, where $E(z)\equiv H(z)/H_0$, which is what we use in fitting the relation here.

Comparing the resulting fits of equations~\eqref{eq:fretained} and \eqref{eq:universal_params} (coloured dashed lines) with those of $z=0$ (dashed grey lines), we see from the left-hand panel that the relation evolves only slightly with redshift. At $z=1$, this evolution is a lot stronger. However, as feedback has had less time to drive gas out of large regions, the covered range in corrected baryon fractions is significantly reduced, which leads to misleadingly large deviations between the $z=1$ and $z=0$ fits at low $f_\mathrm{bc}$. Only the innermost region, $\Delta=\mathrm{2500c}$, truly shows strong evolution with redshift for $z>0.5$, though we note that this region is very poorly resolved for low-mass haloes.

That there exists a redshift dependence at all, hints at other dependencies of the $f_\mathrm{ret}-f_\mathrm{bc}$ relation. \citet{Elbers2025} show that the retained mass fraction, there indicated as $R_\mathrm{b}$, depends on halo concentration and formation time at fixed DMO mass. While the relation presented in this work likely absorbs some of those dependencies, some residual dependence probably remains.

\section{Sensitivity to changes in resolution}
\label{sec:app_resolution}
\begin{figure*}
\includegraphics[width=\columnwidth, trim=17mm 6mm 25mm 12mm]{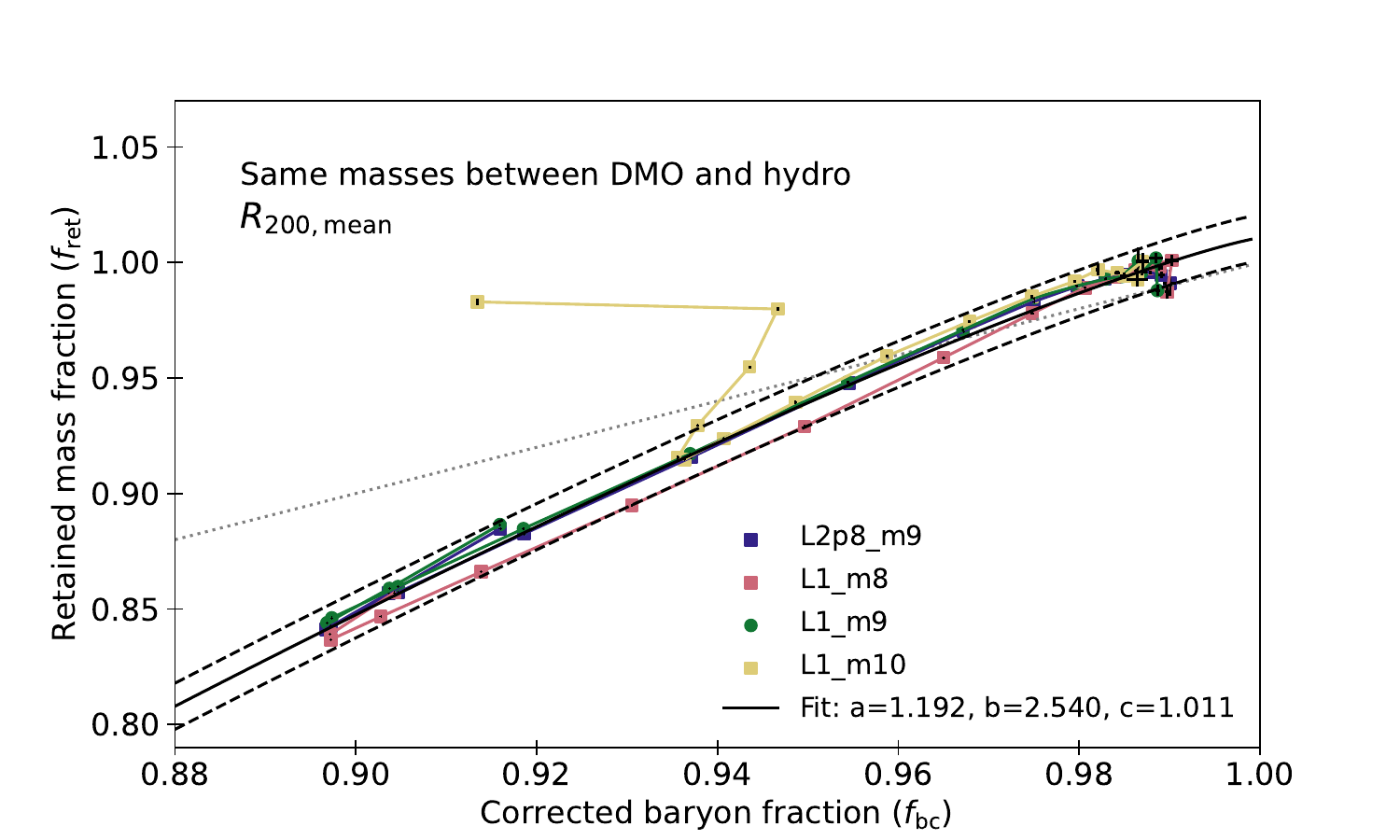}
\includegraphics[width=\columnwidth, trim=8mm 6mm 34mm 12mm]{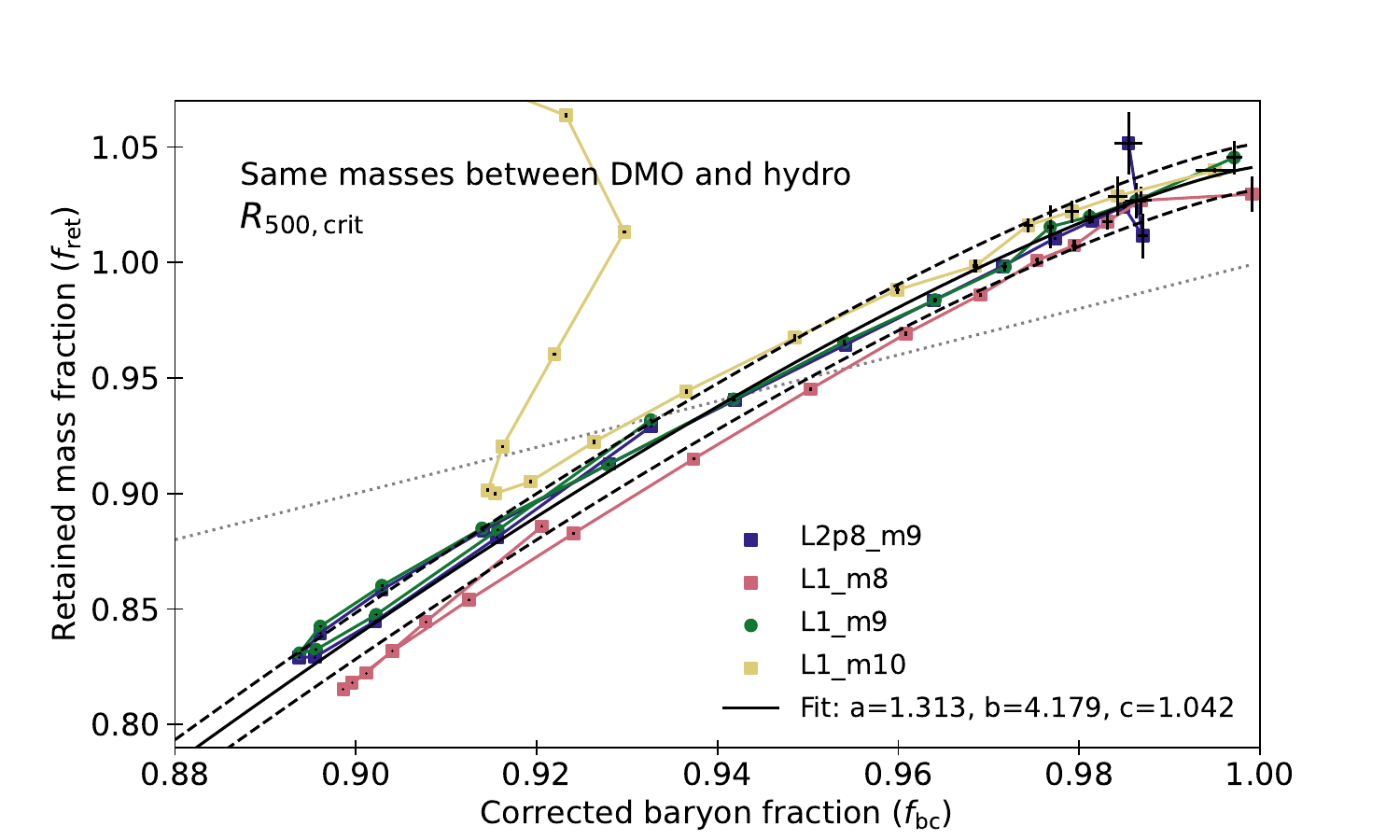}
\caption{As the right panels of Fig.~\ref{fig:fret_fbc_200m_fixedhaloes_fixedmass} and Fig.~\ref{fig:fret_fbc_500c_fixedhaloes_fixedmass}, but now for simulations which vary box size or resolution. A solid line shows a combined fit for equation~\eqref{eq:fretained}. As before, we see that simulations with a different box size at fixed resolution (L1\_m9 and L2p8\_m9) yield practically identical retained mass relations. However, there is a significant dependence on resolution, particularly for $\M$ at all but the highest masses. While the deviation of L1\_m10 at low masses/low $f_\mathrm{bc}$ can be ignored, since the haloes probed there are so poorly resolved, the results for L1\_m8 indicate that the halo relaxation processes are not yet fully converged at the m9 resolution.}
\label{fig:resolution_test_fret}
\end{figure*}
All results presented in this paper were for a fixed simulation mass resolution, called ``m9'' in \flamingo. In Fig.~\ref{fig:resolution_test_fret}, we investigate the effect of changing the resolution on the $f_\mathrm{ret}-f_\mathrm{bc}$ relation at the core of the resummation model. As in the right panels of Fig.~\ref{fig:fret_fbc_200m_fixedhaloes_fixedmass} and Fig.~\ref{fig:fret_fbc_500c_fixedhaloes_fixedmass}, we see that there is no significant difference between simulations that have the same resolution but a different box size: L1\_m9 (green dots) and L2p8\_m9 (blue squares) overlap almost exactly for all halo mass bins. However, when we change the resolution, simulations start to deviate from the relation.

The lower-resolution L1\_m10 simulation agrees with L1\_m9 for large halo masses (large $f_\mathrm{bc}$), but deviates at low mass. This is expected, as low-mass haloes and the relevant physical processes in them are less well resolved. For $\radc$ regions, the differences are larger and occur at higher halo masses than for $\radm$ regions, with the lower-resolution simulation predicting $\approx 1$ percentage point higher retained mass fractions for the same corrected baryon fractions, until the halo mass becomes so low that it deviates from the relation entirely.

The higher-resolution L1\_m8 simulation also agrees for massive haloes. However, as the halo mass decreases it starts to undershoot the m9 simulations by $1$ percentage point consistently for $\radm$ regions. For $\radc$, the deviation is larger, though it still does not exceed $\approx 2$ percentage points at any halo mass. A similar discrepancy between mass ratios in these \flamingo simulations was recently noted by \citet[][their Appendix~B]{Schneider2025}.

\begin{figure*}
\includegraphics[width=\columnwidth, trim=17mm 6mm 25mm 12mm]{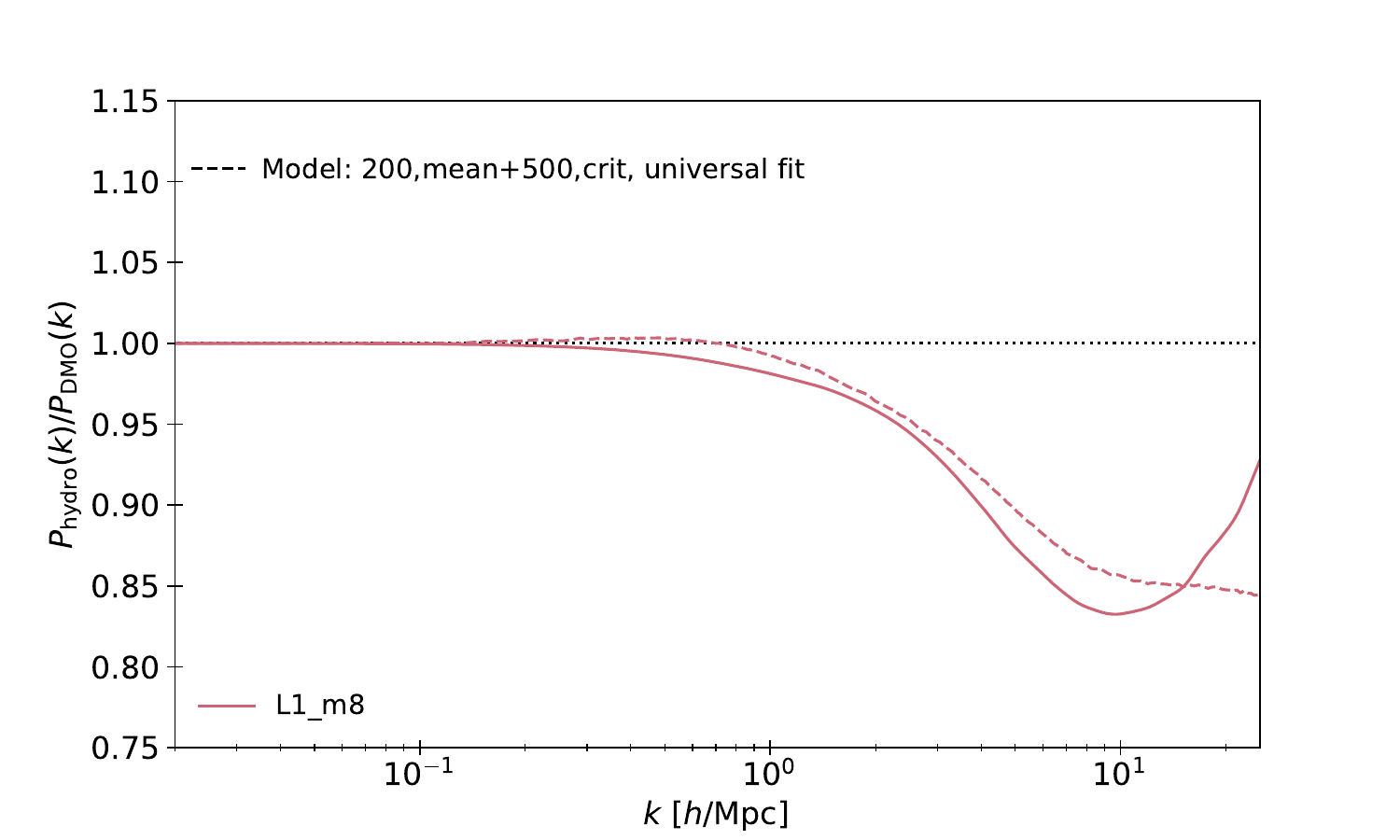}
\includegraphics[width=\columnwidth, trim=8mm 6mm 34mm 12mm]{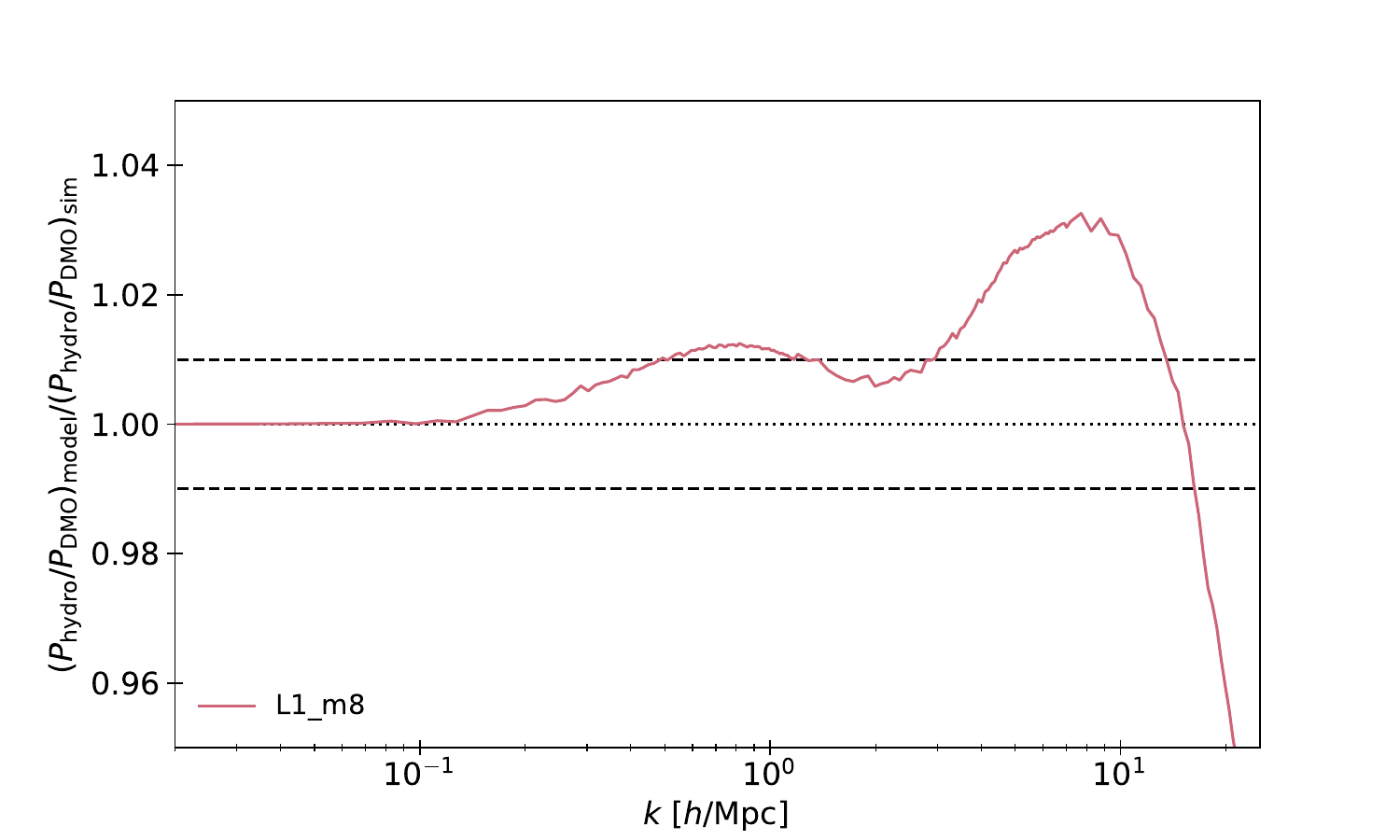}
\caption{The result of applying the $f_\mathrm{ret}-f_\mathrm{bc}$ relation fit to L2p8\_m9 to the higher-resolution L1\_m8 simulation. \textit{Left:} The true suppression signal compared to the model prediction. \textit{Right:} The ratio of the model to true suppression signal. As expected from the results of Fig.~\ref{fig:resolution_test_fret}, the universal relation overpredicts the retained mass fraction for this resolution, although the prediction is still $\lesssim 1\%$ accurate for $\ksc{\lesssim}{3}$.}
\label{fig:resolution_test_dpp}
\end{figure*}

Comparing the same halo mass bins at different resolutions (not shown here), we see that the simulations tend to predict similar retained mass fractions in most bins, but that the corrected baryon fractions move up slightly as the mass resolution improves -- though the retained mass fraction also decreases with resolution at low halo mass. Since simulations with different feedback strengths and cosmologies do follow the same relation within $1$ percentage point of other simulations at fixed resolution, the differences seen here may be due to a resolution dependence of the relevant relaxation processes.

We compare the model prediction for the suppression signal for L1\_m8 to the true one in Fig.~\ref{fig:resolution_test_dpp}, using the $f_\mathrm{ret}-f_\mathrm{bc}$ relation fit to L2p8\_m9. The model tends to underestimate the amount of suppression, as expected from the results of Fig.~\ref{fig:resolution_test_fret}, since the retained mass fraction is overestimated. Still, the prediction is $\lesssim 1\%$ accurate for $\ksc{\lesssim}{3}$, and stays below $4\%$ for all $\ksc{\leq}{10}$.

We note that the resolution is not the only property of the simulation that changes between L1\_m9 and L1\_m8: as noted in \citet{Schaye2023}, two changes were made to the simulation code after most of the \flamingo simulations were completed. First, there was an update to the simulation code that fixed the repositioning of supermassive black holes, to keep them in the centre of their halo's potential. In most \flamingo simulations, the black hole's own potential was erroneously not excluded from the potential calculation (see their \S2.3.5). Second, in these same simulations, particles with exactly zero metallicity had their star formation suppressed, although this only significantly affected the very lowest-mass galaxies (see their footnote 5). Of the simulations explored here, only L1\_m8, L1\_m10 and the jet simulations have the black hole potential correctly subtracted, and the star formation bug fixed. In our results, these are among the most divergent simulations in terms of model performance, which the code changes (particularly the black hole repositioning) could possibly have contributed to.

\section{Using a different DMO simulation}
\label{sec:app_diff_DMO}
\begin{figure*}
\includegraphics[width=\columnwidth, trim=17mm 6mm 25mm 12mm]{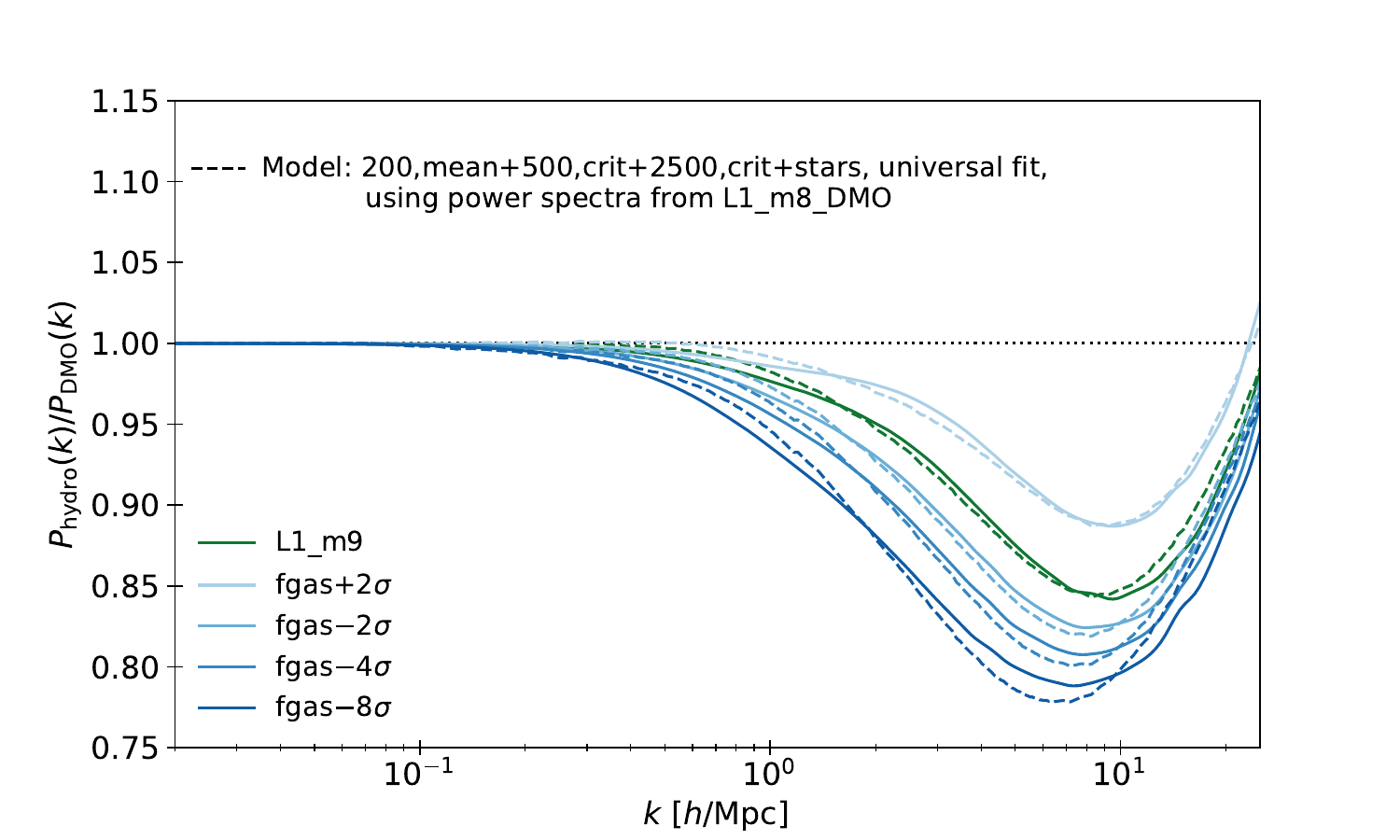}
\includegraphics[width=\columnwidth, trim=8mm 6mm 34mm 12mm]{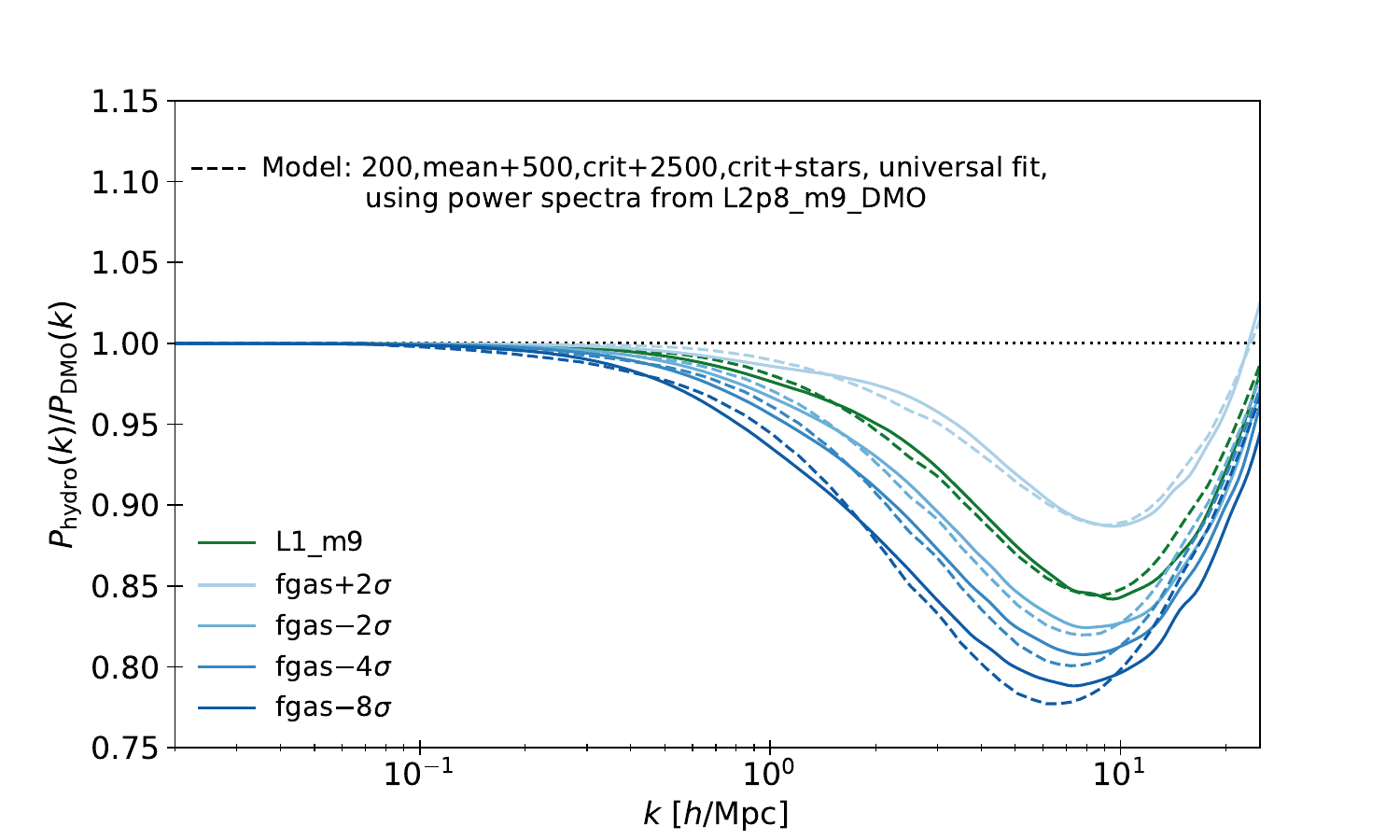}
\caption{The predicted suppression signal for simulations with varying strengths of AGN feedback, but using the power spectra from a different DMO simulation than the one matching the hydrodynamical simulation. \textit{Left:} Using cross power spectra from L1\_m8\_DMO. The cross power spectra probe two mass bins lower than we can reliably measure baryon fractions for, so the corresponding retained mass fractions were set to unity. This does not significantly impact the result. \textit{Right:} Using power spectra from L2p8\_m9\_DMO, which adds one higher-mass bin for which we again set the retained mass fraction to unity, and removes the lowest-mass bin. The results are very similar to those shown in Fig.~\ref{fig:dpp_all_stellar_universal}, despite some of the contributions having been ignored. The L2p8\_m9\_DMO simulation additionally simulates an entirely different patch of universe than L1\_m9\_DMO, showing that there is no significant cosmic variance.}
\label{fig:dpp_AGN_highres}
\end{figure*}
When applying the model to observations, a suitable DMO simulation must be chosen to rescale the cross power spectra of. To demonstrate that the results of the model shown in this work are not due to the DMO simulation being rescaled having the same resolution or initial conditions as the hydrodynamical simulation from which the baryon fractions were determined, we apply the derived retained mass fractions for several simulations to both the higher-resolution L1\_m8\_DMO simulation and the larger-volume L2p8\_m9\_DMO simulation, the latter of which has different initial conditions. The results are shown in Fig.~\ref{fig:dpp_AGN_highres}. For L1\_m8\_DMO, halo cross spectra were calculated for two additional low-mass bins, while for L2p8\_m9\_DMO, one additional high-mass bin was populated while the lowest-mass bin was ignored because of the high computational cost involved. For any mass bin in which baryon fractions were not calculated in L1\_m9, we set the corresponding retained mass fraction to unity. Despite there not being a complete overlap in the halo mass bins for which the baryon fractions were measured and those for which halo cross spectra could be calculated, the results are nearly indistinguishable from those of Fig.~\ref{fig:dpp_all}. For clarity, fewer simulations are shown, but we have checked that the results are comparable for the other feedback variations.

We include the power spectra required to apply the resummation model for L1\_m8\_DMO and L2p8\_m9 in addition to those of the different cosmological variations of L1\_m9\_DMO, as part of the \texttt{resummation} Python package.


\bsp	
\label{lastpage}
\end{document}